\documentclass[pre,amsmath,amssymb,twocolumn, showpacs, superscriptaddress,10pt]{revtex4-2}

\usepackage{amsmath, amsfonts, amssymb}
\usepackage[urlcolor=cyan]{hyperref}
\usepackage{graphicx}
\usepackage{cases}
\usepackage{mathtools}
\usepackage{soul}
\usepackage{upgreek} 
\usepackage{dsfont}
\usepackage{lmodern}
\usepackage{enumitem} 
\usepackage[dvipsnames]{xcolor}
\usepackage{tikz}
\usetikzlibrary{patterns}
\usetikzlibrary{patterns.meta}
\usetikzlibrary{arrows}

\newcommand{\be}{\begin{equation}}
\newcommand{\ee}{\end{equation}}
\newcommand{\bea}{\begin{eqnarray}}
\newcommand{\eea}{\end{eqnarray}}

\newcommand{\I}{\ensuremath{\mathbf{i}}}

\renewcommand{\leq}{\leqslant}
\renewcommand{\geq}{\geqslant}
\newcommand{\EE}{\mathbb E}

\begin{document}

\title{The KPZ equation in a half space with flat initial condition \\
and the unbinding of a directed polymer from an attractive wall} 

\author{Guillaume Barraquand}
\affiliation{Laboratoire  de  Physique  de  l'\'Ecole  Normale  Sup\'erieure,  ENS, CNRS,  Universit\'e  PSL,   Sorbonne  Universit\'e,  Universit\'e de  Paris, 24 rue Lhomond, 75231 Paris, France}

\author{Pierre Le Doussal}
\affiliation{Laboratoire  de  Physique  de  l'\'Ecole  Normale  Sup\'erieure,  ENS, CNRS,  Universit\'e  PSL,  Sorbonne  Universit\'e,  Universit\'e de  Paris, 24 rue Lhomond, 75231 Paris, France}

\date{\today}

\begin{abstract}
   We present an exact solution for the height distribution of the KPZ equation at any time $t$ in a half space with flat initial condition.
   This is equivalent to obtaining the free energy distribution of a polymer of length $t$ pinned at a wall at a single point. In the large $t$ limit a binding transition
   takes place upon increasing the attractiveness of the wall. Around the critical point we find the same statistics as in the Baik-Ben--Arous-P\'ech\'e transition
   for outlier eigenvalues  in random matrix theory. In the bound phase, we obtain the exact measure for the endpoint and the midpoint of the polymer at large time. 
   We also unveil curious identities in distribution between partition functions in half-space and certain partition functions
   in full space for Brownian type initial condition.
\end{abstract}



\maketitle
\section{Introduction.} 
The Kardar-Parisi-Zhang (KPZ) equation \cite{KPZ}, which describes the growth of the height field of an interface driven by white noise in the continuum,
is a paradigmatic example of stochastic non-equilibrium dynamics. It enjoys a remarkable connection to the equilibrium problem of
an elastic line in a random potential, also called directed polymer (DP) \cite{huse1985huse, *kardar1987scaling, *halpinhealy1995kinetic}. In one space dimension, i.e for the DP in dimension $d=1+1$, some exact solutions for the height distribution at all time $t$ have been found in the last ten years. These finite time solutions are valuable since they allow to study the crossover in time from short times, where the growth is in the Edwards-Wilkinson class  \cite{edwards1982surface, hammersley1967harnesses}, to the large time asymptotic
behavior which is common to a large number of systems in the so-called KPZ class \cite{corwin2012kardar,quastel2015one,takeuchi2018appetizer}.
However they have been found only for a few specific 
initial conditions (IC), which are the important ones: the droplet IC (point to point DP) \cite{amir2011probability, dotsenko2010replica, calabrese2010free, sasamoto2010exact}, the flat IC (point to line DP) \cite{calabrese2011exact, ortmann2016exact} and  the Brownian IC (which includes the stationary KPZ) \cite{corwin2013crossover, imamura2011replica, imamura2012exact, imamura2013stationary, borodin2015height}. The KPZ equation on the half-line has also been studied, and is related to
the DP in a half-space with a wall, with a wall parameter $A$, which can be repulsive $A>0$ or attractive $A<0$. { It was found in \cite{kardar1985depinning} that the polymer is bound to the wall for 
$A<-1/2$ and that it unbinds for $A \geq -1/2$ due to the competition with bulk point disorder, a different
mechanism from the usual thermal wetting transition 
\cite{de1985wetting,abraham1980solvable}. It is also different from the 
full space version of the model with a single columnar defect \cite{tang1993directed,basu2014last,soh2017effects}
(slow bond problem) where the DP is always pinned, 
or the case where disorder in only on the column \cite{monthus2000localization,giacomin2006smoothing,toninelli2009localization}.
An experimentally feasible realization of half-line KPZ growth in turbulence liquid
crystal was obtained in \cite{TakeuchiHalf} from a bi-regional geometry with two different growth rates. 
In these types of experiments the aforementioned IC can be easily prepared
\cite{iwatsuka2020direct,takeuchi2012evidence}. 
Although 
the transition} at $A=-1/2$ has been studied in details for other models in  the KPZ class \cite{krug1994disorder, baik2001algebraic, baik2001asymptotics, baik2001symmetrized, baik2018pfaffian, barraquand2018half}, exact finite time solutions for the KPZ equation itself have been
obtained until now only for $A\geqslant -1/2$, for droplet IC \cite{gueudre2012directed, borodin2016directed, barraquand2018stochastic, AlexLD}, and for stationary IC \cite{barraquand2020half}. Furthermore, although it is expected that the height fluctuations are Gaussian at large time in the bound phase, as was found in \cite{deNardisPLDTT} for droplet IC, 
understanding of the fluctuations of the polymer configuration
is still limited, despite the pioneering results of \cite{kardar1985depinning}. 

In this paper, we obtain the ``missing'' exact solution for the KPZ equation in the half-space, that is with flat IC. 
Our solution is valid for any time and any wall parameter $A$, hence it allows for a complete study of the two phases and of the transition. While the
solutions for (i) the flat IC in full space  and (ii) the other IC in half-space, are both complicated,
the combination of flat IC and half-space geometry leads to a remarkable simplification, and to a simpler solution, in terms of a Fredholm determinant. This unveils curious identities in distribution between partition functions in half-space and certain partition functions
   in full space for Brownian type IC.
Around the critical point at $A=-1/2$ we find the same statistics for the height fluctuations at large time as for the outlier eigenvalues in the Baik-Ben Arous-Péché transition \cite{baik2005phase} 
of random matrices. In the bound phase $A < -1/2$, the fluctuations of the height (i.e. the free energy of the polymer) are Gaussian.
To characterize the fluctuations of the polymer configuration we obtain the exact distribution of
its endpoint and of its midpoint for long polymers, and explicit formula for their moments. We predict
an unbinding transition under a force applied to the endpoint. 
Interesting connections with the ground state obtained
in replica Bethe ansatz studies \cite{kardar1985depinning,deNardisPLDTT} of the half-space delta Bose gas are analyzed.

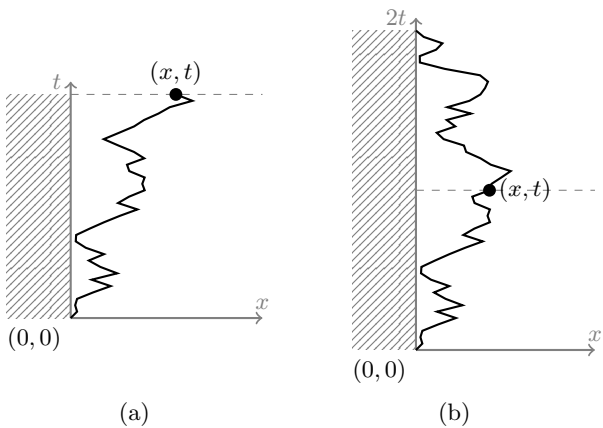
\begin{figure}[ht]
\begin{center}
\begin{tikzpicture}[scale=.85,every text node part/.style={align=center}]
\fill[pattern=north east lines, pattern color=gray] (-1,0) rectangle (0,3.5);
\draw[thick, gray, ->] (0,0) -- (3,0) node[above]{$x$};
\draw[thick, gray, ->] (0,0) -- (0,3.7) node[left]{$t$};
\draw[dashed, gray] (0,3.5) -- (3,3.5);

\draw[thick] (0,0) node[anchor=north east] {$(0,0)$} --  (0.1,0.1) -- (0.07,0.2) -- (0.1,0.3) -- (0.348073,0.4) -- (0.625448,0.5) -- (0.32894,0.6) -- (0.72768,0.7) -- (0.464319,0.8) -- (0.283169,0.9) -- (0.521292,1.) -- (0.235668,1.1) -- (0.0793766,1.2) -- (0.0854228,1.3) -- (0.298271,1.4) -- (0.523455,1.5) -- (0.77003,1.6) -- (1.04369,1.7) -- (0.740922,1.8) -- (0.913704,1.9) -- (1.15725,2.) -- (1.11984,2.1) -- (1.15677,2.2) -- (0.9164,2.3) -- (0.881215,2.4) -- (1.15044,2.5) -- (0.989828,2.6) -- (1.24783-0.5,2.7) -- (1.0175-0.5,2.8) -- (1.25723-0.5,2.9) -- (1.49304-0.5,3.) -- (1.64798-0.5,3.1) -- (1.87179-0.5,3.2) -- (2.04898-0.5,3.3) -- (2.40323-0.5,3.4) -- (2.14578-0.5,3.5)  node[anchor=south] {$(x,t)$};
\fill (2.14578-0.5,3.5) circle(0.1);

\begin{scope}[xshift=5.4cm, yshift=-0.5cm]
\fill[pattern=north east lines, pattern color=gray] (-1,0) rectangle (0,5);
\draw[thick, gray, ->] (0,0) -- (2.8,0) node[above]{$x$};
\draw[thick, gray, ->] (0,0) -- (0,5.2) node[left]{$2t$};
\draw[dashed, gray] (0,2.5) -- (2.8,2.5);
\draw[thick] (0,0) node[anchor=north east] {$(0,0)$} --  (0.1,0.1) -- (0.07,0.2) -- (0.1,0.3) -- (0.348073,0.4) -- (0.625448,0.5) -- (0.32894,0.6) -- (0.72768,0.7) -- (0.464319,0.8) -- (0.283169,0.9) -- (0.521292,1.) -- (0.235668,1.1) -- (0.0793766,1.2) -- (0.0854228,1.3) -- (0.298271,1.4) -- (0.523455,1.5) -- (0.77003,1.6) -- (1.04369,1.7) -- (0.740922,1.8) -- (0.913704,1.9) -- (1.15725,2.) -- (1.11984,2.1) -- (1.15677,2.2) -- (0.9164,2.3) -- (0.881215,2.4) -- (1.15044,2.5)  node[anchor=west] {$(x,t)$};
\draw[thick] (0,5)  -- (0.121845,4.9) -- (0.414127,4.8) -- (0.314959,4.7) -- (0.059995,4.6) -- (0.0616275,4.5) -- (0.451204,4.4)  -- (1.00224,4.3) -- (1.12661,4.2) -- (1.08153,4.1) -- (1.00892,4) -- (0.836694,3.9) -- (0.491767,3.8) -- (0.860705,3.7) -- (0.480596,3.6) -- (0.614963,3.5) -- (0.325605,3.4) -- (0.433176,3.3) -- (0.751403,3.2) -- (0.784852,3.1) -- (1.00935,3.0) -- (1.17107,2.9) -- (1.48252,2.8) -- (1.33514,2.7) -- (1.18167,2.6) -- (1.15044,2.5); 
\fill (1.15044,2.5) circle(0.1);
\end{scope}

\begin{scope}[yshift=-1.5cm]
\draw (1,0) node{(a)};
\draw (6,0) node{(b)};
\end{scope}
\end{tikzpicture}
\end{center}
\caption{(a) Directed polymer path in $1+1$ dimensions in a half-space, pinned at the boundary wall at the point $(0,0)$, with free endpoint $(x,t)$ (point to line problem)
in presence of a bulk white noise random potential $\sqrt{2} \xi(x,t)$. We will discuss the fluctuations of the position of the polymer endpoint. By reversing time, this corresponds to the line to point partition function $Z_A^f(y=0,t)$ in  \eqref{ZAfree}, which  maps to the KPZ field at time $t$ with flat IC. (b) We will also consider the position of the midpoint of a DP with both endpoints pinned at the boundary.}
\label{fig:halfspacepolymer}
\end{figure}


This paper is organized as follows. In Section \ref{sec:halfspaceKPZ} we recall the definitions of the KPZ equation on the half-line 
and of the related model of the continuum directed polymer in the half space. 
In Section \ref{sec:flatinitialcondition}
we derive the exact solution for all times of the KPZ equation with flat initial conditions. Some details are provided in the Appendix \ref{appendix:detailsflat}.
We also obtain the large time asymptotics, the details being provided in Appendix \ref{appendix:asymptotics}. 
In Section \ref{sec:stationaryendpoint}
we study the stationary measure of the KPZ equation on the half-line and apply it to obtain a detailed
description of the statistics of the endpoint in the polymer problem. The connection with the
replica method is given in Appendix \ref{appendix:betheansatz} and the calculations of the mean
endpoint probability and its correlations using Liouville quantum mechanics are described in Appendix \ref{appendix:Liouville}. 
In Section \ref{sec:identities} we generalize the identity in distribution obtained in Section \ref{sec:flatinitialcondition}, 
relating solutions of half-space KPZ equation to solutions with the full-space KPZ equation with different initial data, some of the details are presented in Appendix \ref{appendix:generalization}. The matching with full space KPZ equation distributions uses the statistical tilt symmetry recalled in Appendix \ref{appendix:sts}. In Section \ref{sec:universality} we point out the features which we believe are universal
near the unbinding transition and in relation to the conjectured half-space KPZ fixed point.

\section{Half-space KPZ equation}
\label{sec:halfspaceKPZ}

Let us recall the KPZ equation for the height $h(x,t)$ field of an interface
\be \label{eq:KPZhalfspace}
\partial_t h(x,t) = \nu \partial_x^2 h + \frac{\lambda}{2} (\partial_x h)^2 + \sqrt{D} \xi(x,t)
\ee
where $\xi(x,t)$ is a space-time white noise.
We use space-time units so that $\nu=1$ and $\lambda=D=2$. Here we study the problem in a half-line $x \geq 0$
with boundary conditions $\partial_x h=A$ depending on a parameter $A$. From the Cole-Hopf mapping it can be equivalently defined
as $h(x,t)=\log Z_A(x,t)$ where $Z_A(x,t)$ satisfies the stochastic heat equation (SHE)
\be
\partial_t Z_A(x,t) = \partial^2_{x} Z_A(x,t) + \sqrt{2} Z_A(x,t) \xi(x,t), \;x\geqslant 0, 
\label{eq:mSHEhalf-space}
\ee 
with boundary condition $\partial_x Z_A(x,t)=A Z_A(x,t)$ at $x=0$ and as yet
unspecified initial condition at $t=0$. 

Let us denote by $Z_A(x,t\vert y,0)$ the partition function of a continuous DP of length $t$ in a white noise random potential in dimension $d=1+1$ (at unit temperature), in the half-space $x \geq 0$, with endpoints at $(y,0)$ and $(x,t)$, see Fig. \ref{fig:halfspacepolymer}. In particular $Z_A(x,t\vert y,0)$ satisfies the equation \eqref{eq:mSHEhalf-space} with initial condition at $t=0$ given by a delta mass at the point $y$.

\section{Flat initial condition}
\label{sec:flatinitialcondition}
\subsection{Finite-time solution}
In this Section, we are interested in the solution of the KPZ equation \eqref{eq:KPZhalfspace} with a flat initial condition $h(x,0)=0$. This is equivalent
to studying \eqref{eq:mSHEhalf-space} with $Z_A(x,0)=1$, i.e. a polymer with one fixed endpoint at $(y,t)$ and one free endpoint, of partition function
\be  \label{ZAfree} 
Z_A^f(y,t) = \int_{0}^{\infty} Z_A(y,t\vert x,0)dx.
\ee
the KPZ field being retrieved as $h(y,t)= \log Z^f_A(y,t)$, where the superscript $f$ stands for flat IC. 

Consider now the case where the fixed endpoint is at the position of the wall $y=0$. We will calculate the
moments of $Z_A^f(0,t)$ which will allow us to obtain an expression for the Laplace transform of its distribution.
From \eqref{ZAfree} and by symmetry, we can write the $n$-th integer moment of the partition sum as
\begin{equation}
\mathbb E\left[Z_A^f(0,t)^n\right] = n! \int_{ x_1 \geq \dots \geq x_n\geq 0} \mathbb E \left[\prod_{i=1}^n Z_A(x_i,t\vert 0,0)\right] . 
\end{equation}
where here and below, $\mathbb E$ denote expectation with respect to the noise $\xi$. 
As shown in \cite{borodin2016directed} the moments appearing in the RHS 
can be expressed as a multiple contour integral. 
We have that, for general endpoint positions $x_1\geq \dots \geq x_n \geq 0$, 
\begin{multline}
\mathbb E\left[\prod_{i=1}^n Z_{A}(x_i,t\vert 0,0)\right] = 2^n
\int_{r_1+\I\mathbb R}\frac{\mathrm{d}z_1}{2\I\pi} \dots \int_{r_n+\I\mathbb R}\frac{\mathrm{d}z_n}{2\I\pi}   \\ 
\prod_{i=1}^n \frac{z_i}{z_i+A}  e^{tz_i^2 - x_i z_i} \!\!\!\prod_{1\leqslant a<b\leqslant n} \frac{z_a-z_b}{z_a-z_b-1}\, \frac{z_a+z_b}{z_a+z_b-1},
\label{eq:nestedmoments}
\end{multline}
where the contours are chosen so that $r_1>r_2+1>\dots > r_n+n-1>\max\lbrace n-1-A, n-1\rbrace$, i.e.
all contours are to the right of $-A$. To obtain the result for the flat initial condition we must now integrate over the endpoints $x_i$
over the positive real axis. To this aim we use the identity
\begin{equation}
\int_{ x_1 \geq \dots \geq x_n\geq 0} \prod_{i=1}^n e^{-x_i z_i} = \prod_{i=1}^n \frac{1}{z_1+\dots+z_i},
\label{eq:integration}
\end{equation}
which is a convergent integral, since all ${\rm Re}(z_i) >0$ from our choice of contour.
One thus obtains 
\begin{multline}
\mathbb E\left[Z_{A}^f(0,t)^n\right] = n! 2^n \int_{r_1+\I\mathbb R}\frac{\mathrm{d}z_1}{2\I\pi} \dots \int_{r_n+\I\mathbb R}\frac{\mathrm{d}z_n}{2\I\pi} \\ \prod_{1\leqslant a<b\leqslant n} \frac{z_a-z_b}{z_a-z_b-1} F(\vec z),
\label{eq:nestedmoments2}
\end{multline}
where 
\begin{equation}
    F(\vec z) = \prod_{1\leq a<b\leq n}\frac{z_a+z_b}{z_a+z_b-1}\prod_{i=1}^n \frac{e^{tz_i^2} }{z_1+\dots+z_i} \frac{z_i}{z_i+A}.
\end{equation}
It is convenient to 
deform all the contours to the single contour $r+\I\mathbb R$ with $r> \max(-A,0)$. 
During the 
deformation, one encounters many poles of the integrand whose residues need to be taken into account. 
This is done in a systematic way using \cite[Proposition 5.1]{borodin2016directed} (see also \cite{borodin2014macdonald}). We obtain for \eqref{eq:nestedmoments2}
\begin{multline}
   \sum_{\ell=1}^n  \frac{n! 2^n}{\ell !} \sum_{\vec m}  \prod_{i=1}^{\ell}\int_{r+\I\mathbb R} \!\!\!\frac{dw_i}{2\I\pi} 
\det\left(\frac{1}{w_i+m_i-w_j}\right)_{i,j=1}^{\ell}  \\
\times 
E(w_1,w_1+1,\dots,w_1+m_1-1,\dots,w_\ell,\dots,w_{\ell}
+m_{\ell}-1)
\label{eq:determinantalmoments}
\end{multline}
where the sum over $\vec m$ is a sum  over integers $m_i \geq 1$  with $\sum_{i=1}^{\ell} m_i =n$, and
\bea
E(\vec z) = \sum_{\sigma \in S_n} \prod_{1 \leq b < a \leq n} \frac{z_{\sigma(a)}- z_{\sigma(b)}-1}{z_{\sigma(a)}- z_{\sigma(b)}}
F(\sigma(\vec z)) .
\label{eq:symetrization}
\eea 
This sum over the symmetric group can be simplified. Note that the factor $\prod_{1\leq a<b \leq n}\frac{z_a+z_b}{z_a+z_b-1}$ in $F(\vec z)$ is symmetric, so it can be factored out. The remaining symmetrization can be performed as in the solution for the full space flat initial condition \cite{le2012kpz} 
(see also \cite{lee2010distribution}) where it was found that 
\begin{multline}
 \sum_{\sigma \in S_n} \prod_{1 \leq b < a \leq n}  \frac{z_{\sigma(a)}- z_{\sigma(b)}-1}{z_{\sigma(a)}- z_{\sigma(b)}} \prod_{i=1}^n\frac{1}{z_{\sigma(1)}+\dots +z_{\sigma(i)}} \\= \prod_{1\leq a<b\leq n} \frac{z_a+z_b-1}{z_a+z_b} \prod_{i=1}^n \frac{1}{z_i}. 
 \label{eq:symmetrizationperformed}
\end{multline}
Hence, the products over $a<b$ perfectly cancel each other, and we are left with the remarkably simple expression
\begin{equation} \label{product} 
    E(\vec z) = \prod_{i=1}^n \frac{e^{tz_i^2}}{z_i+A}. 
\end{equation}
Due to the product structure of the function $E$, it can be factored inside the determinant in \eqref{eq:determinantalmoments}. 
This leads to an explicit formula for the integer moments, which we sum over $n$ to obtain 
the moment generating series, leading to the following Fredholm determinant expression 
(see Appendix \ref{appendix:detailsflat} for details) for $u>0$ 
\begin{equation}
\mathbb E[e^{-u Z_A^f(0,t)e^{\frac{t}{12}}}] = \det(I-K_{u,t})_{\mathbb L^2(0,+\infty)},
\label{eq:Laplacetransform}
\end{equation}
where the kernel is given by
\begin{align} \label{eq:KernelFinitetime} 
K_{u,t}(v,v') &= \int_{\mathbb R}  \frac{2u \; \mathrm d r}{ e^{-r} +2u} 
\phi_{A,t}(v+r) \psi_{A,t}(v'+r), \\
\phi_{A,t}(v) &=  \int_{a_z+\I\mathbb R}\frac{dz}{2\I\pi} 
 \frac{e^{t \frac{z^3}{3} -vz} }{\Gamma (A+\frac{1}{2} + z )}, \label{eq:PhiA}\\
\psi_{A,t}(v) &= 
\int_{\mathcal C_{a_w}}\frac{dw}{2\I\pi} e^{-t \frac{w^3}{3} 
+ v w} \Gamma (A+ \tfrac{1}{2} +w).
\label{eq:PsiA}
\end{align}
The contour for $z$ is a vertical line with real part $a_z>0$, and the contour for $w$, denoted $\mathcal C_{a_w}$, is the union of two semi-infinite rays leaving the point $ a_w>-\left(A+\frac{1}{2} \right)$ in the direction $\pm 2\pi/3$ to ensure convergence. This expression is one of our main result and is valid for any value of the wall parameter $A$
and for all time $t>0$. 

\subsection{Large-time limit}
\label{sec:largetimelimit}
From the Laplace transform formula one can extract the probability density function (PDF) of the KPZ height field $h(0,t)=\log Z_A^f(0,t)$ at arbitrary time.
Let us now discuss its large time limit, which depends on the value of $A$. The height 
takes the form as $t \to +\infty$
\be
h(0,t) = \log Z_A^f(0,t)  \simeq v_\infty(A) t + t^\beta \chi 
\ee
where the free energy per unit length exhibits a transition 
\begin{equation}
    v_{\infty}(A) = \begin{cases} - \frac{1}{12} &\text{when }A\geqslant - \frac{1}{2},\\
    - \frac{1}{12} + \left(A+\frac{1}{2} \right)^2 &\text{when }A< - \frac{1}{2}, \end{cases}
    \label{eq:vinfty}
\end{equation}
$\chi$ is an $O(1)$ random variable, and $\beta$ the growth fluctuation 
exponent
\be
\beta= \frac{1}{3} ~ \text{for} ~ A \geq - \frac{1}{2}, \quad  \quad\beta= \frac{1}{2} ~ \text{for} ~ A < - \frac{1}{2}
\ee
Let us turn to the distribution of $\chi$. 

\paragraph{Case $A>\frac{-1}{2}$:} We scale $u$ as $u=e^{- t^{1/3}s}$ with fixed $s$, so that 
\begin{equation}
\lim_{t\to\infty} \mathbb E[e^{-u Z_A^f(0,t)e^{\frac{t}{12}}}]  =  \mathbb P(\chi\leq s).
\end{equation}
The limit of the Fredholm determinant in \eqref{eq:Laplacetransform} is obtained by the change of variables $v=t^{1/3}\tilde v$, $v'=t^{1/3}\tilde v'$ and $r=t^{1/3}\tilde r$ in \eqref{eq:KernelFinitetime}, $z=t^{-1/3}\tilde z$ in \eqref{eq:PhiA}, and $w=t^{-1/3}\tilde w$ in \eqref{eq:PsiA} so that 
 \begin{equation} \label{eq:KAi} 
\lim_{t\to\infty} t^{\frac{1}{3}} K_{u,t}(t^{\frac{1}{3}} \tilde v, t^{\frac{1}{3}} \tilde v' ) = \int_{s}^{\infty} \mathrm{Ai}(\tilde r+\tilde v)\mathrm{Ai}(\tilde r+\tilde v')d\tilde r,
\end{equation}
leading to
\be \label{eq:F2}
\mathbb P(\chi\leq s) = \det(I-K_{\rm Ai})_{\mathbb L^2(s,+\infty)} = F_2(s)
\ee 
where $K_{\rm Ai}$ is the Airy kernel and $F_2(s)$ is the cumulative distribution function (CDF) of the Tracy-Widom distribution for the largest eigenvalue of a GUE random matrix.

\paragraph{Case  $A<\frac{-1}{2}$: } The condition that $a_{w}>-(A+\frac 1 2 )$ in \eqref{eq:PsiA} forbids to use the same change of variables. In this case we scale $u=e^{-(A+\frac 1 2)^2 t -t^{1/2}s}$, and use the change of variables $v=t^{1/2} \tilde v$ (and likewise for the variables $v'$), $r=(A+\frac 1 2 )^2 t+t^{1/2} \tilde r$, $z=-(A+\frac 1 2 ) + t^{-1/2} \tilde z$ in  \eqref{eq:PhiA} and in  \eqref{eq:PsiA} we evaluate the integral by residues (the residue at $w=-(A+\frac 1 2)$ is dominant). We obtain 
\begin{equation}
\lim_{t\to\infty} t^{\frac{1}{2}} K_{u,t}(t^{\frac{1}{2}} \tilde v, t^{\frac{1}{2}} \tilde v' ) =
\frac{1}{\sqrt{ 4 \pi |A+\frac{1}{2}|}} e^{- \frac{(s + \tilde v)^2}{4 |A+\frac{1}{2}|} },
\end{equation}
which implies that $\chi$ has a Gaussian distribution with variance $2\left\vert A+\frac 1 2 \right\vert$, see details in  Appendix \ref{appendix:asymptotics}. 

\paragraph{Near the critical point}: We may also scale $A$ close to the critical point as $A=\frac{-1}{2}+ a t^{-1/3}$. The asymptotics are similar as in the case $A>\frac{-1}{2}$, except that (see Appendix \ref{appendix:asymptotics})
\be
\mathbb P(\chi\leq s) = \det(I-K^{\rm BBP}_a)_{\mathbb L^2(s,+\infty)} = F^{\rm BBP}_a(s)
\label{eq:defBBP}
\ee 
where the CDF $F^{\rm BBP}_a(s)$ was introduced in \cite{baik2005phase} and governs the fluctuations of the eigenvalues of spiked Hermitian matrices. It was also found to arise in the context of the KPZ universality class in full-space models for half-Brownian type IC \cite{corwin2013crossover, imamura2011replica, borodin2012free} and in other contexts \cite{baik2005phase, baik2006painleve, barraquand2014phase, krajenbrink2020tilted}. In particular, for $a=0$, $F^{\rm BBP}_0(s)=(F_1(s))^2$ where $F_1$ is the Tracy-Widom distribution function for the largest eigenvalue of a GOE random matrix.

\section{Stationary endpoint distribution}
\label{sec:stationaryendpoint}
\subsection{Endpoint distribution}
The previous results describe the behavior of the partition function of a polymer of arbitrary length $t$ in a white noise random potential,
with one endpoint fixed
at $x=0$ and another endpoint free to move, see Eq. \eqref{ZAfree}. It is natural to ask about the distribution of the distance
of the endpoint to the wall. This information is contained in the endpoint PDF $\mathcal P_{A}(x,t)$ in a given disorder realization, i.e. 
\be
{\mathcal P}_{A}(x,t) = \frac{Z_{A}(x,t|00)}{\int dy Z_{A}(y,t|00) }  = \frac{Z_{A}(x,t|00)}{Z^f_{A}(0,t) } 
\ee 
and in its average $P_A(x,t)=\mathbb E[\mathcal P_A(x,t)]$. Direct calculation of this quantity is not available, however
one can obtain it in the limit of a large polymer length $t$ in the bound phase $A < - \frac{1}{2}$. 
For a large class of IC (see below),  the distribution of ratios $Z_A(x,t)/Z_A(y,t)$ converges at large time $t$ to a stationary distribution of polymer partition function ratios such that 
\begin{equation}
x\mapsto \frac{Z_A(x,t)}{Z_A(0,t)} = x\mapsto e^{\mathcal B(x)+(A+\frac{1}{2}) x}. 
\label{eq:stationary}
\end{equation}
It is stationary in the sense that if at a time $t$ the field of partition function ratios is distributed as \eqref{eq:stationary} where $\mathcal B(x)$ is a standard Brownian motion, then at any later time $\tilde t$, the field of partition function ratios will still be distributed as \eqref{eq:stationary}, with a new Brownian motion $\tilde{\mathcal B}(x)$ depending nontrivially on $\mathcal B(x)$ and the disorder $\xi(x,s)$ for $t<s<\tilde t$. Therefore, at large time, 
\begin{equation}
  \lim_{t\to\infty}\mathcal P_{A}(x,t)  = p_A(x) := \frac{e^{\mathcal{B}(x)+(A+\frac{1}{2})x}}{\int_0^{+\infty} dy e^{\mathcal{B}(y)+(A+\frac{1}{2})y}}, 
  \label{eq:gibbs1} 
\end{equation}
in the sense that both sides have the same multipoint distribution.

This result allows to obtain formulas for the moments of the endpoint position which become time independent
at large $t$ for fixed $A<-1/2$. One denotes the thermal
average in a given disorder configuration as $\langle O(x) \rangle = \int_0^{+\infty} dx O(x) {\cal P}_{A}(x,t)$,
and the thermal cumulants as usual e.g. $\langle x^2 \rangle_c=\langle x^2 \rangle - \langle x \rangle^2$. Interpreting $p_A(x)$ in \eqref{eq:gibbs1}
as the Gibbs measure of a particle (the endpoint) in a
1d Brownian random potential (at unit temperature), it is natural to introduce \cite{monthus2004low}
$\mathsf Z(v) = \int_0^{+\infty} dy e^{\mathcal{B}(y)+vy}$, for $v<0$, the generating function
of the thermal cumulants, such that $\langle x^p \rangle_c =  \partial_v^p \log \mathsf Z(v)|_{v=A+\frac{1}{2}}$.
It is well-known that the random variable $\mathsf Z(v)$ is distributed as the inverse of a Gamma variable,
$\mathsf Z(v) = 1/\Gamma(-2 v,\frac{1}{2})$ \cite{BouchaudPLD1990,Dufresne1990}. In particular, using
$\mathbb E[\log \mathsf Z(v)]= \log 2 - \psi(-2 v)$,
one obtains the disorder averaged thermal cumulants of the polymer endpoint 
\be  \label{eq:gencum}
 \mathbb E [ \langle x^p \rangle_c ] = - (-2)^p \psi^{(p)}(-2 A-1) 
\ee
for $p \geq 1$, where $\psi(z)$ is the digamma function, e.g. 
\bea
&& \mathbb E [ \langle x \rangle ] =  2 \psi'(-2 A -1) \simeq \frac{2}{(2 A+1)^2},  \label{cum1} \\
&& \mathbb E[\langle x^2 \rangle_c] = - 4 \psi''(-2 A -1) \simeq \frac{8}{|2 A+1|^3},  \label{cum2}
\eea
where we indicated the leading behavior for $A\to -1/2^-$, using $\psi(x) \sim \frac{-1}{x}$ at small $x$.
In the bound phase but near the transition, i.e. for $\epsilon=-(A+\frac{1}{2})>0$ and small,
the endpoint wanders very far. In terms of the rescaled endpoint position $y=\epsilon^2 x$ one has
$e^{\mathcal B(x) - \epsilon x} = e^{ \frac{1}{\epsilon} (\tilde{\mathcal B}(y) - y ) }$,
i.e. $\epsilon$ can be interpreted as an effective temperature which tends to zero.
The PDF of $y$ thus concentrates
around the optimum
$y_m = {\rm argmax}_{z>0} (\tilde{\mathcal B}(z) - z )$. The explicit PDF of $y_m$ is known \cite{ProbaMax}, and reads
$P(y) = \sqrt{\frac{2}{\pi y}} e^{-y/2} - {\rm Erfc}(\sqrt{\frac{y}{2}})$,
which implies the leading behavior of the moments as $A \to -1/2^-$
\begin{equation} \label{momcrit} 
\mathbb E \langle x^n \rangle \simeq 
c_n \big\vert A+\tfrac{1}{2}\big\vert^{-2 n} ,~ ~ c_n = 
\frac{2^n \Gamma(n+ \frac{1}{2})}{(n+1) \sqrt{\pi}} 
\end{equation}
where $c_1=\frac{1}{2}$ agrees with \eqref{cum1}. The PDF $P(y)$ is expected to
be the limit $a \to -\infty$ of a family of distributions indexed by $a$, universal within the KPZ class, which describes
the endpoint distribution around the critical point (see Section \ref{sec:universality}). 
As expected, the thermal fluctuations are subdominant as compared
to the disorder. Recall that for the polymer in full space, with one endpoint fixed at $0$, the cumulants are time dependent with
$\mathbb E [ \langle x^p \rangle_c ] = t \delta_{p,2}$
for any $t$, see Appendix \ref{appendix:sts}, a behavior very different from \eqref{eq:gencum}. For the half-space problem, near the transition the first two average cumulants
are expected to take the following time dependent scaling form 
$\mathbb E \langle x \rangle \simeq t^{2/3} f_1( a )$ and $\mathbb E \langle x^2 \rangle_c = t f_2(a)$, where
$a=t^{1/3} (A+\frac{1}{2})$ is the critical scaling variable, with the asymptotics $f_1(a) \simeq \frac{1}{2 a^2}$ 
and $f_2(a) \simeq 1/a^3$ for $a \to -\infty$, from \eqref{cum1},\eqref{cum2}. 
Finally, the result \eqref{momcrit} is expected to hold provided $t^{-1/3} \ll - (A + \frac{1}{2}) \ll 1$.

The polymer in the half-space can be also studied by the replica method. 
It uses the relation between the $n$-th moment of the partition sum and the Lieb-Liniger
Hamiltonian ${\cal H}_n$ for $n$ bosons on the half-line, solvable via the Bethe ansatz. 
This was pioneered by Kardar \cite{kardar1985depinning} who proposed an 
ansatz for the ground state $\Psi_0$ of ${\cal H}_n$, which is a bound state to the wall for
$A < -1/2$, and used it to predict $\mathbb E [ \langle x \rangle ]$.
This calculation assumes that the limits $n \to 0$ and $t \to +\infty$ 
commute. We checked that the result of \cite{kardar1985depinning}
agrees with \eqref{cum1} (using $\kappa=1/2$ and $\lambda=-A$ there),
which indicates that this assumption 
holds in the bound phase (while it {\it does not} hold in the unbound phase, or in the full space). In
Appendix \ref{appendix:betheansatz} we provide a detailed comparison of the two methods (replica ground state dominance,
and Brownian stationary measure \eqref{eq:gibbs1}) and more  
details on the replica approach. Note that the full spectrum of ${\cal H}_n$ is quite complicated and was
obtained recently in \cite{deNardisPLDTT}, see also \cite{AlexLD}, 
which confirms \cite{kardar1985depinning},
and may allow to obtain subleading large time behavior.

It is also possible to obtain exact formula for the $m$-point averages $\mathbb E[ p_A(x_1)\dots p_A(x_m)]$ 
of the Gibbs measure $p_A(x)$, using the Liouville
quantum mechanics developed in \cite{[{For review, see }]TexierComtetSUSY}, 
\cite{monthus1994flux, broderix1995thermal, comtet1998exponential, monthus2002localization}. 
The detailed calculations are presented in Appendix \ref{appendix:Liouville}.
For instance, in the case $m=1$, we obtain
\bea \label{eq:meanpx20}
&& \mathbb{E} [p_A(x)] = \frac{1}{4 \Gamma(2w) } \int_{-\I\infty}^{\I\infty} \frac{dz (w^2-z^2)}{2\I\pi \Gamma(2z)\Gamma(-2z)} e^{\frac{-x}{2}(w^2-z^2)} \nonumber \\
&& \times \Gamma(w+z)^2\Gamma(w-z)^2\Gamma(1-w-z)\Gamma(1-w+z).
\eea 
which is valid for $0< w=-(A + \frac{1}{2}) <1$ and can be analytically continued to all $w>0$ (see Appendix \ref{appendix:Liouville}
where we also checked that \eqref{eq:meanpx20} is normalized to unity and reproduces the first moment \eqref{cum1}).

\subsection{Midpoint probability and unbinding by a force} 

Consider now a
polymer with both endpoints fixed near the wall at times $t=0$ and $2t$, see Fig. \ref{fig:halfspacepolymer}. One may ask 
about the PDF of the midpoint position $x=x(t)$ for a long polymer, i.e. for 
large $t$. In the bound phase $A<-1/2$, it is proportional to (up to a normalization factor)
\begin{equation}
    e^{\mathcal B_1(x)+\mathcal B_2(x)+2 (A+\frac 1 2 )x},
\end{equation}
where $\mathcal B_1, \mathcal B_2$ are independent Brownian motions. Since $\mathcal B_1(x)+\mathcal B_2(x)$ has the same distribution as $\sqrt{2} \mathcal B(x) \overset{(d)}{=}\mathcal B(2x)$ ($\mathcal B$ being a standard Brownian motion), the PDF of the midpoint equals $2p_A(2x)$,
i.e. the midpoint position is distributed as half of the endpoint position \footnote{$x(2 b t)$ for any fixed $0<b<1$ has the same distribution for $t \to +\infty$.}.

Applying now a force ${\sf f}$ on the polymer endpoint 
in Fig. \ref{fig:halfspacepolymer}
results in the change $Z_A(x,t) \to e^{{\sf f} x} Z_A(x,t)$ to its partition function
(only at the final time $t$). In the stationary large time limit \eqref{eq:gibbs1} 
it amounts to shift $A \to A + {\sf f}$ in all the above results for the endpoint. 
An unbinding transition thus occurs at ${\sf f}={\sf f}_c=- (A+1/2)$, with the same
behavior as the transition at ${\sf f}=0$ upon varying $A$ (this
is also equivalent to tilting the wall). If the force is instead applied
on the midpoint in Fig. \ref{fig:halfspacepolymer}, the unbinding transition
then occurs at ${\sf f}_c=-2 (A+\frac{1}{2})$.

\subsection{Convergence to the stationary distribution}

\begin{figure}
    \centering
    \begin{tikzpicture}[scale=1.2, every text node part/.style={align=center}]
    \fill[blue!20] (-1,-1) --(2,2) -- (2,4.5) -- (-1,4.5) -- cycle;
    \fill[green!20] (-1,-1) --(2,2) -- (4.5,2) -- (4.5,-1) -- cycle;
    \fill[yellow!30] (2,2) --(2,4.5) -- (4.5,4.5) -- (4.5,2) -- cycle;
    \draw[thick, black!70, ->] (-1,0) -- (4.6,0) node[anchor=north] {$A$ \\ boundary parameter};
    \draw[thick, black!70, ->] (0,-1) -- (0,4.6) node[anchor=north east] {$B$ \\ drift \\parameter};
    \draw[thick] (-0.5,4.5) -- (4.5,-0.5);
    \draw[thick] (-1,-1) -- (2,2);
    \draw[thick] (2,4.5) -- (2,2);
    \draw[thick] (4.5,2) -- (2,2);
    \draw[dashed] (2,0) -- (2,2);
    \draw[dashed] (0,2) -- (2,2);
    \draw (2,0.1) -- (2,-0.1) node[anchor=north] {$\frac{-1}{2}$};
    \draw (0.1,2) -- (-0.1,2) node[anchor=east] {$\frac{-1}{2}$};
    \draw (0.6,2.5) node{$\mu_{A, -1-A}$};
    \draw (2.5,0.3) node{$\mu_{A,B}$};
    \draw (3.2,3) node{$\mu_{A,\frac{-1}{2}}$};
    \end{tikzpicture}
    \caption{Domains of attraction of KPZ (conjectural) stationary measures. The horizontal coordinate $A$ is the boundary parameter in the half-space SHE \eqref{eq:mSHEhalf-space}. The vertical coordinate means that we start from an initial condition which behaves as a Brownian motion with drift $-(B+\frac 1 2)$ at infinity. Then, the ratios of partition function $Z(x,t)/Z(0,t)$ converge to one of the stationary measures $\mu$. In the blue region (low density phase in ASEP context) these ratios converge to $\mu_{A, -1-A}$, as stated in the text. In the green region (high density phase) the ratios converge to $\mu_{A,B}$, and in the yellow region (maximal current phase), they converge to $\mu_{A, \frac{-1}{2}}$. Along the antidiagonal line $B=-1-A$, the ratios always converge to the Brownian stationary measures \eqref{eq:stationary}.}
    \label{fig:phasediagram}
\end{figure}
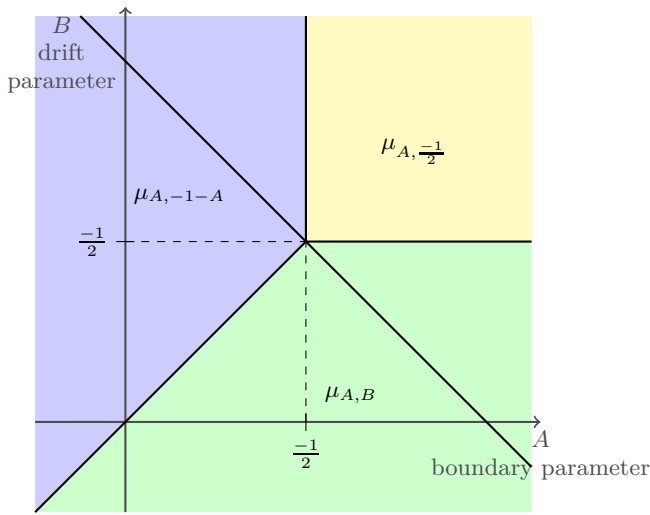

So far we have not justified why the ratios of partition functions converge to \eqref{eq:stationary} in the bound phase. 
It was shown in \cite{barraquand2020half} that this distribution of ratios is indeed stationary. 
When $A\leqslant \frac{-1}{2}$, we claim that for a large class of initial condition such that the drift at infinity is less than $-(A+\frac{1}{2})$, that is 
\begin{equation}
   \log Z_A(x+y,0)-\log Z_A(x,0) \simeq -(B+\tfrac{1}{2})y,
\end{equation}
for large $x$ and $y$, with $B\geqslant A$ (and even for more general random initial conditions), then the ratios of partition functions converge to \eqref{eq:stationary}. This class of initial conditions includes the flat IC for the KPZ equation, as well as the fixed endpoint for the DP (equivalently the droplet IC for KPZ). It is important to note that there exist more general stationary distributions of partition function ratios than the ones described in \eqref{eq:stationary}. They can be parametrized by $(A,B)$, and denoted $\mu_{A,B}$, where $A$ is the boundary parameter, and $B$ is a drift parameter meaning that $\log Z_A(x+y,t)-\log Z_A(x,t)$ behaves as a Brownian motion in $y$ with drift $-(B+\frac{1}{2})$ for large $x$. In the case $B=-1-A$, the stationary measure $\mu_{A,-1-A}$ is exactly the one described in \eqref{eq:stationary}. Depending on the value of $A$ and the drift at infinity of the initial condition, we expect that the partition function ratios converge to one of these stationary measures according to the phase diagram in Fig. \ref{fig:phasediagram}, based on a similar analysis performed for the asymmetric simple exclusion process (ASEP) \cite{liggett1975ergodic, derrida1993exact}. 
\medskip 

\section{Identities in distribution} 
\label{sec:identities}
The formula \eqref{eq:Laplacetransform}, which characterizes the distribution of $Z_A^f(0,t)$, matches with a known formula characterizing the distribution of another quantity. 
Let $Z^{(A)}(x,t)$ be the solution to the stochastic heat equation
\be
\partial_t Z(x,t) = \partial^2_{x} Z(x,t) + \sqrt{2} Z(x,t) \xi(x,t), \;x\in \mathbb R, 
\label{eq:mSHEfull-space}
\ee 
\emph{on the full line}, with ``half-Brownian'' initial condition given by $\mathds{1}_{x\geqslant0}e^{\mathcal B(x)-(A+\frac{1}{2}) x}$, where $\mathcal B$ is a standard Brownian motion. It was shown in \cite{imamura2011replica} (see also \cite{corwin2013crossover, borodin2012free}) that the Laplace transform of  $Z^{(A)}(0,t)$  is given by the same Fredholm determinant as the one that we obtained in \eqref{eq:Laplacetransform}. Matching parameters and notations between \cite{imamura2011replica} and the present paper  (see Appendix \ref{appendix:generalization}) we find the following surprising identity in distribution: for all fixed $t>0$ and any $A\in \mathbb R$, 
\begin{equation}
    Z_A^f(0,t) = 2 Z^{(A)}(0,t), 
    \label{eq:identityinlaw}
\end{equation}
where $Z_A^f(0,t)$ was defined in \eqref{ZAfree}. Remarkably, an identity of a similar flavour as \eqref{eq:identityinlaw} can be deduced from \cite[Eq. (7.59)]{baik2001algebraic} for a model of last passage percolation \footnote{Assuming universality, both identities would imply in the large time limit the same relation between full-space and half-space KPZ fixed point.}.   
We stress that \eqref{eq:identityinlaw} is also valid in the phase $A<\frac{-1}{2}$, where in the l.h.s., the polymer is bound to the wall. In the RHS, $Z^{(A)}(0,t)$ is the partition function of polymer paths in the full space, weighted by $\mathds{1}_{x\geqslant0}e^{\mathcal B(x)-(A+\frac{1}{2}) x}$ ($x$ being the starting point) which, for $A<-1/2$, is dominated by $x=O(t)$ so that the fluctuations of the Brownian motion are dominant over KPZ-type fluctuations.
We do not know whether \eqref{eq:identityinlaw} extends at several times. Nevertheless, we may generalize \eqref{eq:identityinlaw} by introducing a spatial parameter, though we cannot simply replace the point $0$ by an arbitrary point $X>0$ in \eqref{eq:identityinlaw}. Let us define 
\begin{equation}
Z_A^{\rm shifted}(X,t) =\int_{ x\geq X} Z_A(0,t\vert x,0)dx.
\label{eq:shiftedpartitionfunction}
\end{equation}
This corresponds to the value at the origin of the solution to the half-line SHE \eqref{eq:mSHEhalf-space} with initial condition $Z(x,0)=\mathds{1}_{x\geqslant X}$. Then, we may readily adapt Eqs. \eqref{eq:integration}, \eqref{eq:nestedmoments2}, \eqref{eq:determinantalmoments}, \eqref{product} and \eqref{eq:Laplacetransform} (see details in Appendix \ref{appendix:generalization}) and match the result with \cite{imamura2011replica}. We obtain that for any fixed time $t>0$, $X\geqslant 0$ and $A\in \mathbb R$, we have the identity in distribution 
\begin{equation}
    Z_A^{\rm shifted}(X,t) = 2 Z^{(A)}(-X,t). 
    \label{eq:identityinlaw2}
\end{equation}
Note that $Z^{(A)}(-X,t)$ has the same law as $e^{- \frac{X^2}{4 t} } Z^{(A+ \frac{X}{2 t})}(0,t)$
from the tilt symmetry
 (see Appendix \ref{appendix:sts}), hence we also have $Z_A^{\rm shifted}(X,t)=e^{- \frac{X^2}{4 t} } Z_{A+\frac{X}{2t}}^{\rm shifted}(0,t)$ in law.

An even more general identity in distribution holds. Let us denote $Z_{A,B}(x,t)$ the solution to \eqref{eq:mSHEhalf-space} on the half line $x\geqslant 0$ with initial condition given by $e^{\mathcal B(x)-(B+\frac 1 2) x}$. The moments of $Z_{A,B}(x,t)$ are given in \cite[Sec. 4.4]{barraquand2020half} in a very similar form as in \eqref{eq:nestedmoments}. Thus we may still apply the same steps: we define 
\begin{equation}
Z_{A,B}^{\rm shifted}(X,t) =\int_{ x\geq X} Z_{A,B}(x,t)dx,
\label{eq:shiftedpartitionfunctionAB}
\end{equation} 
and compute the Laplace transform of $Z_{A,B}^{\rm shifted}(X,t)$. 
Then, for $t>0$, $X\geqslant 0$, and parameters $A,B$ such that $A+B+1>0$ and $B>\frac{-1}{2}$, we have the identity in distribution
\begin{equation}
        Z_{A,B}^{\rm shifted}(X,t) = 2 Z^{(B\vert A,B)}(-X,t), 
    \label{eq:identityinlaw3}
\end{equation}
where the quantity $Z^{(B\vert A,B)}(-X,t)$ is again the solution to full-line SHE \eqref{eq:mSHEfull-space} with some specific IC, that we obtain in Appendix \ref{appendix:initialdata} using exact formulas valid for the exactly solvable log-gamma polymer model. To describe it,  let $\mathcal W_1, \mathcal W_2, \mathcal W_3$ be three independent  Brownian motions with respective drifts $-(B+\frac 1 2 )$, $-(A+\frac 1 2 )$ and $-(B+\frac 1 2 )$. Let $w$ be an independent inverse Gamma random variable with parameter  $(2B+1)$. Then for $x\leqslant 0$, 
\begin{equation}
Z^{(B\vert A,B)}(x,0) = w e^{\mathcal W_1(-x)},
    \label{eq:complicatedinitialdataleft}
\end{equation}
and for $x\geqslant 0$, 
\begin{equation}
Z^{(B\vert A,B)}(x,0) = e^{\mathcal W_3(x)}\left(w + \int_0^x e^{\mathcal W_2(y)-\mathcal W_3(y)}dy\right).
    \label{eq:complicatedinitialdataright}
\end{equation}
When $B\to+\infty$, then $B\, Z_{A,B}^{\rm shifted}(X,t)$ goes to $Z_{A}^{\rm shifted}(X,t)$ and $B\, Z^{(B\vert A,B)}(X,t)$ goes to $ Z^{(A)}(X,t)$ (see Appendix \ref{appendix:degenerationBtoinfty}), so that we recover \eqref{eq:identityinlaw2}. 
When $A\to+\infty$, we obtain yet another identity in law 
\be
Z_{+\infty,B}^{\rm shifted}(X,t) = 2 w  \tilde Z^{(B\vert B)}(-X,t)
\ee 
where the l.h.s. 
is related to half-line solution 
to \eqref{eq:mSHEhalf-space} with $e^{\mathcal B(x)-(B+\frac 1 2 )x}$ IC 
and Dirichlet boundary condition (this solution was studied in  \cite{parekh2019positive, AlexLD}), and $\tilde Z^{(B\vert B)}(X,t)$ is the full-line solution to \eqref{eq:mSHEfull-space} with $e^{\mathcal B(x)-(B+\frac 1 2 )|x|}$ IC, 
independent on $w$. This solution was studied in \cite{imamura2012exact, imamura2013stationary,borodin2015height} and one checks that the Fredholm determinant obtained 
here characterizing the law of $Z_{A,B}^{\rm shifted}(X,t)$ for $A\to+\infty$ matches the one in \cite{imamura2013stationary}, see Appendix \ref{appendix:generalLaplace}.
Going back to the solution $ Z^{(B\vert A,B)}(-X,t)$ 
in \eqref{eq:identityinlaw3}, its
distribution was
not obtained in the literature, 
though its moments can be computed using known methods, and we explain in Appendix \ref{appendix:initialdata} appendix how they match with the moments of $Z_{A,B}^{\rm shifted}(X,t)$ for generic parameters $A,B$. 

\section{Universality}
\label{sec:universality}
In this paper we studied the continuum directed polymer model. We found that in the bound phase, taking the limit $t \to +\infty$ first, 
very near the transition, with $0< \epsilon=- (A+1/2) \ll 1$, the PDF of the scaled endpoint position $y= x (A+ 1/2)^2$ concentrates
around the optimum $y_m = {\rm argmax}_{z>0} (\tilde{\mathcal B}(z) - z )$. Hence in that limit this scaled position is
distributed with 
\begin{equation}
    P(y) = \sqrt{\frac{2}{\pi y}} e^{-y/2} - {\rm Erfc}(\sqrt{\frac{y}{2}}),
\end{equation}
a PDF which behaves as $P(y) \simeq \sqrt{\frac{2}{\pi y}}$ for $y \ll 1$ and
as $P(y) \simeq \sqrt{\frac{2}{\pi}} y^{-3/2} e^{-y/2}$ for $y \gg 1$.

One can argue that this result holds for finite but very large time as well, as long as the critical 
parameter $a = (A+\frac{1}{2}) t^{1/3} $ is very large negative $- a \gg 1$. In fact one can surmise
that there is a scaling function which describes the endpoint position $x(t)$ in the critical region
as follows 
\be \label{PDFuniversal} 
 \lim_{t \to +\infty}   {\rm Prob}\left( x(t) (A+ 1/2)^2 < y \Big\vert A+ \tfrac{1}{2} = \frac{a}{t^{1/3}} \right) =  {\cal P}(y , a) 
\ee 
To match the previous result one would need that $\lim_{a \to -\infty} {\cal P}(y , a) = \int_0^{y} dz P(z)$.
\medskip 
On the other hand one can conjecture the existence of a (critical) half-space Airy process, denoted ${\cal A}_a(\hat x)$
continuously depending on the parameter $a$. It would describe in particular the simultaneous limit $t \to +\infty$
and $A \to -1/2$ of the continuum directed polymer model, with fixed $a=t^{1/3} (A + \frac{1}{2})$ 
\be \label{airya} 
 \log Z_A(x,t|0,0) \simeq t^{1/3} ({\cal A}_a(\hat x) - \hat x^2), \quad \quad \hat x= \frac{x}{2 t^{2/3}}
\ee
By universality, this process should be the same as the limit process obtained from half-space last passage percolation, so that the finite dimensional marginals of $\mathcal A_a(\hat x)$ are described in \cite[Theorem 1.7]{baik2018pfaffian}. 
The main result of this paper (solution for the flat IC) can be stated in terms of this process $\mathcal A_a(\hat x)$
\bea 
\max_{\hat x>0} [ ({\cal A}_a(\hat x) - \hat x^2) ] \overset{(d)}{=}  \mathrm{BBP}_a
\eea 
where $\overset{(d)}{=}$ denotes the equality of distributions and $\mathrm{BBP}_a$ denotes the BBP distribution defined in \eqref{eq:defBBP}. More generally, from the
identity \eqref{eq:shiftedpartitionfunction}
obtained in this paper, one would conclude that, for fixed $\hat X\geqslant 0$, 
\bea 
&& \max_{\hat x> \hat X} [ ({\cal A}_a(\hat x) - \hat x^2) ] \overset{(d)}{=} a^2+2a \hat X \\
&& + [ {\cal A}_{2 \to BM}(-\hat X-a) - (\hat X + a)^2], 
\eea 
where ${\cal A}_{2 \to BM}$ was introduced in \cite{imamura2004fluctuations} (see also \cite{corwin2010limit, corwin2013crossover}). 
For \eqref{airya}  to match our results on the stationary large time limit requires that for $\hat x \ll 1$
\bea 
 {\cal A}_a(\hat x) \simeq \sqrt{2} B(\hat x) + 2 a \hat x.
\eea

Finally, the endpoint PDF scaling function ${\cal P}(y , a)$ defined in \eqref{PDFuniversal} would be
obtained from this process as the PDF of $y = {\rm argmax} [ ({\cal A}_a(\hat x) - \hat x^2) ] $. It is then natural to conjecture the universality 
of the above distributions at a half-space KPZ fixed point.

\bigskip 
In the context of last passage percolation, an identity reminiscent of \eqref{eq:identityinlaw} relating the distribution of the point to point energy in a full-space model and the point to line energy in a half-space model, was stated as \cite[Eq. (7.59)]{baik2001algebraic}. In the large scale limit (studied in \cite{baik2001asymptotics}) both distributions converge to the BBP distribution (the limiting distribution function was denoted $F^{\boxtimes}(x;w)$ in \cite{baik2001asymptotics}, it coincides with the BBP distribution defined later in \cite{baik2005phase}). Our asymptotic results at large time for the KPZ equation in Section \ref{sec:largetimelimit} thus confirm universality predictions. Let us stress, however, that the identity in distribution from \cite{baik2001algebraic} cannot be scaled to the KPZ equation: one cannot deduce from it our finite time identities in distribution \eqref{eq:identityinlaw}, \eqref{eq:identityinlaw2}, \eqref{eq:identityinlaw3} and  \eqref{eq:identityinlaw4}.

\section{Conclusion}
We obtained the solution for all times to the KPZ equation on a half-line with flat IC, i.e.
the distribution of the height at the origin for any wall parameter $A$. 
Thanks to remarkable algebraic cancellations it is 
simpler than the solution for flat IC on the full line. In fact, we find that it is related to the
half-Brownian IC on the full line, and uncover further curious relations between full and half-line problems.
Equivalently it gives the free energy of a DP of any length $t$ in a half-space,
with one free endpoint and the other pinned at the wall. We showed that its critical behavior
at the unbinding transition at $A=-1/2$ is identical to the BBP critical behavior for outliers of
GUE random matrices. For $A<-1/2$ the polymer is bound to the wall and at large $t$ its endpoint
position fluctuates as a particle at equilibrium in a one-sided Brownian plus linear confining potential.
This considerably extends early predictions within the replica Bethe ansatz.
These results open questions such as generalization of the aforementioned identities to several
points or times, studying the rate of convergence to the stationary measure $\mu_A$ that we determined,
possibly in relation to excited states within the replica Bethe ansatz,
and properties of the non Gaussian stationary measures $\mu_{A,B}$ (their analogues in finite volume were recently studied in  \cite{corwin2021stationary}).
Our results near criticality are part of a larger universal KPZ fixed point structure in half-space, yet to be
fully characterized.



\let\oldaddcontentsline\addcontentsline
\renewcommand{\addcontentsline}[3]{}
\let\addcontentsline\oldaddcontentsline






\setcounter{secnumdepth}{2}

\appendix

\begin{widetext}

\section{Flat initial condition: details }
\label{appendix:detailsflat}

\subsection{Explicit formula for the moments} 
Computing the factor $E(w_1,w_1+1,\dots,w_1+m_1-1,\dots,w_\ell,\dots,w_{\ell}
+m_{\ell}-1)$ in \eqref{eq:determinantalmoments} using \eqref{product} one finds 
\be
\mathbb E\left[Z_A^f(0,t)^n\right] =  n!2^n \sum_{\ell=1}^n  \frac{1}{\ell !} \sum_{\vec m : \sum_{i=1}^\ell m_i=n}  \prod_{i=1}^{\ell}\int_{r+\I\mathbb R} \frac{dw_i}{2\I\pi} 
\det\left(\frac{1}{w_i+m_i-w_j}\right)_{i,j=1}^{\ell} \, \prod_{i=1}^{\ell} \frac{ \Gamma (A+w_i)}{\Gamma (A+w_i+m_i )} e^{t (G(w_i+m_i)-G(w_i))}
\label{eq:determinantalmoments2} 
\ee
where we used that $\sum_{k=0}^{m-1} (w+k)^2=G(w+m)-G(w)$ and $G(w)$ is defined as $G(w) := \frac{w^3}{3} - \frac{w^2}{2} + \frac{w}{6}$.
We recall that the real part of integration contours is such that $r> -A$. One can check that the integrals over $w_j$ are convergent. 

\subsection{Laplace transform}
\label{appendix:Laplacetransform} 

Let us now consider the generating function $1 + \sum_{n=1}^{+\infty} \frac{(-u)^n}{n!} \mathbb E\left[Z^f_A(0,t)^n\right]$. 
The summation over $n$ allows to eliminate the constraint $\sum_{i=1}^\ell m_i=n$ in the sum over the variables $m_i$ in \eqref{eq:determinantalmoments2}.
Although it is a divergent series, after rearrangements of the terms and use of a Mellin Barnes representation of the sums, it yields an expression
for the Laplace transform $\mathbb E\left[e^{- u Z^f_A(0,t) } \right]$ which in all known cases has given the correct result. 
The Mellin-Barnes representation replaces the sum over the integer variables $m_i$ by integrals, for each $i$ 
\be 
\sum_{m=1}^{+\infty} (-1)^m f(m) = \int_{\mathcal C} \frac{dz}{2\I\pi} \frac{\pi}{\sin (-\pi z)} f(z)
\ee 
where $\mathcal C = a + \I \mathbb{R}$ with $0<a<1$, oriented from bottom to top. Introducing a variable $z_j$ for each $m_j$ and
performing the change from $z_j$ to $s_j=w_j+z_j$, this leads to
\begin{multline} \label{eq:multiple} 
	\EE[e^{-u Z^f_A(0,t)} ] = \sum_{\ell=0}^{+\infty} \frac{1}{\ell!} 
  \int_{\mathcal C_{a_w}}\frac{dw_1}{2\I\pi} 
 \dots \int_{\mathcal C_{a_w}}\frac{dw_{\ell}}{2\I\pi}  
  \int_{\mathcal C_{a_s}[w_1]}\frac{ds_1}{2\I\pi} 
 \dots \int_{\mathcal C_{a_s}[w_{\ell}]}\frac{ds_{\ell}}{2\I\pi}  
\det\left(\frac{1}{s_i-w_j}\right)_{i,j=1}^{\ell}  \\
\times \prod_{j=1}^{\ell} \left[ \frac{e^{t G(s_j) }}{
e^{t G(w_j) }} \frac{\pi}{\sin(-\pi(s_j-w_j)} (2u)^{s_j-w_j}\right]
\prod_{j=1}^{\ell}  \frac{\Gamma (A+w_j)}{\Gamma (A+s_j )}.
 \end{multline}
The contour for variables $w_i$, denoted $\mathcal C_{a_w}$, is the union of two semi-infinite rays leaving the point $ a_w>-A$ in the direction $\pm 2\pi/3$, oriented from bottom to top.  The contour for variables $s_i$, denoted  $\mathcal C_{a_s}[w]$ is formed by the union of the vertical line  $a_s+\I\mathbb R $ and the union of negatively oriented circles around the poles at $w+1, w+2, \dots$ when these lie to the left of the vertical line (see Fig. \ref{fig:contours}). The vertical line is oriented from bottom to top.  Furthermore, a sufficient condition for the integrals over $s_i$ to be convergent is that $a_s-1/2>0$. 
\begin{figure} 
\begin{tikzpicture}[scale=1.2]
\draw[thick, gray, ->] (-2,0) -- (2,0);
\draw[thick, gray, ->] (0,-2) -- (0,3);
\draw[gray] (0,0) node[anchor=north west] {$0$};
\draw[thick] (-1/2,0) -- +(120:3.3);
\draw[thick, ->] (-1/2,0) -- +(120:1);
\draw[thick] (-1/2,0) -- +(-120:2.2);
\draw[] (0.5,-0.1) -- (0.5,0.1) node[anchor=south] {$1/2$};
\draw[] (-1.5,-0.1) -- (-1.5,0.1) node[anchor=south] {$-A$};
\draw[] (-0.5,-0.1) -- (-0.5,0.1) node[anchor=south] {$a_w$};
\draw[] (1,0) node[anchor=south west] {$a_s$};
\draw[thick] (1,-2) -- (1,3);
\draw[thick, ->] (1,-2) -- (1,-1);
\fill (-1/2,0)+ (120:2) circle(0.07);
\fill (1/2,0)+ (120:2) circle(0.07);
\fill (3/2,0)+ (120:2) circle(0.07);
\fill (5/2,0)+ (120:2) circle(0.07);
\fill (-1/2,0.3)+ (120:2) node{$w$};
\fill (1/2,0.4)+ (120:2) node{$w+1$};
\fill (3/2,0.4)+ (120:2) node{$w+2$};
\fill (5/2,0.3)+ (120:2) node{$w+3$};
\draw[thick, ->] (0.75,0)+ (120:2) arc(0:-360:0.25);
\draw[thick, ->] (1.75,0)+ (120:2) arc(0:-360:0.25);
\draw (-1.4,-1) node {$\mathcal C_{a_w}$};
\draw (1.5,-1) node {$\mathcal C_{a_s}[w]$};
\end{tikzpicture}
\caption{The contours  $\mathcal C_{a_w}$ and $\mathcal C_{a_s}[w]$ are shown in the figure. The contour $\mathcal C_{a_s}[w]$ depends on the location of $w$. For the $w$ depicted in the figure, the contour consists of the union of the vertical line with real part $a_s$ and small negatively oriented circles around $w+1$ and $w+2$, since $w+1$ and $w+2$ lie to the left of the vertical line with real part $a_s$. } 
\label{fig:contours}
\end{figure}
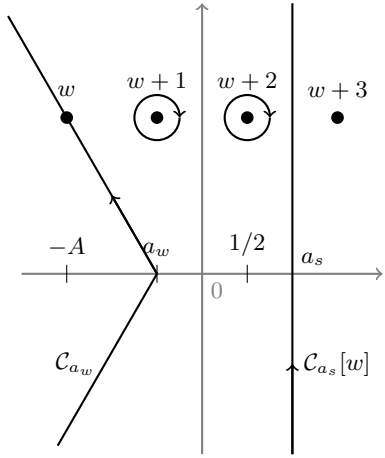
We choose the real numbers $a_s$ and $a_w$ 
so that
\begin{equation}
-A< a_w<a_s .
\end{equation}
Since from the choices of integration contours one has ${\rm Re}(s_i-w_j) >0$ in \eqref{eq:multiple}, we can use the representation
\bea
\frac{1}{s_i-w_j} = \int_0^{+\infty} dv e^{- v (s_i - w_j)} 
\eea 
inside the determinant. After some simple manipulations one recognizes the expansion of a Fredholm determinant 
\bea
\EE[e^{-u Z^f_A(0,t)} ] 
= {\rm Det}( I + \overline K_{u,t})_{\mathbb{L}^2(\mathbb{R}^+)}
\eea
with the kernel 
\bea
&& \overline K_{u,t}(v,v') = 
\int_{\mathcal C_{a_w}}\frac{dw}{2\I\pi}  
  \int_{\mathcal C_{a_s}[w]}\frac{ds}{2\I\pi} 
   \frac{\pi}{\sin(-\pi(s-w))} (2u)^{s-w} e^{- v s + v' w}  \frac{e^{t G(s)}}{
e^{t G(w)}} \frac{\Gamma (A+w)}{ \Gamma (A+s )}. 
\eea
This provides an expression of the generating function in terms of a kernel involving two contour integrals. We will
now transform this formula to obtain an alternative expression in terms a second kernel, as given in the main text in Section \ref{sec:flatinitialcondition}. 
Let us use the identity for $\Re \lambda >0$
\bea \label{MB} 
 (2u)^\lambda \frac{\pi}{\sin(- \pi \lambda)} 
= - \int_\mathbb{R} d r \frac{2u}{ e^{-r} + 2u} e^{- \lambda r}  
\eea 
and perform the shift $w_j \to w_j+1/2$ and change of variables $s_j = z_j + 1/2$.
This leads to our final formula
\begin{equation}
\mathbb E[e^{-u Z_A^f(0,t)e^{\frac{t}{12}}}] = \det(I-K_{u,t})_{\mathbb L^2(0,+\infty)},
\label{eq:Laplacetransformsuppmat}
\end{equation}
with the kernel 
\begin{align}  \label{Kut1} 
K_{u,t}(v,v') &= \int_{\mathbb R} \mathrm d r \frac{2u}{ e^{-r} +2u} 
\phi_{A,t}(v+r) \psi_{A,t}(v'+r) \\
\phi_{A,t}(v) &=  \int_{a_z+\I\mathbb R}\frac{dz}{2\I\pi} 
 \frac{e^{t \frac{z^3}{3} -vz} }{\Gamma (A+\frac{1}{2} + z )} \label{Kut2}  \\
\psi_{A,t}(v) &= 
\int_{\mathcal C_{a_w}}\frac{dw}{2\I\pi} e^{-t \frac{w^3}{3} 
+ v w} \Gamma (A+ \tfrac{1}{2} +w) \label{Kut3} 
\end{align}
where the contour for $z$ is a vertical line such that  $a_z>0$ 
and the contour for $w$, denoted $\mathcal C_{a_w}$, is the union of two semi-infinite rays leaving the point $ a_w>-\left(A+\frac{1}{2} \right)$ in the direction $\pm 2\pi/3$.

\section{Asymptotics: details} 
\label{appendix:asymptotics}

\subsection{Case $A> \frac{-1}{2}$}

Using the rescalings indicated in Section \ref{sec:flatinitialcondition}, we have the limits
\begin{equation}
    \lim_{t\to\infty} \Gamma(A+ \frac{1}{2}) t^{1/3} \phi_{A,t}(t^{1/3}\tilde v) = \lim_{t\to\infty} \frac{t^{1/3}\psi_{A,t}(t^{1/3}\tilde v)}{\Gamma(A+ \frac{1}{2})} = \mathrm{Ai}(\tilde v) = \int_{1+\I\mathbb R}\frac{dz}{2\I\pi} 
 e^{\frac{z^3}{3} -\tilde v z}.
\end{equation}
We also have, with $u=e^{-t^{1/3} s}$ and $r=t^{1/3}\tilde r$, 
\be
\frac{2u}{ e^{-r} +2u} = \frac{2}{2 + e^{t^{1/3}(s-\tilde r)} } \xrightarrow[t\to\infty]{} \theta(\tilde r - s)
\ee 
which leads to Eqs. \eqref{eq:KAi} and \eqref{eq:F2} in the main text. 

\subsection{Case $A<\frac{-1}{2}$}  

When $A<\frac{-1}{2}$, we use the following change of variables in Eqs. \eqref{Kut1}-\eqref{Kut3}
\be
r = \left(A + \frac{1}{2}\right)^2 + t^{1/2} \tilde r, \quad  \quad v = t^{1/2} \tilde v, 
\quad  \quad z = - \left(A + \frac{1}{2}\right) + t^{-1/2} \tilde z
\ee
so that 
\begin{equation}
\phi_{A,t}\left( \left(A+\frac 1 2 \right)^2 t + t^{1/2}(\tilde v+ \tilde r) \right) 
\simeq \frac{1}{t} e^{\frac{2t(A+\frac 1 2)^3}{3}} e^{(A+\frac 1 2) t^{1/2}(\tilde v+ \tilde r)}  
\int  \frac{d\tilde z}{2\I\pi} \tilde z  e^{- (A+\frac 1 2 ) \tilde z^2 - (\tilde v + \tilde r)\tilde z}.
\end{equation}
In  order to take the asymptotics of $\psi_{A,t}$, we first shift the contour to the left of the pole at $w=-(A+\frac{1}{2})$ and we obtain 
\begin{equation}
\psi_{A,t}(v) =  e^{\frac{t(A+\frac 1 2 )^3}{3} - v(A+\frac 1 2 ) } + 
\int_{\mathcal C_{a_w'}}\frac{dw}{2\I\pi} e^{-t \frac{w^3}{3} 
+ v w} \Gamma (A+ \tfrac{1}{2} +w)
\end{equation}
where now, $\mathcal C_{a_w'}$ is the union of two semi-infinite rays in direction $\pm \phi$ where $\phi\in (\frac{\pi}{2}, \frac{5\pi}{6})$, (so that
${\rm Re}[ w^3] >0$), which intersect the horizontal axis at $a_w'$ with $ -A-\frac 3 2 < a_w' < -(A+\frac 1 2)$. We obtain 
\begin{equation}
\psi_{A,t} \left( \left(A+\frac 1 2 \right)^2 t + t^{1/2}(\tilde v'+\tilde r) \right) \simeq  e^{-\frac{2t(A+\frac 1 2)^3}{3}} e^{-(A+\frac 1 2) t^{1/2}(\tilde v'+\tilde r)} \left( 1+
\int_{\mathcal C_{a_{\tilde w}}} \frac{d\tilde w}{2 \I\pi} \frac{e^{ (A+\frac 1 2 ) \tilde w^2 + (\tilde v' + \tilde r)\tilde w} }{\tilde w}\right)
\label{eq:remainderterm}
\end{equation}
where we have chosen $\phi\in (\frac{3\pi}{4}, \frac{5\pi}{6})$, 
(so that 
${\rm Re} [\tilde w^2 ]>0$), and we have  used the change of variables $w=-(A+\frac 1 2)+ t^{-1/2}\tilde w$ and we scale $a_w'=-(A+\frac 1 2) +a_{\tilde w}t^{-1/2}$ with  $a_{\tilde w}<0$. Notice that in the integral in the RHS of \eqref{eq:remainderterm}, the contour can be freely shifted to the left to $-\infty$, so that the integral is zero. 
We now set $u= e^{-(A+ \frac{1}{2})^2 t - s t^{1/2}}$ so that
\be
\frac{2u}{ e^{-r} +2u} = \frac{2}{2 + e^{t^{1/2}(s-\tilde r)} }  \xrightarrow[t\to\infty]{} \theta(\tilde r - s)
\ee 
and 
\begin{equation}
    \mathbb E\left[e^{-u Z_A^f(0,t)e^{\frac{t}{12}}}\right]  \xrightarrow[t\to\infty]{} \mathbb P \left( \frac{\log Z_A^f(0,t)+ \frac{t}{12}- (A+\tfrac{1}{2} )^2t}{t^{1/2}} \leqslant s\right). 
\end{equation}
Putting all terms together, there are cancellations of the prefactors from $\psi_{A,t}$ and $\phi_{A,t}$, so that $t^{1/2}K_{u,t}(v,v')\xrightarrow[t\to\infty]{} \widetilde K_s(\tilde v, \tilde v')$ and we obtain that 
\begin{equation}
   \lim_{t\to\infty} \det(I-K_{u,t})_{\mathbb L^2(0,+\infty)} = \det(I-\widetilde K_{s})_{\mathbb L^2(0,+\infty)},
\end{equation}
where 
\begin{equation}
    \widetilde K_s(\tilde v, \tilde v') = \int_{s}^{+\infty} d \tilde r \int  \frac{d\tilde z}{2\I\pi} \tilde z  e^{- (A+\frac 1 2 ) \tilde z^2 - (\tilde v + \tilde r)\tilde z} = \int  \frac{d\tilde z}{2\I\pi}   e^{- (A+\frac 1 2 ) \tilde z^2 - (\tilde v + s)\tilde z} = 
    \frac{1}{\sqrt{ 4 \pi |A+\frac{1}{2}|}} e^{- \frac{(s + \tilde v)^2}{4 |A+\frac{1}{2}|} }
\end{equation}
so that, at large time
\be
\log Z_A^f(0,t)  \simeq \left(- \frac{1}{12} + \left(A+\frac{1}{2} \right)^2 \right)  t + t^{1/2} \chi 
\ee 
with 
\begin{equation} 
\label{FD1} 
  \mathbb{P}(\chi<s) =   \det(I-\widetilde K_{s})_{\mathbb L^2(0,+\infty)}   =1 - \int_0^{\infty} \widetilde K_s(\tilde v,\tilde v)d\tilde v =  \mathbb P\left(G \leqslant s   \right)
\end{equation}
where $G$ is a centered Gaussian random variable with variance $2\left\vert A+\frac 1 2 \right\vert$, as announced in the main text. In \eqref{FD1}
the Fredholm determinant is simple to evaluate since $\widetilde K_{s}$ is a rank one kernel (i.e. a projector). Note that the variance of the random variable $\chi$ has the same value as for the droplet IC, as found by a more heuristic method in \cite{deNardisPLDTT}.

\subsection{Critical case $A=\frac{-1}{2}+ a t^{-1/3}$.}
We use the same scalings as indicated in Section \ref{sec:flatinitialcondition} for $A>\frac{-1}{2}$, and we obtain 
\begin{eqnarray}
    \lim_{t\to\infty} t^{2/3}\phi_{A,t}(t^{1/3}\tilde v) &= \phi_{a}(\tilde v) &:= \int_{1+\I\mathbb R}\frac{d\tilde z}{2\I\pi}  (\tilde z+a)
 e^{\frac{\tilde z^3}{3} -\tilde v \tilde z}, \label{eq:phiBBP}
 \\
 \lim_{t\to\infty} \psi_{A,t}(t^{1/3}\tilde v) &=   \psi_{a}(\tilde v) &:= \int_{\mathcal C_{a_{\tilde w}}}\frac{d\tilde w}{2\I\pi} 
 \frac{e^{\frac{-\tilde w^3}{3} +\tilde v \tilde w}}{\tilde w+a} 
 , \label{eq:psiBBP}
\end{eqnarray}
where the  contour $\mathcal C_{a_{\tilde w}}$  in \eqref{eq:psiBBP} is the union of two semi-infinite rays in direction $\pm 2\pi/3$  which intersect the horizontal axis at $a_{\tilde w}$ is such that $a_{\tilde w}>-a$.  Putting all together we obtain, as announced in the main text, 
\be \label{eq:defBBPsuppmat}
\mathbb P(\chi\leq s) = \det(I-K^{\rm BBP}_a)_{\mathbb L^2(s,+\infty)} = F^{\rm BBP}_a(s)
\ee 
with the kernel
\be
K^{\rm BBP}_a( v, v') = \int_{0}^{+\infty} \mathrm d  r 
\phi_{a}( v+ r) \psi_a( v'+ r).
\ee 
where the functions $\phi_a$ and $\psi_a$ are defined in \eqref{eq:phiBBP} and \eqref{eq:psiBBP}. This distribution was introduced in \cite{baik2005phase} and the form that we obtained in \eqref{eq:defBBP} (i.e. \eqref{eq:defBBPsuppmat} above) can be matched with the original definition from \cite{baik2005phase} using e.g. \cite{borodin2012free}. 

\section{Generalization to the shifted partition function with $X\geqslant 0$ and arbitrary parameters  $A,B$}
\label{appendix:generalization}
\subsection{Moment formulas}
The starting point is the following formula from \cite[Sec. 4.4]{barraquand2020half} for the moments of $Z_{A,B}(x,t)$, that is the solution to the half-space SHE  \eqref{eq:mSHEhalf-space} with Brownian IC $e^{\mathcal B(x)-(B+\frac 1 2 )}$. For $B>n-1$, $A+B>n-1$, and $x_1 \geq x_2 \geq \dots \geq x_n$
\begin{multline}
\mathbb E\left[\prod_{i=1}^n Z_{A,B}(x_i,t)\right] = 2^n \frac{\Gamma(A+B+1)}{\Gamma(A+B+1-n)}\int_{r_1+\I\mathbb R}\frac{\mathrm{d}z_1}{2\I\pi} \dots \int_{r_n+\I\mathbb R}\frac{\mathrm{d}z_n}{2\I\pi}
 \prod_{1\leqslant a<b\leqslant n} \frac{z_a-z_b}{z_a-z_b-1}\frac{z_a+z_b}{z_a+z_b-1}\\ \times \prod_{i=1}^n  \frac{z_i}{z_i+A} \frac{1}{B^2-z_i^2}  e^{tz_i^2 - x_i z_i} ,
\label{eq:nestedmomentsgeneral}
\end{multline}
where the contours are chosen so that $B> r_1>r_2+1>\dots > r_n+n-1>\max\lbrace n-1-A, n-1\rbrace$, i.e.
all contours are to the right of $-A$ and to the left of $B$. Recall the definition of $Z_{A,B}^{\rm shifted}(X,t)$ from \eqref{eq:shiftedpartitionfunctionAB}. Using the same steps as in the main text around Eq. \eqref{eq:integration}, we have 
\begin{equation}
\mathbb E\left[Z_{A,B}^{\rm shifted}(X,t)^n\right] = n! 2^n \frac{\Gamma(A+B+1)}{\Gamma(A+B+1-n)}\int_{r_1+\I\mathbb R}\frac{\mathrm{d}z_1}{2\I\pi} \dots \int_{r_n+\I\mathbb R}\frac{\mathrm{d}z_n}{2\I\pi}
 \prod_{1\leqslant a<b\leqslant n} \frac{z_a-z_b}{z_a-z_b-1} F(\vec z),
\label{eq:nestedmomentsgeneral2}
\end{equation}
where now, 
\begin{equation}
    F(\vec z) = \prod_{1\leq a<b\leq n}\frac{z_a+z_b}{z_a+z_b-1}\prod_{i=1}^n \frac{e^{tz_i^2-X z_i} }{z_1+\dots+z_i} \frac{z_i}{z_i+A}\frac{1}{B^2-z_i^2}.
    \label{eq:defFcompliquee}
\end{equation}
As in the main text, we may use  \cite[Proposition 5.1]{borodin2016directed} to obtain that \eqref{eq:nestedmomentsgeneral} becomes
\begin{multline}
   n!2^n \frac{\Gamma(A+B+1)}{\Gamma(A+B+1-n)} \sum_{\ell=1}^n  \frac{1}{\ell !} \sum_{\vec m : \sum m_i=n}  \int_{r+\I\mathbb R} \frac{dw_1}{2\I\pi} \dots \int_{r+\I\mathbb R} \frac{dw_{\ell}}{2\I\pi} 
\det\left(\frac{1}{w_i+m_i-w_j}\right)_{i,j=1}^{\ell}  \\ \times 
E(w_1,w_1+1,\dots,w_1+m_1-1,\dots,w_\ell,\dots,w_{\ell}
+m_{\ell}-1)
\label{eq:determinantalmomentsgeneral}
\end{multline}
where now, the real part of contours is such that $\max\lbrace -A,0\rbrace <r< B$ and the function $E$ is given in terms of the function $F$ in \eqref{eq:defFcompliquee} by the same formula as in as in Eq. \eqref{eq:symetrization}. It is computed as in the main text using the same symmetrization formula \eqref{eq:symmetrizationperformed}, and we obtain that 
\begin{equation}
    E(\vec z) = \prod_{i=1}^n \frac{e^{tz_i^2-Xz_i}}{z_i+A}\frac{1}{B^2-z_i^2}. 
\end{equation}

At this point, we may use \cite[Proposition 5.1]{borodin2016directed} backwards, and obtain that the moments of $Z_{A,B}^{\rm shifted}(X,t)$ are given by the relatively simple nested contour formula: 
\begin{equation}
\mathbb E\left[Z_{A,B}^{\rm shifted}(X,t)^n\right] = 2^n \frac{\Gamma(A+B+1)}{\Gamma(A+B+1-n)} 
\int_{r_1+\I\mathbb R}\frac{\mathrm{d}z_1}{2\I\pi} \dots \int_{r_n+\I\mathbb R}\frac{\mathrm{d}z_n}{2\I\pi}  \prod_{1\leqslant a<b\leqslant n} \frac{z_a-z_b}{z_a-z_b-1} \prod_{i=1}^n \frac{z_i}{z_i+A} \frac{1}{B^2-z_i^2}  e^{tz_i^2 - X z_i} ,
\label{eq:nestedmomentsfromnonnested}
\end{equation}
where the contours are chosen so that $B> r_1>r_2+1>\dots > r_n+n-1>\max\lbrace n-1-A, n-1\rbrace$.

\subsection{Laplace transform} 
\label{appendix:generalLaplace}
We may also use \eqref{eq:determinantalmomentsgeneral} to form the moment generating series as in Appendix \ref{appendix:Laplacetransform} above and obtain a Fredholm determinant formula. The main difference with Appendix \ref{appendix:detailsflat} is that there is a prefactor $\frac{\Gamma(A+B+1)}{\Gamma(A+B+1-n)} $ in the moment formula \eqref{eq:determinantalmomentsgeneral}. This is the reason why it is convenient to introduce an inverse gamma random variable $W$ with parameter $A+B+1$, i.e. of PDF $P(w)=\frac{1}{\Gamma(A+B+1)} w^{-A-B-2} e^{-1/w} \theta(w)$ and moments $\mathbb{E}[W^n]=\frac{\Gamma(A+B+1-n)}{\Gamma(A+B+1)} $, independent from $Z_{A,B}^{\rm shifted}(X,t)$, so that the moments of $W Z_{A,B}^{\rm shifted}(X,t)$ satisfy the same formula as \eqref{eq:determinantalmomentsgeneral} without the ratio of Gamma functions.  At this point we may reproduce the steps detailed above in Appendix \ref{appendix:Laplacetransform}. We use 
$\prod_{i=0}^{m-1} \frac{1}{w+i+A} = \frac{\Gamma (A+w)}{\Gamma (A+w+m )}$ and $\prod_{i=0}^{m-1}  \frac{1}{B^2-(w+i)^2}  
   = \frac{\Gamma (B+w) \Gamma (B-w-m +1)}{\Gamma (B-w+1) \Gamma (B+w+m )}$. We obtain 
\begin{multline} \label{eq:multiplegeneral} 
	\EE[e^{-u W Z_{A,B}^{\rm shifted}(X,t)} ] = \sum_{\ell=0}^{+\infty} \frac{1}{\ell!} 
  \int_{\mathcal C_{a_w}}\frac{dw_1}{2\I\pi} 
 \dots \int_{\mathcal C_{a_w}}\frac{dw_{\ell}}{2\I\pi}  
  \int_{\mathcal D_{a_s}[w_1]}\frac{ds_1}{2\I\pi} 
 \dots \int_{\mathcal D_{a_s}[w_{\ell}]}\frac{ds_{\ell}}{2\I\pi}  
\det\left(\frac{1}{s_i-w_j}\right)_{i,j=1}^{\ell}  \\
\times \prod_{j=1}^{\ell} \left[ \frac{e^{t G(s_j) -X \frac{(s_j-1/2)^2}{2}}}{
e^{t G(w_j) -X \frac{(w_j-1/2)^2}{2}}} \frac{\pi}{\sin(-\pi(s_j-w_j)} (2u)^{s_j-w_j}\right]
\prod_{j=1}^{\ell}  \frac{\Gamma (A+w_j)\Gamma(B+w_j)\Gamma(B-s_j+1)}{\Gamma (A+s_j )\Gamma(B+s_j)\Gamma(B-w_j+1)}.
 \end{multline}
 The contours are chosen similarly as in Appendix \ref{appendix:Laplacetransform}. More precisely,  the contour for variables $w_i$, denoted $\mathcal C_{a_w}$, is the union of two semi-infinite rays leaving the point $ a_w>\max \lbrace -A, -B \rbrace $ in the direction $\pm 2\pi/3$.  The contour for variables $s_i$, denoted  $\mathcal D_{a_s}[w]$ is formed by the union two parts: (1) a wedge shaped contour, that is the union of two semi-infinite rays leaving the point $a_s$ with  $a_w<a_s< B+1$, and (2) the union of negatively oriented circles around the poles at $w+1, w+2, \dots$ when these lie to the left of the wedge, see Fig. \ref{fig:contours2}. All infinite contours are oriented from bottom to top. 
 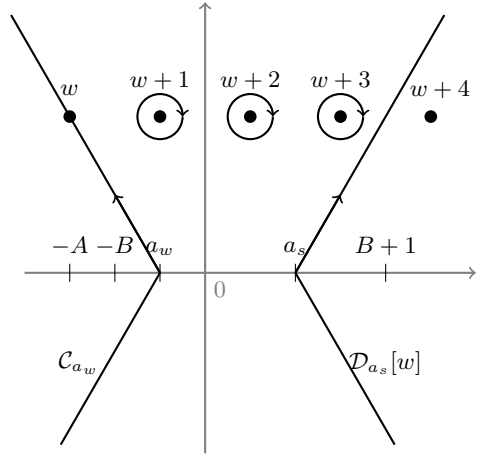
\begin{figure} 
\begin{tikzpicture}[scale=1.2]
\draw[thick, gray, ->] (-2,0) -- (3,0);
\draw[thick, gray, ->] (0,-2) -- (0,3);
\draw[gray] (0,0) node[anchor=north west] {$0$};
\draw[thick] (-1/2,0) -- +(120:3.3);
\draw[thick, ->] (-1/2,0) -- +(120:1);
\draw[thick] (-1/2,0) -- +(-120:2.2);
\draw[] (-1.5,-0.1) -- (-1.5,0.1) node[anchor=south] {$-A$};
\draw[] (2,-0.1) -- (2,0.1) node[anchor=south] {$B+1$};
\draw[] (-1,-0.1) -- (-1,0.1) node[anchor=south] {$-B$};
\draw[] (-0.5,-0.1) -- (-0.5,0.1) node[anchor=south] {$a_w$};
\draw[] (1,-0.1) -- (1,0.1) node[anchor=south] {$a_s$};
\draw[thick] (1,0) -- +(60:3.3);
\draw[thick, ->] (1,0) -- +(60:1);
\draw[thick] (1,0) -- +(-60:2.2);
\fill (-1/2,0)+ (120:2) circle(0.07);
\fill (1/2,0)+ (120:2) circle(0.07);
\fill (3/2,0)+ (120:2) circle(0.07);
\fill (5/2,0)+ (120:2) circle(0.07);
\fill (7/2,0)+ (120:2) circle(0.07);
\draw (-1/2,0.3)+ (120:2) node{$w$};
\draw (1/2,0.4)+ (120:2) node{$w+1$};
\draw (3/2,0.4)+ (120:2) node{$w+2$};
\draw (5/2,0.4)+ (120:2) node{$w+3$};
\draw (3.6,0.3)+ (120:2) node{$w+4$};
\draw[thick, ->] (0.75,0)+ (120:2) arc(0:-360:0.25);
\draw[thick, ->] (1.75,0)+ (120:2) arc(0:-360:0.25);
\draw[thick, ->] (2.75,0)+ (120:2) arc(0:-360:0.25);
\draw (-1.4,-1) node {$\mathcal C_{a_w}$};
\draw (2,-1) node {$\mathcal D_{a_s}[w]$};
\end{tikzpicture}
\caption{The contours  $\mathcal C_{a_w}$ and $\mathcal D_{a_s}[w]$ are shown in the figure. The contour $\mathcal D_{a_s}[w]$ depends on the location of $w$. For the $w$ depicted in the figure, the contour consists of the union of the wedge crossing the real axis at the point $a_s<B+1$ and small negatively oriented circles around $w+1$, $w+2$ and $w+3$, since $w+1$, $w+2$ and $w+3$ lie to the left of the wedge. } 
\label{fig:contours2}
\end{figure}
The moment formula \eqref{eq:multiplegeneral}  leads to 
\begin{equation}
\mathbb E[e^{-u W Z_{A,B}^{\rm shifted}(X,t)e^{\frac{t}{12}}}] = \det(I-K^{A,B}_{u,t})_{\mathbb L^2(0,+\infty)},
\label{eq:Laplacetransformgeneral}
\end{equation}
with the kernel 
\begin{align} \label{eq:KernelFinitetimegeneral} 
K^{A,B}_{u,t}(v,v') &= \int_{\mathbb R} \mathrm d r \frac{2u}{ e^{-r} +2u} 
\phi_{A,B,t}(v+r) \psi_{A,B,t}(v'+r) \\
\phi_{A,B,t}(v) &=  \int_{\mathcal D_{a_z}}\frac{dz}{2\I\pi} 
 e^{t \frac{z^3}{3} -X \frac{z^2}{2}-vz} \frac{ \Gamma(B+\frac{1}{2}-z)}{\Gamma (A+\frac{1}{2} + z )\Gamma(B+\frac 1 2 +z)}   \\
\psi_{A,B,t}(v) &= 
\int_{\mathcal C_{a_w}}\frac{dw}{2\I\pi} e^{-t \frac{w^3}{3} +X\frac{w^2}{2}
+ v w} \frac{\Gamma (A+ \tfrac{1}{2} +w) \Gamma(B+\frac{1}{2}+w)}{\Gamma(B+\frac{1}{2}-w)} 
\end{align}
where the contours for $z$, denoted  $\mathcal D_{a_z}$ is the union of two semi-infinite rays leaving the point $a_z< B+\frac{1}{2} $ in the direction $\pm \pi/3$,  and the contour for $w$, denoted $\mathcal C_{a_w}$, is the union of two semi-infinite rays leaving the point $ a_w>\max\left\lbrace -\left(A+\frac{1}{2}\right), -\left(B+\frac{1}{2}\right) \right\rbrace$ in the direction $\pm 2\pi/3$.

\bigskip 
{\bf Limit as $A\to+\infty$. } In this limit, $Z_{A,B}(x,t)$ converges to $Z_{Dir,B}(x,t)$, the solution to the half-space SHE  \eqref{eq:mSHEhalf-space} with Brownian IC $e^{\mathcal B(x)-(B+\frac 1 2 )x}$ and Dirichlet boundary condition, that is it satisfies the boundary condition $Z_{Dir, B}(0,t)=0$ for all $t>0$. Then, defining
\begin{equation}
    Z_{Dir,B}^{\rm shifted}(X,t) = \int_{ x\geq X} Z_{Dir,B}(x,t)dx,
\end{equation}
as in \eqref{eq:shiftedpartitionfunctionAB},  Equations
\eqref{eq:Laplacetransformgeneral} and  \eqref{eq:KernelFinitetimegeneral} become
\begin{equation}
\mathbb E[e^{-u  Z_{Dir,B}^{\rm shifted}(X,t)e^{\frac{t}{12}}}] = \det(I-K^{\infty,B}_{u,t})_{\mathbb L^2(0,+\infty)},
\label{eq:Laplacetransformgeneralinfty}
\end{equation}
where 
\begin{align} \label{eq:KernelFinitetimegeneralinfty} 
K^{\infty,B}_{u,t}(v,v') &= \int_{\mathbb R} \mathrm d r \frac{2u}{ e^{-r} +2u} 
\phi_{\infty,B,t}(v+r) \psi_{\infty,B,t}(v'+r)\\
 \phi_{\infty,B,t}(v) &=  \int_{a_z+\I\mathbb R}\frac{dz}{2\I\pi} 
 e^{t \frac{z^3}{3} - X\frac{z^2}{2} -vz}  \frac{\Gamma (B+\frac{1}{2} - z )}{\Gamma (B+\frac{1}{2} + z )} \label{eq:PhiinftyB}\\
\psi_{\infty,B,t}(v) &= 
\int_{\mathcal C_{a_w}}\frac{dw}{2\I\pi} e^{-t \frac{w^3}{3} + X\frac{w^2}{2} 
+ v w} \frac{\Gamma (B+ \tfrac{1}{2} +w)}{\Gamma (B+ \tfrac{1}{2} -w)}.
\label{eq:PsiinftyB}
\end{align}
Note that when performing the limit $A\to +\infty$ the prefactor $\frac{\Gamma(A+B+1)}{\Gamma(A+B+1-n)} $ in the moment formula \eqref{eq:determinantalmomentsgeneral}
is replaced by $A^n$ which is compensated by the total factor $A^{-n}$ from the Gamma functions inside the integrals. Hence there is no need anymore for the
variable $W$ and one obtains the finite limit in \eqref{eq:Laplacetransformgeneralinfty}.

This kernel $K^{\infty,B}$ already appeared in \cite[Prop. 1]{imamura2013stationary} and \cite[Th. 2.9]{borodin2015height} in the context of the full space KPZ with two sided Brownian IC. This implies that we have the equality in distribution, for fixed $X,t$ and $B>\frac{-1}{2}$, 
 \begin{equation}
    Z_{Dir,B}^{\rm shifted}(X,t) = 2 Z^{(B\vert B)}(-X,t)  = 2 w \tilde Z^{(B\vert B)}(-X,t),
    \label{eq:identityinlaw4} 
\end{equation}
where $Z^{(B\vert B)}(X,t)$ is the solution to the full-space SHE \eqref{eq:mSHEfull-space} with initial condition $w e^{\mathcal B(x)-( B+\frac 1 2)\vert x \vert }$ where $\mathcal B(x)$ is a two-sided Brownian motion with $\mathcal B(0)=0$ and $w$ is an independent inverse Gamma random variable with parameter $2B+1$. The last identity is trivial since
$w$ can be put in factor at all $x,t$ and $\tilde Z^{(B\vert B)}(-X,t)$ has the same IC but without the $w$ factor, as defined in the main text. As we explained in the main text, the identity in law \eqref{eq:identityinlaw4} can be seen as the limit of \eqref{eq:identityinlaw3} as $A$ goes to infinity.

\subsection{Mapping to full-space KPZ with specific initial condition}
 \label{appendix:initialdata}
In this Section, we explain the identity in law \eqref{eq:identityinlaw3} (which in particular implies the identity in distribution \eqref{eq:identityinlaw2} after letting $B$ go to $+\infty$). Recall the definition of $Z^{(B\vert A,B)}(X,t)$, that is the solution to full-line SHE \eqref{eq:mSHEfull-space} with IC depending on parameters $A,B$ and  specified by \eqref{eq:complicatedinitialdataleft} and \eqref{eq:complicatedinitialdataright}. It seems that for this quantity, moment formulas have not been written down previously, nor a Fredholm determinant representation for the moment generating function. Thus, we cannot immediately compare its distribution with the formulas \eqref{eq:nestedmomentsfromnonnested} or \eqref{eq:Laplacetransformgeneral}.  

Nevertheless, the moments of $Z^{(B\vert A,B)}(X,t)$ can be obtained from methods that are available in the literature. For this, we first need to establish a moment formula for certain partition functions of the log-gamma polymer, a directed polymer model on the lattice $\mathbb Z^2$ introduced in \cite{seppalainen2012scaling}. Then we will take the continuous limit to the KPZ equation along the lines of \cite{barraquand2020half}, and find that the moments match with \eqref{eq:nestedmomentsfromnonnested}, up to a factor $2$ that accounts for the factor $2$ in \eqref{eq:identityinlaw3}. 

\bigskip 
\textbf{Remark: }Note that for the same log-gamma polymer model, Fredholm determinant formulas are available in \cite{borodin2013log, borodin2015height, barraquand2020fluctuations}, and after taking the limit to the KPZ equation, this should allow to match with  \eqref{eq:Laplacetransformgeneral} but this route is more technical and we will not pursue it here. 

\bigskip 
We need to briefly define the log-gamma polymer partition function that we will be working with, and we refer to \cite{seppalainen2012scaling, corwin2014tropical} for details. Consider a sequence of random variables $(w_{i,j})_{i,j\geq 1}$ distributed as independent inverse Gamma random variables with parameter $\alpha_i+\beta_j$, where $\alpha_i$ and $\beta_j$ are arbitrary sequences of real numbers such that $\alpha_i+\beta_j>0$. We define the partition function 
\begin{equation}
    Z(n,m) = \sum_{\pi:(1,1)\to(n,m)} \prod_{(i,j)\in \pi} w_{i,j}, 
\end{equation}
where the sum runs over up-right paths $\pi$ in $\mathbb Z^2$ going from $(1,1)$ to $(n,m)$.  For $n_1 \geq \dots\geq n_k\geq 1 $ and $1\leq m_1\leq \dots\leq m_k,$ we have 
	\begin{equation}
	\EE\left[ \prod_{i=1}^k Z(n_i,m_i)\right]  =\oint\frac{\mathrm{d}w_1}{2\I\pi}\dots \oint\frac{\mathrm{d}w_k}{2\I\pi} \prod_{a<b} \frac{w_a-w_b}{w_a-w_b-1} \prod_{j=1}^k \left( \prod_{i=1}^{n_j}\frac{1}{\alpha_i-1/2+w_j}\prod_{i=1}^{m_j}\frac{1}{\beta_i-1/2-w_j}\right),
	\label{eq:momentformulaloggamma}
	\end{equation}
	where all integration contours are positively oriented and enclose the $-\alpha_i+1/2$ but not the $\beta_j-1/2$, and are nested such that for $i<j$, the $w_i$-contour encloses the $w_j$-contour shifted by $1$. These conditions can be satisfied only for small enough $k$. The contours may be taken as closed curves or be deformed to become infinite vertical lines.	It seems that the formula \eqref{eq:momentformulaloggamma} has not been written anywhere in the literature (though Fredholm determinant formulas for the Laplace transform are given in \cite{borodin2013log, borodin2015height, barraquand2020fluctuations}). This moment formula can be obtained by  taking  appropriate specializations and limits in \cite[Theorem 4.6]{borodin2016observables} (the appropriate specializations and limits that one needs to take are explained in many references, see e.g. \cite[Sec. 4 and 5.3]{borodin2014macdonald}).

Now, we are ready to take the continuous limit. The fact that the partition function $Z(n,m)$ converges to the solution to the SHE \eqref{eq:mSHEfull-space}  was originally proved in \cite{alberts2012intermediate, alberts2014intermediate} for general directed polymer models (see also \cite{corwin2017intermediate} for the application to the log-gamma polymer), but we will follow the arguments from the physics work \cite[Sec. 4]{barraquand2020half}. 
Assume that we scale $\alpha_i$ and $\beta_j$ such that 
\begin{equation}
\beta_1=\frac 1 2 + B, \;\;\beta_i= \frac{1}{2}+ \sqrt{n}, \;\; (i\geq 2),
\label{eq:scalings1}
\end{equation}
and 
\begin{equation}
\alpha_1=\frac 1 2 + A, \alpha_2=\frac{1}{2} + B, \;\;\alpha_i= \frac{1}{2}+ \sqrt{n}, \;\; (i\geq 3),
\label{eq:scalings2}
\end{equation}
  The rescaled partition function 
\begin{equation}
\mathcal Z_n(x,t) = n^{tn} Z(tn-x\sqrt{n}/2, tn+x\sqrt{n}/2)
\label{eq:scalings3}
\end{equation}
converges as $n$ goes to infinity \cite[Claim 4.6]{barraquand2020half} to the solution of \eqref{eq:mSHEfull-space}
with initial condition given as follows. Let $\mathcal W_1, \mathcal W_2$ and $\mathcal W_3$ be independent Brownian motions with respective drifts $-(B+\frac 1 2), -(A+\frac 1 2), -(B+\frac 1 2)$. Let $w_{11}$ be an inverse Gamma random variable with parameter $A+B+1$ and $w_{21}$ be an inverse Gamma random variable with parameter $2B+1$.  For $x\leq 0$, the initial condition is given by
\begin{equation}
Z(x,0) = w_{11}w_{21} e^{\mathcal W_1(-x)}.
\label{eq:boundarynegative}
\end{equation}
and for $x\geq 0$, 
\begin{equation}
Z(x,0) = w_{11}w_{21} e^{\mathcal W_3(x)} + w_{11} \int_0^{x} e^{\mathcal W_2(y)+\mathcal W_3(x)-\mathcal W_3(y)}\mathrm dy. 
\label{eq:boundarypositive}
\end{equation} 
Let us briefly explain how this initial condition is obtained. Note that under \eqref{eq:scalings1} and \eqref{eq:scalings2}, the weights $w_{11}$ and $w_{22}$ are independent and inverse Gamma distributed with parameters $A+B+1$ and $2B+1$.  We have that for $i\geqslant 3$ $\alpha_i+\beta_1 = B+1+\sqrt{n}$, so that under the scaling given in \eqref{eq:scalings3}, products of weights along the first row converge to $e^{\mathcal W_1(-x)}$, where $x<0$ and $\mathcal W_1$ has drift $-(B+\frac 1 2)$ (see \cite[Eq. (4.12)]{barraquand2020half} for details).
Since paths need to go through the vertices $(1,1)$ and $(2,1)$ before continuing along the first row until location $(-x\sqrt{n}, 1)$, this explain the expression \eqref{eq:boundarynegative}. Along the first columns, 
we have that for $j\geq 2$, $\alpha_1+\beta_j=A+1+\sqrt{n}$, and $\alpha_2+\beta_j=B+1+\sqrt{n}$, so that products of weights along the first row converge to $e^{\mathcal W_2(x)}$ and products of weights along the second row converge to $e^{\mathcal W_3(x)}$. We need to consider two types of paths: those going through vertices $(1,1)$, $(2,1)$, and collecting a number of weights on the second column until the location $(1,x\sqrt{n})$, hence the first term in \eqref{eq:boundarypositive}; those going through vertex $(1,1)$, then collecting a number of weights along the first column until a  location close to  $(1,y\sqrt{n})$  then collecting a number of weights on the second columns between locations  $(1,y\sqrt{n})$ and  $(1,x\sqrt{n})$, hence the second term in \eqref{eq:boundarypositive}.

\bigskip

Note that the weight $w_{11}$ is in factor of the IC $Z(x,0)$ for any $x$, so that the solution of  the full-space SHE  \eqref{eq:mSHEfull-space} with such initial condition that we have obtained as a limit of the log-gamma polymer model can be written as $Z(x,t) = w_{11}Z^{(B\vert A,B)}(x,t)$, and $w_{21}=w$, using the notations in the main text in Section \ref{sec:identities}.  
Assuming the convergence of moments, and taking the limit of the integral formula \eqref{eq:momentformulaloggamma} under the scalings \eqref{eq:scalings1}, \eqref{eq:scalings2} and \eqref{eq:scalings3}, we obtain the  following moment formula: For $x_1\leq \dots \leq x_n$, 
\begin{equation}
\EE\left[ \prod_{i=1}^n Z^{(B \vert A,B)}(x_i,t)\right]  = \frac{\Gamma(A+B+1)}{\Gamma(A+B+1-n)}  \int_{r_1+\I\mathbb R}\frac{\mathrm{d}z_1}{2\I\pi}\dots \int_{r_n+\I\mathbb R}\frac{\mathrm{d}z_n}{2\I\pi} \prod_{a<b} \frac{z_a-z_b}{z_a-z_b-1} \prod_{j=1}^n \left( \frac{1}{A+z_j}\frac{1}{B^2-z_j^2}e^{tz_j^2+x_j z_j}\right),
\label{eq:momentformulafullspace}
\end{equation}
where the contours are such that 
\begin{equation}
B> r_1 >r_2+1 > \dots >r_k+k-1, \;\;\;\text{ with } r_k>-A,-B, 
\end{equation}
and we have used that the moments of $w_{11}$ are given by 
$\EE\left[ w_{11}^k\right] = \frac{\Gamma(A+B-k+1)}{\Gamma(A+B+1)}$. Comparing with \eqref{eq:nestedmomentsfromnonnested}, we have that for $x\geq 0$, and any integer $n\geq 1$,  
\begin{equation}
2^n \mathbb E\left[ Z^{(B\vert A,B)}(-x,t)^n \right] = \EE\left[ Z^{\rm shifted}_{A,B}(x,t)^n\right],
\label{eq:equalitymoments}
\end{equation}
from which we deduce the equality in distribution \eqref{eq:identityinlaw3} (strictly speaking, an equality of moments does not imply an equality in distribution but we will ignore this mathematical subtlety).

\subsection{Degeneration as $B\to +\infty$}
\label{appendix:degenerationBtoinfty}
In the  $B \to +\infty $ limit,  we need to multiply both members of \eqref{eq:equalitymoments} by $B^n$ before taking the limit. Then, the full space solution in the l.h.s. of \eqref{eq:equalitymoments} has half-Brownian IC (in the limit), as can be seen from  \eqref{eq:complicatedinitialdataleft} and \eqref{eq:complicatedinitialdataright}, that is, on $\mathbb R_+$ the initial condition is the exponential of a Brownian motion with drift $-(A+\frac 1 2)$ and on $\mathbb R_+$ the initial condition is zero. The half-space solution involved in the RHS of \eqref{eq:equalitymoments} has Robin type boundary condition with parameter $A$, and delta at $0$ IC in the limit (i.e. droplet initial condition). Hence, we obtain the identity in distribution \eqref{eq:identityinlaw2}. 

\bigskip 

The identity in distribution can also be obtained by a comparison of Fredholm determinant formulas. Indeed, 
the kernel $K_{u,t}$ already appeared in \cite{imamura2011replica}. This paper was considering the solution $Z^{(A)}(x,t)$ to the full-space SHE \eqref{eq:mSHEfull-space} with half-Brownian IC $Z^{(A)}(x,0) = e^{\mathcal B(x)- (A+\frac 1 2 ) x}$  for $x>0$ and $Z^{(A)}(x,0) = 0$ for $x<0$. In fact the solution was obtained there for any $x$ but in the absence of the drift (i.e. for $A=-1/2$),
however it is immediate to extend it to arbitrary drift, using the statistical symmetry (see section \ref{appendix:sts}). 
Comparing the formula \eqref{eq:Laplacetransformsuppmat} 
with \cite[Prop. 2]{imamura2011replica}, we obtain the identity in distribution \eqref{eq:identityinlaw}. The correspondence of notations is as follows:
one must set $\alpha=1$, $\gamma_t=t^{1/3}$ there and here $u=e^{-t^{1/3} s}$

\section{Tilt symmetry} 
\label{appendix:sts} 

Let us consider the SHE on the full line \eqref{eq:mSHEfull-space} with standard space time white noise $\xi(x,t)$.
Suppose that $\{ Z(x,t) \}_{x \in \mathbb{R} , t > 0}$ is a solution with IC $Z(x,0)=Z_0(x)$. Consider now for any fixed real $a$
\be
\tilde Z(x,t)= e^{a x + a^2 t} Z(x + 2 a t,t) .
\ee
Since $\tilde \xi(x,t)= \xi(x+2 a t,t)$ is also a standard space time white noise, 
$\tilde Z(x,t)$ is also a solution of the SHE \eqref{eq:mSHEfull-space}, with IC $\tilde Z(x,0)=\tilde Z_0(x)= e^{a x} Z_0(x)$
in another realization of the noise. Hence the (statistical) tilt symmetry (STS) relates the statistics of the solutions of the SHE with
``tilted'' initial conditions. In the particular case of the droplet IC, $\tilde Z_0(x)=Z_0(x)=\delta(x)$ these IC are identical
and the statistics of $\{ Z(x,t) \}_{x \in \mathbb{R} , t > 0}$ and $\{ \tilde Z(x,t) \}_{x \in \mathbb{R} , t > 0}$
are thus identical (as space time processes).
\\

{\bf Half-Brownian IC}. Let us denote now $Z_v(x,t)$ the solution with the half Brownian IC with drift $v$, i.e. $Z_v(x,0)= e^{B(x) + v x} \theta(x)$.
The solution $Z_{v+a}(x,t)$ with a half Brownian IC with drift $v+a$, i.e. $Z_{v+a}(x,0)= e^{B(x) + (v+a) x} \theta(x)$
can thus be constructed using the STS, i.e. one has in law
\be
Z_{v+a}(x,t) = e^{a x + a^2 t} Z_v(x + 2 a t,t) .
\ee
This is an identity between space time processes. We want now to focus only on the distribution at a single fixed
space-time point $(x,t)$. Then we can choose $a=-x/(2 t)$ and $v = w -a $ and obtain the equality in law
\be
Z_{w}(x,t) = e^{- \frac{x^2}{4 t}} Z_{w +\frac{x}{2t} } (0,t) .
\ee
Setting $w=-(A+\frac{1}{2})$ and $x=-X$ we obtain the identity in law $Z^{(A)}(-X,t) = e^{- \frac{X^2}{4 t} } Z^{(A+ \frac{X}{2 t})}(0,t)$ given in the main text.
\\

{\bf Droplet IC}. Consider now the solution $Z(x,t)=Z(x,t|0,0)$ of the full line SHE with the droplet IC. Let us define $G(\mathsf f,t)=\int_{-\infty}^{+\infty} dx e^{\mathsf f x} Z(x,t)$.
It is the partition sum of a directed polymer with one fixed endpoint at $(0,0)$ and one free endpoint $(x,t)$ but with 
an applied force $\mathsf f$ on that endpoint. From the STS property we have that $G(\mathsf f,t)$ has the same distribution as 
\be 
\int_{-\infty}^{+\infty} dx e^{\mathsf f x} e^{a x + a^2 t} Z(x + 2 a t,t) = G(\mathsf f+a,t) e^{- (a^2 + 2 a \mathsf f) t}.
\ee 
Hence, choosing $a=-\mathsf f$
\be 
\mathbb{E} \log G(\mathsf f,t) = \mathbb{E} \log G(0 ,t) + \mathsf f^2 t .
\ee 
It follows, by differentiation, that the averaged thermal cumulants of the free endpoint $(x,t)$ in the absence of the force, i.e. for $\mathsf f=0$,
are simply $\mathbb{E} \langle x^p \rangle_c = \partial_{\mathsf f}^p \mathbb{E} \log G(\mathsf f,t)
= 2 t \delta_{p,2}$ on the full line, as mentioned in the main text. Similar remarkable identities for thermal fluctuations occur in a larger class of 
disordered models \cite{schulz1988thermal}. While it is valid for any $t$, for large $t$ this result is usually 
interpreted within the droplet picture \cite{fisher1991directed,monthus2004low}.
The typical Gibbs measure of the endpoint is localized, i.e. the thermal fluctuations of the endpoint are typically $\delta x = O(1)$.
However, with probability $p(t) \sim T/t^{1/3}$ (where the temperature is $T=1$ here) there exists two distant states, almost degenerate
in energy (within $O(T)$): the Gibbs measure is splitted between them and that leads to a much larger $\delta x \sim t^{2/3}$. Putting
these factors together leads to $\mathbb{E} \langle x^2 \rangle_c \sim T t^{-1/3} t^{4/3} \sim T t$

Note that by the tilt symmetry the polymer configurations are mapped into each others. In the case of the half-space the STS maps 
a problem with a vertical wall to a problem with a tilted wall, so a priori one cannot readily use it. In the main text we have 
found the curious relation \eqref{eq:identityinlaw2} and used on the r.h.s. the STS for the half-Brownian in full space
(shown above) to deduce the equality in law $Z_A^{\rm shifted}(X,t)=e^{- \frac{X^2}{4 t} } Z_{A+\frac{X}{2t}}^{\rm shifted}(0,t)$.
Although it has a flavor of STS in half space, it is not, and in fact there is no simple correspondence between the polymer
trajectories on both sides of this relation. 

A similar puzzle occurs upon applying a force $\mathsf f$ to the endpoint in the half space. A tilt transformation which removes the 
force would also tilt the wall, so no obvious consequence can be obtained. Nevertheless, as shown in the main text, the
force induces an additional drift in the drifted Brownian stationary measure leading to the simple shift $A \to A+\mathsf f$.
There also, it does not seem to exist any simple picture in terms of tilted polymer paths.

\section{Replica Bethe ansatz approach}
\label{appendix:betheansatz}
In this Appendix we explore the interplay between the energy spectrum of the replica delta Bose gas in the half-space and the stationary measure of increments
of partition function that is used in the main text in Section \ref{sec:stationaryendpoint} to study endpoint distributions of polymers.

\subsection{Moments of partition sum} 

The replica Bethe ansatz method (RBA) allows to write the multipoint equal time moments of $Z(x,t)$, solution of the SHE equation (in full or half space), as
a quantum mechanical expectation (denoting $\vec x=(x_1,\dots,x_n)$)
\be \label{eq:sum}
\mathbb E [ Z(x_1,t) \cdots Z(x_n,t) ] = \langle \vec x | e^{- t H_n} |\Psi(t=0) \rangle = 
\sum_\mu \Psi_\mu(x_1,\dots,x_n) \langle \Psi_\mu | \Psi(t=0) \rangle \frac{e^{- t E_\mu}}{||\Psi_\mu||^2} .
\ee
i.e. a sum over the unnormalized eigenfunctions $\Psi_\mu$ (of norm denoted $||\Psi_\mu ||$) of the $n$-body Lieb-Liniger (LL) Hamiltonian 
\begin{equation}H_{n} = -\sum_{j=1}^n \frac{\partial^2}{\partial {x_j^2}}  - 2  \sum_{1 \leqslant  i<j \leqslant n} \delta(x_i - x_j),\end{equation}
with eigenenergies $E_\mu$. In \eqref{eq:sum}, we have denoted $\Psi_\mu(x_1,\dots,x_n) = \langle \vec x |\Psi_\mu \rangle$ and the initial state $|\Psi(t=0) \rangle$ encodes the initial condition of the SHE, with 
$\langle \vec x | \Psi(t=0) \rangle = \prod_{i=1}^n Z(x_i,0)$ for a deterministic IC and 
$\langle \vec x | \Psi(t=0) \rangle = \mathbb E[ \prod_{i=1}^n Z(x_i,0) ]$ for a random IC. These initial conditions
being symmetric in $(x_1,\cdots,x_n)$ only the symmetric, i.e. bosonic, eigenstates contribute to the sum in \eqref{eq:sum}. 
The representation \eqref{eq:sum} is valid in full space and half space. The sum \eqref{eq:sum} is weighted by the overlaps
$\langle \Psi_\mu | \Psi(t=0) \rangle = \int \prod_i dx_i \Psi_\mu^*(\vec x)  \langle \vec x | \Psi(t=0) \rangle$. 

\bigskip 
In the half space case $H_n$ acts on wave-functions
which satisfy the boundary conditions $\partial_{x_i} \Psi(\vec x)|_{x_i=0} = A \Psi(\vec x)|_{x_i=0}$, $i=1\dots ,n$
(hence the eigenstates $\Psi_\mu$ satisfy this condition). The $n$ body spectrum of the half-space problem is complicated and was obtained from the Bethe ansatz in \cite{deNardisPLDTT}. We also refer
to \cite{deNardisPLDTT} for a more detailed presentation and for the references to the literature on the 
Bethe ansatz for the half-space LL model. In addition to the usual bulk string bound states which exist in the full space problem and have an arbitrary center of mass momentum, 
there are also $n$ body boundary bound states which are localized at the boundary.

The ground state (lowest energy state) was found by Kardar in \cite{kardar1985depinning}, and this was confirmed in \cite{deNardisPLDTT}. One has

\begin{enumerate} 
\item For $n\leqslant 1+2 A$ the ground state is a state made of a single bulk string with a vanishing momentum. Far from the boundary
its wave-function behaves like the ground state of the full-space problem 
\begin{equation}\Psi_0(\vec x) \sim e^{-\frac{1}{2} \sum_{1\leq i < j \leqslant n} |x_i - x_j| }  = e^{\frac{1}{2} \sum_{j=1}^n (n+1-2 j) x_j}, \;\;\text{ for large }x_1 \leq \dots \leq x_n. 
\label{eq:groundstateapprox}
\end{equation}
The ground state energy is $E_0(n) = - \frac{1}{12} n (n^2-1)$. 
\item  For $n > 1 + 2 A$ the ground state is made of a single boundary string. 
The ground state energy is $E_0(n) = - n \left(A + \frac{1}{2} (1-n)\right)^2 - \frac{1}{12} n (n^2-1)$. It has the form
\be 
\Psi_0(\vec x) \propto e^{ \sum_{j=1}^n ( A - j+1 ) x_j  }, \;\;\text{ for any }  0 \leq x_1 \leq x_2 \leq \cdots \leq x_n,  
\label{eq:groundstatebound}
\ee 
and its norm was computed in \cite{deNardisPLDTT}. Note that shifting all $x_i \to x_i + \bar x$ the wave-function decays as $e^{- \frac{\bar x}{2} n (n - (1+ 2 A))}$.
\end{enumerate}
 
Exactly at the transition for $n = 1 + 2A$, the two states are identical as the RHS of \eqref{eq:groundstateapprox} and \eqref{eq:groundstatebound} match.  
This state should be considered as a bulk string ground state as its center of mass is delocalized in the full volume. The ground state energy is continuous across the transition.

In the limit $t \to +\infty$, for any fixed positive integer $n$, the sum over states is dominated by the ground state $\Psi_0(\vec x)$. It is thus
tempting to follow the following two steps: 
\begin{enumerate}
\item[(i)] write 
\begin{eqnarray} \label{eq:limit}
\mathbb E [ Z_A(x_1,t) \cdots Z_A(x_n,t) ]  \simeq \frac{\Psi_0(\vec x)}{||\Psi_0||^2}  \langle \Psi_0 | \psi(t=0) \rangle e^{- E_0(n) t}
\end{eqnarray}
where $Z_A(x,t)$ solves the half-space SHE \eqref{eq:mSHEhalf-space}, 
\item[(ii)] postulate that the form of the expression found for positive integer $n$ can be extended to real $n>0$ in the limit $n\to 0 $ to calculate moments.

\end{enumerate}

This is what was done by Kardar in \cite{kardar1985depinning}
to predict $\mathbb E (\langle x \rangle)$ in the bound phase. 
In the following it will be convenient to rewrite \eqref{eq:limit} as 
\begin{eqnarray} \label{eq:limit2}
\mathbb E [ Z_A(x_1,t) \cdots Z_A(x_n,t) ] \simeq  c_n(t) \tilde \Psi_0(\vec x)  , \quad \quad c_n(t)=\mathbb E [ Z_A(0,t)^n ]  , \quad \quad
\tilde \Psi_0(\vec x)|_{x_1 \leq \cdots \leq x_n} = e^{ \sum_{j=1}^n ( A - j+1 ) x_j  } .
\end{eqnarray}
 In that form it is clear that the continuation of $c_n(t)$ to $n=0$ is simply unity. 

\bigskip 
It is  well known that in the full space problem (and we can expect the same in the unbound phase for the half-space problem), in step (i) the amplitude in \eqref{eq:limit} is not correct since the spectrum is gap-less and one must further integrate over the low lying center of mass excitations, but this integration can be performed for a given initial condition. In step 
(ii), more severely, the limit $t \to +\infty$ and $n \to 0$ do not commute (i.e. one would need to perform the limit $n \to 0$ on the full sum and then take the limit $t \to +\infty$). 
However, once these two issues are addressed, this program enables to obtain the right tails of the free energy $\log Z$ \cite{le2016large,de2017mutually}. 

\bigskip 
In the bound phase however, for $A<-1/2$, step (i) is more reasonable as there are no center of mass excitations and there is a finite gap between 
the ground state and the excited states \cite{deNardisPLDTT}, so \eqref{eq:limit} should give the correct asymptotics. In step (ii) it is quite likely that the 
limits $t \to +\infty$ and $n \to 0$ commute in that case: for $A<-1/2$ the ground state holds for any $n >0$, and it is a system of effectively finite size.
Indeed our results below confirm that. 

\subsection{Thermal cumulants of the endpoint position via the RBA} 

Let us recall the definition of the endpoint distribution (in a given noise realization) and of the thermal averages 
\be 
{\cal P}(x,t) = \frac{Z_A(x,t)}{Z} , \quad  \quad Z=\int_0^{+\infty} dy Z_A(y,t) , \quad  \quad \langle O(x) \rangle = \int_0^{+\infty} dx  O(x) {\cal P}(x,t),
\ee 
where the time dependence of the averages is implicit.
Note that we have not specified the initial condition, so it can be fixed endpoint at $t=0$ in some position, or more general.
Let us first calculate the thermal cumulants.
\\

The generating function $f(v)$ of the averaged thermal cumulants can be written as
\begin{align}
f(v)= \mathbb E \left[ \log \int_0^{+\infty} dx e^{v x} Z_A(x,t) \right]  &= \partial_n \mathbb E \left[ \left(\int_0^{+\infty} dx e^{v x} Z_A(x,t)\right)^n \right]\Bigg\vert_{n \to 0}\\ 
&= \partial_n  \int_0^{+\infty} d\vec x e^{v \sum_{i=1}^n x_i } \mathbb E \left[ Z_A(x_1,t) \cdots Z_A(x_n,t) \right]\Big\vert_{n \to 0}.
\end{align}
Thus in the large time limit (under the above assumptions) it becomes
\be
\mathbb E [ \langle x^p \rangle_c ] = \partial_{v}^p f(v)\big\vert_{v=0} , \quad \quad f(v) = \partial_n ( c_n(t) I_n(v) )\big\vert_{n=0}  , \quad \quad  I_n(v) := \int d\vec x e^{v \sum_{i=1}^n x_i } \tilde \Psi_0(\vec x).
\ee
Let us calculate the integral 
\be
 I_n(v) 
= n! \int_{0<x_1<\dots<x_n} e^{\sum_{j=1}^n ( A + v - j+1 ) x_j } 
= n! \prod_{j=1}^n \frac{-2}{j(1 + 2 A + 2 v + j-2 n)} 
= \frac{2^n \Gamma (-2 A+n-2 v-1)}{\Gamma (-2 A+2 n-2 v-1)} 
\ee
where we have used the identity
\begin{eqnarray}
 \int_{0<y_1<\dots<y_p} e^{\sum_{j=1}^p z_j y_j } = \prod_{j=1}^p \frac{-1}{z_p+\dots+z_{p-j+1}}
\end{eqnarray}
We see that $I_0(v)=1$ as expected, and recall that $c_0(t)=1$, so that  
\be
f(v) = \partial_n c_n(t) + \partial_n I_n(v) \big\vert_{n=0}= \mathbb E[ \log Z_A(0,t) ] + \log (2)-\psi(-2 A-2 v-1).
\ee
Note that this is exactly compatible with what was obtained in the main text around \eqref{eq:gencum} since 
\be 
\mathbb E \left[ \log \int_0^{+\infty} dx e^{v x} Z_A(x,t) \right] - \mathbb E\left[ \log Z_A(0,t) \right] = \mathbb E \left[ \log \int_0^{+\infty} dx e^{v x} \frac{Z_A(x,t)}{Z_A(0,t)} \right] 
\ee 
where the limit $\frac{Z_A(x,t)}{Z_A(0,t)}$ was shown to converge to the exponential of the Brownian with drift $A+\frac{1}{2}$. 
So the RBA method reproduces exactly the result \eqref{eq:gencum} for the general averaged thermal cumulant. In the 
case $p=1$ this is the result obtained by Kardar \cite{kardar1985depinning}.
This coincidence between the results of the RBA and of the method used in the main text appears to extend to all moments 
as we now discuss.

\subsection{Comparison of the two methods and general moments}

Let us put side by side the results of the two methods. In the method based on stationary measures of increments described in the main text in Section \ref{sec:stationaryendpoint}, one states that 
\be \label{resstat}
\mathbb E \left[ \frac{Z_A(x_1,t)}{Z_A(0,t)} \cdots \frac{Z_A(x_n,t)}{Z_A(0,t)} \right] \xrightarrow[t\to+\infty]{} 
\Phi_0(\vec x) :=
\mathbb E \left[  e^{\sum_{i=1}^n {\cal B}(x_i) + (A+ \frac{1}{2}) x_i }\right] ~,~
\Phi_0(\vec x)|_{x_1 \leq \cdots \leq x_n} = e^{ \sum_{j=1}^n  (n - j + \frac{1}{2}) x_j + (A+ \frac{1}{2}) x_j} 
\ee
while in the RBA method, one obtains
\be \label{resRBA} 
\mathbb E [ Z(x_1,t) \cdots Z(x_n,t) ]  \simeq_{t \to +\infty} \mathbb E [ Z(0,t)^n ] \, \tilde \Psi_0(\vec x) \quad , \quad  
\tilde \Psi_0(\vec x)|_{x_1 \leq \cdots \leq x_n} = e^{ \sum_{j=1}^n ( A - j+1 ) x_j  }  
\ee
The functions $\Phi_0(\vec x)$ and $\tilde \Psi_0(\vec x)$ are both fully symmetric in their arguments and very similar, although different. 
\\

{\bf General moments}.
The multipoint average of the endpoint distribution can be written as
\be \label{120} 
 \mathbb E [{\cal P}(x_1,t) \cdots {\cal P}(x_p,t) ] =  
\mathbb E \left[ \frac{1}{Z^p} Z_A(x_1,t) \cdots Z_A(x_p,t)\right] = \lim_{n \to 0} \mathbb E \left[ \int_0^{+\infty} dx_{p+1} \dots \int_0^{+\infty} dx_n Z_A(x_1,t) \cdots Z_A(x_n,t) \right]
\ee 
where we used that $\frac{1}{Z^p} = \lim_{n \to 0} Z^{n-p}$, and we recall that $Z=\int_0^{+\infty} dy Z_A(y,t)$.

Using the RBA result \eqref{resRBA}, in the large time limit it thus becomes
\be \label{eq:gengibbs} 
\mathbb E [{\cal P}(x_1,t) \cdots {\cal P}(x_p,t) ]  \simeq \lim_{n \to 0} \int_0^{+\infty} dx_{p+1} \dots \int_0^{+\infty} dx_n \, \tilde \Psi_0(\vec x) 
\ee
Upon multiplication by $O_1(x_1) \dots O_p(x_p)$ and integration, it implies in particular for the most general type of moment
\be  \label{eq:genmom} 
\mathbb E [ \langle O_1(x) \rangle \cdots \langle O_p(x) \rangle] \simeq \lim_{n \to 0} \int_0^{+\infty} dx_1 \dots \int_0^{+\infty} dx_n 
O_1(x_1) \dots O_p(x_p) \, 
\tilde \Psi_0(\vec x)
\ee

Within the method based on the stationary measure, one has from \eqref{eq:gibbs1}, $\lim_{t\to\infty}\mathcal P_{A}(x,t)  = p_A(x)$ with $p_A(x)= \frac{1}{{\sf Z}} e^{\mathcal{B}(x)+(A+\frac{1}{2})x}$
and ${\sf Z}={\int_0^{+\infty} dy e^{\mathcal{B}(y)+(A+\frac{1}{2})y}}$. Using the same steps as in \eqref{120} with $\frac{1}{{\sf Z}^p} = \lim_{n \to 0} {\sf Z}^{n-p}$
one obtains using \eqref{resstat}
\be \label{121} 
 \mathbb E [p_A(x_1) \cdots p_A(x_p) ] =  \lim_{n \to 0} \int_0^{+\infty} dx_{p+1} \dots \int_0^{+\infty} dx_n \Phi_0(\vec x)
\ee 
which provides a starting formula for the evaluations of the many point correlations of the stationary measure via the replica method.

We can now compare \eqref{121} and \eqref{eq:gengibbs} and use that $\Phi_0(\vec x)$ and $\tilde \Psi_0(\vec x)$ are very similar. More
precisely, $\Phi_0(\vec x) = e^{ n \sum_{i=1}^n x_i} \tilde \Psi_0(\vec x)$, i.e. they differ by a term which contains $n$ explicitly
and vanishes at $n=0$. This indicates that the results of the two methods for
the endpoint probability correlations, and thus for all the moments, are the same. 
\\

{\bf Correlation between endpoint positions and free energy}. The RBA ground state $\tilde \Psi_0(\vec x)$ thus seems to contain the information about the stationary measure of the partition sum ratios. However it should contain
more, and encode also for some information about the correlations between $Z(0,t)$ and these ratios, or with the endpoint distribution, which remains to be explored. Indeed putting together 
\eqref{resstat} and \eqref{resRBA} we obtain that at large time
\be 
\mathbb{E}[ Z(0,t)^n Z_t(x_1) \dots Z_t(x_n) ] \simeq e^{ - n \sum_{i=1}^n x_i}  \mathbb{E}[ Z(0,t)^n ] \, \mathbb{E}[Z_t(x_1) \dots Z_t(x_n) ]
\ee 
where we denoted the ratios as $Z_t(x)=Z(x,t)/Z(0,t)$. Taking $x_i=x$ for $i=1,\dots,p$ and $x_i=0$ for $i=p+1,\dots,n$, it gives 
$\mathbb{E}[ Z(0,t)^n Z_t(x)^p] = e^{ - n p x }  \mathbb{E}[ Z(0,t)^n ] \, \mathbb{E}[Z_t(x)^p ]$ for all positive integers $n \leq p$.
This suggest that at large time the only non zero joint cumulant between $\log Z(0,t)$ and $\log Z_t(x)$ is the two point covariance
\be  \label{eq:cov} 
{\rm Cov}( \log Z(0,t) , \log Z_t(x) ) = -  x .
\ee 
It is possible to obtain some understanding of how this relation (which is a conjecture at this stage) could come about. At large $t$ and large $x$ 
we may approximate the polymer partition functions by the exponential of the energy collected along geodesics (i.e. paths with maximal energy).
Consider the geodesics $\gamma_1(t)$ from $(0,0)$ to $(0,t)$ and $\gamma_2(t)$ from $(0,0)$ to $(x,t)$. They first coincide
and then split at some point $M$ of coordinate $(x',t-\tau)$, see Fig. \ref{fig:geodesics}. 
\begin{figure}
    \centering
    \begin{tikzpicture}[scale=.8,every text node part/.style={align=center}]
\fill[pattern=north east lines, pattern color=gray] (-1,0) rectangle (0,8);
\draw[thick, gray, ->] (0,0) -- (4,0);
\draw[thick, gray, ->] (0,0) -- (0,8.2);
\draw[dashed, gray] (0,8) -- (4,8);

\draw[thick] (0,0) node[anchor=north east] {$(0,0)$} -- (0.0448824,0.1) -- (0.1397217,0.2) -- (0.1541936,0.3) -- (0.1215552,0.4) -- (0.0506304,0.5) -- (0.0541346,0.6) -- (0.0441635,0.7) -- (0.0332063,0.8) -- (0.100197,0.9) -- (0.0509484,1.) -- (0.0165682,1.1) -- (0.0179117,1.2) -- (0.063864,1.3) -- (0.0953191,1.4) -- (0.049668,1.5) -- (0.0585695,1.6) -- (0.0722703,1.7) -- (0.0539764,1.8) -- (0.0566139,1.9) -- (0.0218539,2.) -- (0.0715931,2.1) -- (0.0701884,2.2) -- (0.0965294,2.3) -- (0.0860858,2.4) -- (0.0572306,2.5) -- (0.0398316,2.6) -- (0.0134104,2.7) -- (0.0103617,2.8) -- (0.0512422,2.9) -- (0.0634241,3.) -- (0.10637,3.1) -- (0.145781,3.2) -- (0.136951,3.3) -- (0.131881,3.4) -- (0.154119,3.5) -- (0.114936,3.6) -- (0.135136,3.7) -- (0.119598,3.8) -- (0.0974727,3.9) -- (0.133238,4.) -- (0.139568,4.1) -- (0.102674,4.2) -- (0.0619342,4.3) -- (0.022134,4.4) -- (0.0565248,4.5) -- (0.0420673,4.6) -- (0.0288595,4.7) -- (0.0730709,4.8) -- (0.121093,4.9) -- (0.120799,5.) -- (0.119571,5.1) -- (0.144967,5.2) -- (0.137048,5.3) -- (0.152181,5.4) -- (0.162283,5.5) -- (0.178031,5.6) -- (0.211436,5.7) -- (0.189178,5.8) -- (0.160759,5.9) -- (0.204035,6.) -- (0.198828,6.1) -- (0.155113,6.2) -- (0.105908,6.3) -- (0.0681244,6.4) -- (0.0436301,6.5) -- (0.00308055,6.6) -- (0.0207256,6.7) -- (0.0695581,6.8) -- (0.0401852,6.9) -- (0.0520343,7.) -- (0.0431129,7.1) -- (0.0749229,7.2) -- (0.0927092,7.3) -- (0.0677219,7.4) -- (0.074485,7.5) -- (0.0875026,7.6) -- (0.0951467,7.7) -- (0.111307,7.8) -- (0.061556,7.9) -- (0,8.)   node[anchor=south east] {$(0,t)$};
\fill (0,0) circle(0.08);
\fill (0,8) circle(0.08);
\fill (0.204035,6.) circle(0.08);
\fill (3.12485,8.) circle(0.08);
\draw[thick] (0.204035,6.) node[anchor= north west] {$M=(x', t-\tau)$} -- (0.334491,6.1) -- (0.503763,6.2) -- (0.789787,6.3) -- (1.04203,6.4) -- (1.2609,6.5) -- (1.56604,6.6) -- (1.69513,6.7) -- (1.83813,6.8) -- (1.9276,6.9) -- (2.24149,7.) -- (2.25485,7.1) -- (2.41854,7.2) -- (2.50315,7.3) -- (2.76659,7.4) -- (2.73451,7.5) -- (2.73089,7.6) -- (2.84652,7.7) -- (2.89343,7.8) -- (2.96084,7.9) -- (3.12485,8.) node[anchor=south] {$(x,t)$};
\draw (1.5,3) node{$\gamma_{1}(s)=\gamma_2(s)$};
\draw (0.6,7) node{$\gamma_1(s)$};
\draw (2.7,6.7) node{$\gamma_2(s)$};
\draw[gray, <->] (4,6) -- (4,8);
\draw[black!70] (5.2,7) node{$\tau=\frac{x}{2\vert A+\tfrac{1}{2}\vert}$};
\draw (7,1.5) node{$E_{12}= \int_0^{t-\tau} \xi(\gamma_1(s))ds$};
\draw (7,3) node{$E_{1}= \int_{t-\tau}^t \xi(\gamma_1(s))ds$};
\draw (7,4.5) node{$E_{2}= \int_{t-\tau}^t \xi(\gamma_2(s))ds$};
    \end{tikzpicture}
    \caption{Illustration of the geodesics $\gamma_1= (\gamma_1(s),s)_{0\leqslant s\leqslant t}$ and $\gamma_2 = (\gamma_2(s),s)_{0\leqslant s\leqslant t}$.}
    \label{fig:geodesics}
\end{figure}
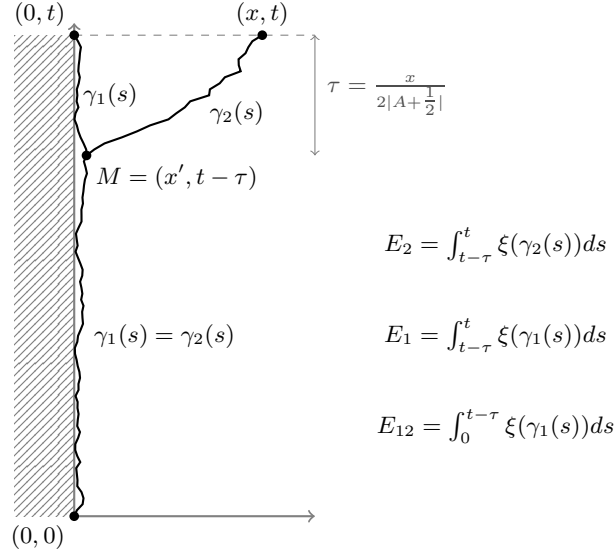
In the bound phase $A<-1/2$, $\gamma_1(t)$ remains close to the wall, and $x'$ remains
bounded. Let us call $E_{12}$ the energy of the common segment (that is $E_{12}= \int_0^{t-\tau} \xi(\gamma_1(s))ds$), $E_1$ the one of the segment from $M=(x',t-\tau)$ to $(0,t)$, and $E_2$ from $M$ to $(x,t)$. 
At large $t$, one has $\log Z(0,t) \simeq E_1 + E_{12}$ and $\log Z(x,t) \simeq E_2 + E_{12}$. Conditionally
on the position of $M$, $E_{12}$ is independent from $E_1$ and $E_2$, and for large $t$ and $x$, $E_1$ and $E_2$ are asymptotically independent. Hence the left hand side of \eqref{eq:cov} is thus 
\be 
{\rm Cov}(  E_{12}+ E_1 , E_2 - E_1 ) = {\rm Cov}( E_1 , E_2 - E_1) \simeq - {\rm Var}(E_1). 
\label{eq:estimategeodesics}
\ee 
We have obtained in the main text, see also \cite{deNardisPLDTT},  that for large $\tau$, we have  ${\rm Var} E_1 \simeq 2 \tau |A+ \frac{1}{2}|$. Further one expects that for large $\tau$ and  $x$,  the length $\tau$ is proportional to $x$. If one equate the elastic energy $x^2/(4 \tau)$ with $(A+\frac{1}{2})^2 \tau$ one obtains $\tau \simeq x/(2 |A+ \frac{1}{2}|)$,
which makes \eqref{eq:estimategeodesics} consistent with \eqref{eq:cov} in the limit of large $x$. A similar understanding when $x$ is not going to infinity seems more difficult.

\bigskip

{\bf Remark: midpoint}. The partition function with two fixed endpoints in $x=0$ at times $0$ and $2 t$ and a given midpoint position $x(t)=x$
is given by 
$Z(0,2t|x,t) Z(x,t|0,0)$, which has the same law as two independent copies of $Z(x,t|0,0)$. The moments of this partition function will
thus be the square of the moments of $Z(x,t|0,0)$. Within the RBA it will thus amount to the same formula as above, replacing 
$\tilde \Psi_0(x) \to \tilde \Psi_0(x)^2$. From \eqref{resRBA} it is an exponential 
linear in the $x_i$, hence this replacement amounts to change $x_i \to 2 x_i$. This agrees with the result of the main text.

\bigskip 

{\bf Remark: basin of attraction}. We expect that the property of ``ground state dominance'' \eqref{eq:limit} in the RBA will hold in the bound phase upon some condition on the initial condition
(i.e. on the behavior of the overlaps). The condition for the convergence to the stationary measure for the ratios was discussed in the main text in Section \ref{sec:stationaryendpoint}. It would be interesting to see whether 
it can also be obtained with the RBA. 

\section{Correlations of $p_A(x)$ via Liouville quantum mechanics}
\label{appendix:Liouville}

The stationary measure for the endpoint position at large time in the bound phase is $p_A(x)$ given in \eqref{eq:gibbs1}. It is possible to compute its
$m$-point correlations (and therefore all the moments) using methods developped in \cite{monthus1994flux,comtet1998exponential} and in \cite{TexierComtetSUSY} 
based on stochastic processes, replica, and most notably on Liouville quantum mechanics. Other works adressed similar questions
in various contexts \cite{shelton1998effective,nagar2006strong,quinn2015scaling}, often motivated by multifractal properties of eigenfunctions
of random Dirac type operators. Although it is a simple extension of these works (which often focus on periodic boundary conditions) 
the formula for the case of the Brownian (i.e. with free boundary conditions) and in presence of a drift have not been given, so we 
display them here (for details of the method we refer to \cite{TexierComtetSUSY} and \cite{monthus1994flux}). 

\subsection{Moments of $p_A(x)$} 
Let us denote $w=-(A+ \frac{1}{2}) >0$. As in these works we consider a finite $L$ truncation, denoting
$Z_L^w = \int_0^L dx e^{{\mathcal B}(x) - w x}$, and we take $L \to +\infty$ at the end. A simple but useful observation \cite{TexierComtetSUSY,comtet1998exponential} is 
that $p_A(x)$ in \eqref{eq:gibbs1} can be rewritten as 
\be
 p_A(x)= \lim_{L\to\infty } \frac{1}{\int_0^L dy e^{{\mathcal B}(y) - {\mathcal B}(x) - w (y-x)} } = \lim_{L\to\infty} \frac{1}{ Z_x^{-w} + \tilde Z_{L-x}^w } =  \frac{1}{ Z_x^{-w} + \tilde Z_{\infty}^w } 
\ee
where $\tilde Z$ contains an independent realisation of the Brownian. This is obtained splitting the integral over $y$ on the two
interval $[0,x]$ and $[x,L]$ and performing the change of variable $y=x-z$ on the first and $y=x+z$ on the second, with $z$ positive.
In the last equation we used the property that $Z_L^w$ converges for $L \to +\infty$ to an inverse Gamma random variable
$\tilde Z_{\infty}^w = \Gamma(2 w,\frac{1}{2})$ where here $1/2$ is the scale parameter, i.e. $z=\tilde Z_{\infty}^w$
has PDF $p(z)=\frac{1}{2 \Gamma(2 w)} (\frac{2}{z})^{1+2 w} e^{-2/z}$. Thus the moments of $p_A(x)$ are given by
\bea \label{pn1} 
&& \mathbb{E} [p_A(x)^n] =
\int_0^{+\infty} dp \frac{p^{n-1}}{\Gamma(n)} \mathbb{E}\left[e^{- p Z^{-w}_x}\right] \,  \mathbb{E} \left[e^{- p Z_{\infty}^w} \right]  = 
 \frac{2}{\Gamma(2 w)}  \int_0^{+\infty} dp \frac{p^{n-1}}{\Gamma(n)} \phi^{\mu=- 2 w}(p,x)
(2 p)^{w} K_{2 w}(2 \sqrt{2 p}) 
\eea
where we inserted the exact result for $Z_{\infty}^w$ and the function $\phi^{\mu}(p,x)$ was obtained in \cite[(3.6)]{monthus1994flux} (setting $\beta=\sigma=1$, $\alpha=1/2$
there) 
\be \label{phi2} 
\phi^{\mu=-2 w}(p,x) = \mathbb{E} \left[e^{- p Z^{-w}_x} \right] = \frac{(2 p)^{-w}}{4 \pi^2} 
 \int_{-\infty}^{+\infty} dq  q \sinh(\pi q) 
 \left\vert\Gamma\left(w + \frac{\I q}{2} \right)\right\vert^2 K_{\I q} (2 \sqrt{2 p}) e^{- \frac{x}{8} (q^2 + 4 w^2)}
 \ee 
Inserting into \eqref{pn1} the factors $(2p)^w$ and $(2p)^{-w}$ cancel. Let us specialize now to $n=1$. For $w<1$
we can 
interchange the integrals and use that
\be
 \int_0^{+\infty} dp 
 K_{2 w}(2 \sqrt{2 p})   K_{\I q} (2 \sqrt{2 p})  = \int_0^{+\infty} \frac{du}{4} u 
 K_{2 w}(u)   K_{\I q} (u) = 
\frac{\pi ^2 \left(q^2+4 w^2\right)}{16 (\cosh (\pi  q)-\cos (2 \pi  w))}
\ee 
leading to our first result, valid for $0< w \leq 1$, $w=-(A+ \frac{1}{2})$
\bea \label{meanpx} 
 \mathbb{E} [p_A(x)] = \frac{1}{32  \Gamma(2 w)}  
 \int_{-\infty}^{+\infty} dq e^{- \frac{1}{8} (q^2 + 4 w^2) x} q \sinh(\pi q) 
  \left\vert\Gamma\left(w + \frac{\I  q}{2} \right)\right\vert^2
 \frac{q^2+4 w^2}{\cosh (\pi  q)-\cos (2 \pi  w)}.
 \eea 
Under the change of variables $\I q=2z$, and after using some trigonometric identities and Euler's reflection formula,  \eqref{meanpx} can be rewritten as 
\begin{equation}\label{eq:meanpx2}
\mathbb{E} [p_A(x)] = \frac{1}{4 \Gamma(2w) } \int_{-\I\infty}^{\I\infty} \frac{dz}{2\I\pi} \frac{(w^2-z^2)e^{\frac{-x}{2}(w^2-z^2)}\Gamma(w+z)^2\Gamma(w-z)^2\Gamma(1-w-z)\Gamma(1-w+z)}{\Gamma(2z)\Gamma(-2z)}.
\end{equation}
We may now analytically continue this formula for all $w>0$ by subtracting and adding the necessary residues when $w>1$ (see Figure \ref{fig:analyticcontinuation}).
\begin{figure}
    \centering
    \begin{tikzpicture}[scale=0.6]
\draw[ultra thick, ->] (0,-4) -- (0,2);
\draw[ultra thick] (0,2) -- (0,4) node[anchor = west]{$\I\mathbb R$};
\draw[gray, thick, dashed] (-4,0) -- (4,0);
\draw (0,5) node{$\boxed{0<w<1}$};
\fill (1.7,0) circle(0.1);
\fill (-1.7,0) circle(0.1);
\draw (1.7,0.6) node{$1-w$};
\draw (-1.7,0.6) node{$w-1$};
\begin{scope}[xshift=9cm]
\draw (0,5) node{$\boxed{1<w<2}$};
 \draw[ultra thick, rounded corners] (0,-4)  -- (0,-0.2) -- (-2,-0.2);
 \draw[ultra thick, rounded corners] (-2,0.2) -- (-0.2,0.2) -- (0.2,-0.2) -- (2,-0.2);
\draw[ultra thick, ->, rounded corners] (2,0.2)  -- (0,0.2) -- (0,2);
\draw[ultra thick] (0,2) -- (0,4) node[anchor = west]{$\mathcal C$};
\draw[ultra thick] (-2,-0.2) arc(330:30:0.4);
\draw[ultra thick] (2,-0.2) arc(-150:150:0.4);
\fill (2.3,0) circle(0.1);
\fill (-2.3,0) circle(0.1);
\draw (2.2,0.6) node{$w-1$};
\draw (-2.2,0.6) node{$1-w$};
\draw[gray, thick, dashed] (-4,0) -- (4,0);   
\end{scope}
\begin{scope}[xshift=18cm]
\draw (0,5) node{$\boxed{1<w<2}$};
 \draw[ultra thick, ->] (0,-4) -- (0,2);
\draw[ultra thick] (0,2) -- (0,4) node[anchor = west]{$\I\mathbb R$};
\draw[gray, thick, dashed] (-4,0) -- (4,0);
\draw[ultra thick, ->] (-2.3,0.5) arc(90:-270:0.5);
\draw[ultra thick, <-] (2.3,0.5) arc(90:-270:0.5);
\draw[gray, thick, dashed] (-4,0) -- (4,0);   
\fill (2.3,0) circle(0.1);
\fill (-2.3,0) circle(0.1);
\end{scope}
    \end{tikzpicture}
    \caption{We consider a function $I(w)$ such that for $0<w<1$, we have $ I(w) = \int_{\I\mathbb R} f(z,w)$, and $f(z,w)$ is analytic 
    in both $z$ and $w$ except at some isolated poles. In particular it has poles at $z=w-1, w-2, w-3, \dots$ and $z=1-w, 2-w, 3-w, \dots$.
    Then, its analytic continuation to $w$ such that $n<w< n+1$ is $\int_{\mathcal C} f(z,w)$ where $\mathcal C$ is the contour shown above (in the case $n=1$). This contour is such that $1-w, \dots, n-w$ still lie on the right of the contour (as when $0<w<1$), and the poles at $w-1, \dots, w-n$ still lie on the left of the contour (as when $0<w<1$). Since the poles do not cross the contour, the formula remains analytic in $w$.  
    This contour can be then deformed to the union of a vertical line and small circles around simple poles whose contribution can be computed by the residue Theorem. }
    \label{fig:analyticcontinuation}
\end{figure}
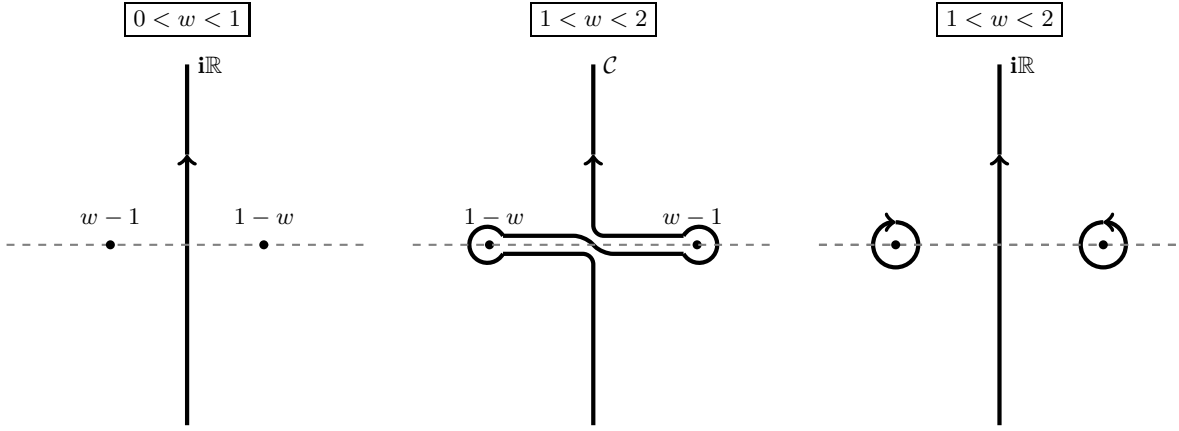

The analytic continuation of the RHS of \eqref{eq:meanpx2} to $w>1$ is 
\begin{equation}
\frac{1}{4 \Gamma(2w) } \int_{-\I\infty}^{\I\infty} \frac{dz}{2\I\pi} \frac{(w^2-z^2)e^{\frac{-x}{2}(w^2-z^2)}\Gamma(w+z)^2\Gamma(w-z)^2\Gamma(1-w-z)\Gamma(1-w+z)}{\Gamma(2z)\Gamma(-2z)}\\
+\frac{1}{4 \Gamma(2w) } \sum_{1\leqslant n<w} \left( R_{w-n}  - R_{n-w}  \right)
\end{equation}
where $R_{w-n}$ and $R_{n-w}$ are residues of the integrand at $z=w-n$ and $z=n-w$ respectively. We have 
\begin{equation}
    R_{w-n} = -R_{n-w}  = 2(-1)^{n-1}n!e^{\frac{-x}{2}(2wn-n^2)}(w-n)\frac{\Gamma(2w)\Gamma(1-2w)}{\Gamma(n-2w)},
\end{equation}
so that, for $w=-(A+ \frac{1}{2})>0$, 
\begin{multline}
\label{eq:meanpxanalytic}
\mathbb{E} [p_A(x)] = \frac{1}{4 \Gamma(2w) } \int_{-\I\infty}^{\I\infty} \frac{dz}{2\I\pi} \frac{(w^2-z^2)e^{\frac{-x}{2}(w^2-z^2)}\Gamma(w+z)^2\Gamma(w-z)^2\Gamma(1-w-z)\Gamma(1-w+z)}{\Gamma(2z)\Gamma(-2z)}\\
+ \sum_{1\leqslant n<w} n! (-1)^{n-1}e^{\frac{-x}{2}(2wn-n^2)}(w-n)\frac{\Gamma(1-2w)}{\Gamma(n-2w)}.
\end{multline}
Note that for large $w$ the finite series dominates and the integral can be neglected for most averages.
\bigskip 

\textbf{Remark: Normalization.} 
Let us first check that the formula \eqref{eq:meanpxanalytic} obeys  the normalization condition $\int_0^{+\infty} dx \mathbb{E} [p_A(x)] = 1$. 
By analyticity, it suffices to check it for $0<w<1$. 
Then, using  \eqref{eq:meanpx2}, this is equivalent to the identity 
\begin{equation}
 \int_{-\I\infty}^{\I\infty} \frac{dz}{2\I\pi} \frac{\Gamma(w+z)^2\Gamma(w-z)^2\Gamma(1-w-z)\Gamma(1-w+z)}{\Gamma(2z)\Gamma(-2z)} = 2\Gamma(2w).
\end{equation}
which is a particular case of the known identity  \cite[Eq. (3.6.1)]{andrews1999special}
\begin{equation}\label{eq:deBranges}
 \int_{-\I\infty}^{\I\infty} \frac{dz}{2\I\pi} \frac{\Gamma(a+z)\Gamma(a-z)\Gamma(b+z)\Gamma(b-z)\Gamma(c+z)\Gamma(c-z)}{\Gamma(2z)\Gamma(-2z)} = 2\Gamma(a+b)\Gamma(a+c)\Gamma(b+c),
\end{equation}
valid for $a,b,c$ with positive real part. 

\medskip 

\textbf{Remark: First moment.} We may also check that \eqref{eq:meanpx2} is consistent with the result \eqref{cum1} for the first thermal cumulant given in Section \ref{sec:stationaryendpoint}, i.e. 
$\mathbb{E} \langle x \rangle = \int_0^{+\infty}  x \mathbb{E} [p_A(x)] dx = 2 \psi'(2 w)$. Indeed, from \eqref{eq:meanpx2}, 
we have
\begin{equation}\label{eq:firstcumulantformula}
\int_0^{+\infty} x \mathbb{E} [p_A(x)] dx  = \frac{1}{ \Gamma(2w) } \int_{-\I\infty}^{\I\infty} \frac{dz}{2\I\pi} \frac{\Gamma(w+z)^2\Gamma(w-z)^2\Gamma(-w-z)\Gamma(-w+z)}{\Gamma(2z)\Gamma(-2z)}. 
\end{equation}
The integral in \eqref{eq:firstcumulantformula} cannot be simplified directly using \eqref{eq:deBranges}, but we may use that 
\begin{equation} 
\eqref{eq:firstcumulantformula} = \lim_{\epsilon\to 0 } \frac{1}{ \Gamma(2w) } \int_{-\I\infty}^{\I\infty} \frac{dz}{2\I\pi} \frac{\Gamma(w+z)^2\Gamma(w-z)^2\Gamma(\epsilon-w-z)\Gamma(\epsilon-w+z)}{\Gamma(2z)\Gamma(-2z)}.
\end{equation} 
When $0<w<1$, this integral above is analytic in $\epsilon\in (0, +\infty)$, 
and the expression for $0<\epsilon <w$ can be obtained by an analytic continuation from the expression when $\epsilon>w$,  
which is given  by  \eqref{eq:deBranges}. More precisely, for $0<\epsilon<w$, 
\begin{equation}
    \frac{1}{ \Gamma(2w) } \int_{-\I\infty}^{\I\infty} \frac{dz}{2\I\pi} \frac{\Gamma(w+z)^2\Gamma(w-z)^2\Gamma(\epsilon-w-z)\Gamma(\epsilon-w+z)}{\Gamma(2z)\Gamma(-2z)} =  \frac{1}{ \Gamma(2w) } \left(  2 \Gamma(2w)\Gamma(\epsilon)^2 + R_{\epsilon-w} - R_{-\epsilon+w}\right)
\end{equation} 
where  $R_{\pm \epsilon\mp w}$ are residues of the integrand at $z=\pm \epsilon\mp w$, which can be computed as
\begin{equation}
R_{\epsilon -w} = -R_{w-\epsilon} =- \frac{\Gamma(2w-\epsilon)^2 \Gamma(\epsilon)^2}{\Gamma(2w-2\epsilon)}.
\end{equation} 
Finally, taking the limit $\epsilon\to 0$, we obtain 
\begin{equation}
\int_0^{+\infty}  x \mathbb{E} [p_A(x)] dx  = \lim_{\epsilon\to 0 } \frac{1}{ \Gamma(2w) } \left(  2 \Gamma(2w)\Gamma(\epsilon)^2 
- 2\frac{\Gamma(2w-\epsilon)^2 \Gamma(\epsilon)^2}{\Gamma(2w-2\epsilon)} \right) = 2 \psi'(2w).
\end{equation}
\medskip 

\textbf{Remark: Value of $\mathbb E[p_A(0)]$.} Letting $x=0$ in \eqref{eq:meanpx2} and using the identity \eqref{eq:deBranges} 
yields  $\mathbb E[p_A(0)]=w$. This result has a simple origin: when $x=0$ $p_A(0) = \left(\int_{0}^{\infty} e^{\mathcal B(x)-wx}\mathrm d x \right)^{-1}\sim\Gamma(2w, 1/2)$ and $\mathbb E[\Gamma(2w, 1/2)]=w$. 

\medskip 
\textbf{Remark: Decay for large $x$.}
Let us start with $0<w<1$. Saddle point analysis and rescaling in formula \eqref{eq:meanpx2} gives that at large 
$x$ the decay is exponential with a $-3/2$ power law prefactor, 
$\mathbb{E} [p_A(x)] \simeq c_w x^{-3/2} e^{- w^2 x/2}$, with $c_w=\frac{\pi ^{3/2} \csc ^2(\pi  w) \Gamma (w+1)^2}{\sqrt{2} \Gamma (2 w)}$.
For $w>1$ however the decay at large $x$ is dominated by the term $n=1$ in the discrete series in \eqref{eq:meanpxanalytic} and
$\mathbb{E} [p_A(x)]  \sim e^{\frac{-x}{2}(2w-1)}$ for $w>1$, i.e. a much slower decay than $e^{\frac{-x}{2} w^2}$. 

\medskip 
{\bf Remark: Limit $w=-(A+\frac{1}{2}) \to 0$.} In the limit $w \to 0$, 
it is easy to check, upon rescaling $q \to w q$ and $x \to y/w^2$ in \eqref{meanpx} that as $w \to 0^+$ one has 
$\mathbb{E} [p_A(x)] dx  \to P(y) dy$ where
$P(y)$ is the probability given in the main text (of moments given by \eqref{momcrit}). 
In that limit the large $x$ tail obtained above matches the tail of $P(y)$ at large $y$,
since $c_w \sim \sqrt{\frac{2}{\pi}}/w$ for $w \to 0$. The $-3/2$ exponent, ubiquitous in this
types of problems \cite{broderix1995thermal,TexierComtetSUSY}, 
is known to originate from quasi-degenerate extrema of the Brownian \cite{laloux1998aging}.

\subsection{$m$-point correlations} 

To compute the $m$ point correlations with $m \geq 2$ one uses the Liouville quantum mechanics. One introduces
the Liouville Hamiltonian $H_p$ on the real axis $U \in \mathbb{R}$, and its eigenfunctions $\psi_k(U)$ which are real and indexed by $k \geq 0$  
\be 
H_p = - \frac{1}{2} \frac{d^2}{dU^2} + p e^U, \quad  \quad H_p \psi_k(U) = \frac{k^2}{8} \psi_k(U) ,
\quad  \quad \psi_k(U)= \frac{1}{\pi} \sqrt{k \sinh( \pi k) } K_{i k}( 2 \sqrt{2 p} e^{U/2}).
\ee 
These eigenfunctions  form a continuum orthonormal basis (we use the conventions in \cite{TexierComtetSUSY} with $\beta=\sigma=1$ and $\alpha=p$). It allows to compute our observables of interest. The first one is
expressed as follows, using the path integral representation for the Brownian motion with drift, $U(x)=\mathcal B(x)-w x$, 
with $U_0=U(0)=0$ and free $U(L)=U_L$, followed by the Feynman-Kac formula
\begin{align} \label{phi3} 
 \phi^{\mu=2 w}(p,L) &= \mathbb{E} [e^{- p Z^{w}_L} ] = e^{- \frac{w^2 L}{2}} \int_{-\infty}^{+\infty} dU_L e^{- w U_L} \langle U_L | e^{- L H_p } | U_0=0 \rangle \\
&=  \int_0^{+\infty} dk \int_{-\infty}^{+\infty} dU_L \psi_k(U_L) \psi_k^*(0) e^{- w U_L - \frac{L}{8} (k^2 + 4 w^2)} \nonumber
\end{align}
where the dependence in the drift $-w$ is made explicit through a trivial shift. In the last equation
we have used the spectral decomposition of $H_p$ in terms of its eigenvectors, $\langle U | k \rangle = \psi_k(U)$ introduced above. For $w<0$ one can use the identity
\be \label{id1} 
 \int_{-\infty}^{+\infty} dU e^{-w U} K_{\I k}(2 \sqrt{2 p} e^{U/2}) 
= \frac{(2 p)^w}{2} \left\vert \Gamma\left( - w + \frac{\I k}{2}\right) \right\vert^2 
\ee
and one checks that \eqref{phi3} yields  \eqref{phi2} after the change  $w \to -w$.

The $m$-point correlation can be written following closely \cite{TexierComtetSUSY} upon adding
the drift $w$. Upon exponentiation of the denominators $\frac{1}{{\sf Z}^m}=
\int_0^{+\infty} dq \frac{q^{m-1}}{\Gamma(m)} e^{- q {\sf Z}}$ and using the same path integral representation, one 
obtains 
for $L \geq x_1 \geq \dots x_m \geq 0$,
\bea
&& \mathbb{E} [p_A(x_1) \dots p_A(x_m) ] = \int_0^{+\infty} dq \frac{q^{m-1}}{\Gamma(m)} 
e^{- \frac{w^2 L}{2}} \int_{-\infty}^{+\infty} dU_L e^{- w U_L} 
\langle U_L | e^{-  H_q (L-x_1) } e^{\hat U} e^{-  H_q (x_1-x_2)} e^{\hat U} \dots   e^{-  H_q x_m} | U_0=0 \rangle \nonumber 
\\
&& 
= \frac{e^{- \frac{w^2 L}{2}} p^m}{\Gamma(m)} 
\int_{-\infty}^{+\infty} dU_0 
 \int_{-\infty}^{+\infty} dU_L e^{- w (U_L-U_0)} 
\langle U_L | e^{-  H_p (L-x_1) } e^{\hat U} e^{-  H_p (x_1-x_2)} e^{\hat U} \dots   e^{-  H_p x_m} | U_0 \rangle
\eea
where $e^{\hat U} = \int_{-\infty}^{+\infty} dU | U \rangle e^{U} \langle U|$. The second line is obtained after the
standard trick in Liouville theory, i.e. the change of variable $q=p e^{U_0}$ followed by the shift 
$U(x) \to U(x)-U_0$. Introducing the eigenbasis of $H_{p}$, and choosing $p=1/2$ for convenience, one obtains (here $w=-(A+ \frac{1}{2}) >0$)
\bea
&& \mathbb{E} [p_A(x_1) \dots p_A(x_m) ] = \frac{1}{2^m \Gamma(m)} e^{- \frac{w^2 L}{2}} 
\int_{-\infty}^{+\infty} dU_0 
 \int_{-\infty}^{+\infty} dU_L e^{- w (U_L-U_0)} \int_0^{+\infty} dk \psi_k(U_L) \\
 && \times \prod_{j=1}^m \int_0^{+\infty} dk_j  F(k,k_1) F(k_1,k_2) \dots F(k_{m-1},k_m) \psi_{k_m}^*(U_0) e^{-\frac{k^2}{8} (L-x_1) - \frac{k_1^2}{8} (x_1-x_2) - \dots -  \frac{k_m^2}{8} x_m} 
\eea 
where we have defined the matrix elements \cite{TexierComtetSUSY} 
\bea
F(k,k') = \langle k | e^{\hat U } | k' \rangle = \int_{-\infty}^{+\infty} dU \psi_k^*(U) e^{U} \psi_{k'}(U) =
\frac{1}{8} \sqrt{k k' \sinh(\pi k) \sinh(\pi k') } \frac{k^2-(k')^2}{\cosh(\pi k) - \cosh(\pi k')} 
\eea 
Examination of the calculations in \cite{monthus1994flux} (Section 5, in particular (5.6)) for a simpler
quantity, indicates that the limit $L \to +\infty$ is controlled by setting $k=-2 \I w$ (the integral 
$\int_{-\infty}^{+\infty} dU_L e^{- w U_L}  \psi_k(U_L)$ is divergent for $w>0$, but one can extract the
residue of its analytic continuation). This amount to use that as $L \to +\infty$ 
\be 
e^{- \frac{w^2 L}{2}} \int_{-\infty}^{+\infty} dU_L e^{- w U_L} \int_0^{+\infty} dk \psi_k(U_L) F(k,k_1) e^{-\frac{k^2}{8} L} \xrightarrow[L\to\infty]{} \frac{2 \pi}{\Gamma(2 w)} 
\frac{F(k,k_1)}{\sqrt{k \sinh(\pi k) }}\bigg\vert_{k=-2 i w} (2 p)^w
\ee 
valid for $p$ arbitrary, and recalling our choice $p=1/2$. Using further \eqref{id1} (with $w \to -w$) to integrate over $U_0$ this leads to the final result, for $0< w \leq 1$
and $x_1 \geq x_2 \dots \geq x_m\geq 0$
\begin{multline} \label{px1px2}
\mathbb{E} [p_A(x_1) \dots p_A(x_m) ]  = \frac{1}{2^{m} \Gamma(2 w) \Gamma(m) }
\prod_{j=1}^m \int_0^{+\infty} \frac{dk_j}{8} k_j \sinh(\pi k_j) \\ 
\times 
\prod_{j=1}^{m-1} \frac{k_j^2-k_{j+1}^2}{\cosh(\pi k_j) - \cosh(\pi k_{j+1})} 
\frac{k_1^2+ 4 w^2}{\cosh(\pi k_1) - \cos(2 \pi w)}  \left\vert \Gamma\left( w + \frac{i k_m}{2}\right) \right\vert^2 
 e^{- \frac{w^2}{2} x_1 - \frac{k_1^2}{8} (x_1-x_2) - \dots -  \frac{k_m^2}{8} x_m} 
\end{multline}
which for $m=1$ agrees with the formula \eqref{meanpx} obtained by a different method. The normalization is checked below.
For $m=2$, we have checked numerically using Mathematica,
using also \eqref{meanpx}, that 
$\int_0^{+\infty} dx x^2 \mathbb{E} [p_A(x)]- \int_0^{+\infty} dx_1 \int_0^{+\infty} dx_2 x_1 x_2 \mathbb{E} [p_A(x_1) p_A(x_2) ]$ 
agrees numerically with the result \eqref{cum2} for the thermal cumulant $\mathbb{E}[\langle x^2 \rangle_c]$.

Using similar manipulations as around \eqref{eq:meanpx2}, we may write (with the convention $x_{m+1}=0$)
\begin{multline} \label{px1px2bis}
\mathbb{E} [p_A(x_1) \dots p_A(x_m) ]  = \frac{1}{2^{2m} \Gamma(2 w) \Gamma(m) }
 \int_{\I\mathbb R} \frac{dz_1}{2\I\pi} \dots \int_{\I\mathbb R} \frac{dz_m}{2\I\pi} \prod_{j=1}^m 
 \frac{e^{\left(\frac{z_j^2}{2} -\frac{w^2}{2}\right) (x_j-x_{j+1})}}{\Gamma(2z_j)\Gamma(-2z_j)}   \\ 
\times \Gamma(w+z_m)\Gamma(w-z_m) (w^2-z_{1}^2)\Gamma(w+z_1)\Gamma(w-z_1)\Gamma(1-w+z_1)\Gamma(1-w-z_1)  
\\ \times \prod_{j=1}^{m-1} \Gamma(1+z_{j+1}-z_j)\Gamma(1+z_{j+1}+z_j)\Gamma(1-z_{j+1}+z_j)\Gamma(1-z_{j+1}-z_j).
\end{multline}
A similar analytic continuation as in \eqref{eq:meanpxanalytic} can be performed on \eqref{px1px2bis} to obtain a formula when $w>1$. 
\bigskip 

\textbf{Verification of the normalization.} 
Let us compute
\begin{equation}
    C_m = m! \int_{x_1 \geq x_2 \dots \geq x_m\geq 0} \mathbb{E} [p_A(x_1) \dots p_A(x_m) ] dx_1\dots dx_m,
\end{equation}
and check that $C_m=1$. We use the change of variables $y_j=x_j-x_{j+1}$ and compute the integrals over $y_1, z_1, y_2,z_2 \dots$  sequentially. 
We will need  the identity  \cite[Th. 3.6.2]{andrews1999special}
\begin{multline}\label{eq:Wilson}
 \int_{\I\mathbb R} \frac{dz}{2\I\pi} \frac{\Gamma(a+z)\Gamma(a-z)\Gamma(b+z)\Gamma(b-z)\Gamma(c+z)\Gamma(c-z)\Gamma(d+z)\Gamma(d-z)}{\Gamma(2z)\Gamma(-2z)} =\\ \frac{2\Gamma(a+b)\Gamma(a+c)\Gamma(a+d)\Gamma(b+c)\Gamma(b+d)\Gamma(c+d)}{\Gamma(a+b+c+d)},
\end{multline}
valid for $a,b,c,d$ with positive real part. Performing the integration over $y_1$ and $z_1$ using \eqref{eq:Wilson} with $\lbrace a,b,c,d \rbrace = \lbrace w,1-w, 1+z_2,1-z_2\rbrace  $, 
we obtain 
\begin{multline}
C_m  = \frac{\Gamma(m+1)}{2^{2(m-1)} \Gamma(2 w) \Gamma(m) }
 \int_{\I\mathbb R} \frac{dz_2}{2\I\pi} \dots \int_{\I\mathbb R} \frac{dz_m}{2\I\pi} \prod_{j=2}^m \frac{e^{\left(\frac{z_j^2}{2} -\frac{w^2}{2}\right) (x_j-x_{j+1})}}{\Gamma(2z_j)\Gamma(-2z_j)}   \\ 
\times \Gamma(w+z_m)\Gamma(w-z_m) (w^2-z_{2}^2)\Gamma(w+z_2)\Gamma(w-z_2)\Gamma(2-w+z_2)\Gamma(2-w-z_2)  \frac{\Gamma(1)\Gamma(2)}{\Gamma(3)}
\\ \times \prod_{j=2}^{m-1} \Gamma(1+z_{j+1}-z_j)\Gamma(1+z_{j+1}+z_j)\Gamma(1-z_{j+1}+z_j)\Gamma(1-z_{j+1}-z_j).
\end{multline}
Then we integrate over $y_2$ and $z_2$ using \eqref{eq:Wilson} with $\lbrace a,b,c,d \rbrace = \lbrace w,2-w, 1+z_3,1-z_3\rbrace  $, we integrate over $y_3$ and $z_3$ using \eqref{eq:Wilson} with $\lbrace a,b,c,d \rbrace = \lbrace w,3-w, 1+z_4,1-z_4\rbrace$,  and we continue until we are left with variables $y_m, z_m$, where we use \eqref{eq:deBranges} with $\lbrace a,b,c\rbrace = \lbrace w , w , m-w \rbrace  $. Keeping track of all the Gamma factors involved at each step, we find that 
\begin{equation}
    C_m = \frac{\Gamma(m+1)}{\Gamma(m)\Gamma(2w)} \times  \frac{\Gamma(2)\Gamma(1)}{\Gamma(3)}\frac{\Gamma(2)\Gamma(2)}{\Gamma(4)}\frac{\Gamma(2)\Gamma(3)}{\Gamma(5)} \dots \frac{\Gamma(2)\Gamma(m-1)}{\Gamma(m+1)} \times \Gamma(2w)\Gamma(m)^2 = 1.
\end{equation}






\end{widetext}
 \subsection*{Acknowledgements:} We thank Alexandre Krajenbrink for many  useful conversations related to half-space solutions of the KPZ equation.  We thank Christophe Texier for an interesting discussion.
 G.B. was partially supported by ANR grant ANR-19-CE40-0012 MicMov. PLD acknowledges support from ANR grant ANR-17-CE30-0027-01 RaMaTraF. 
\bibliography{biblioResubPRE.bib}

\begin{thebibliography}{86}%
\makeatletter
\providecommand \@ifxundefined [1]{%
 \@ifx{#1\undefined}
}%
\providecommand \@ifnum [1]{%
 \ifnum #1\expandafter \@firstoftwo
 \else \expandafter \@secondoftwo
 \fi
}%
\providecommand \@ifx [1]{%
 \ifx #1\expandafter \@firstoftwo
 \else \expandafter \@secondoftwo
 \fi
}%
\providecommand \natexlab [1]{#1}%
\providecommand \enquote  [1]{``#1''}%
\providecommand \bibnamefont  [1]{#1}%
\providecommand \bibfnamefont [1]{#1}%
\providecommand \citenamefont [1]{#1}%
\providecommand \href@noop [0]{\@secondoftwo}%
\providecommand \href [0]{\begingroup \@sanitize@url \@href}%
\providecommand \@href[1]{\@@startlink{#1}\@@href}%
\providecommand \@@href[1]{\endgroup#1\@@endlink}%
\providecommand \@sanitize@url [0]{\catcode `\\12\catcode `\$12\catcode
  `\&12\catcode `\#12\catcode `\^12\catcode `\_12\catcode `\%12\relax}%
\providecommand \@@startlink[1]{}%
\providecommand \@@endlink[0]{}%
\providecommand \url  [0]{\begingroup\@sanitize@url \@url }%
\providecommand \@url [1]{\endgroup\@href {#1}{\urlprefix }}%
\providecommand \urlprefix  [0]{URL }%
\providecommand \Eprint [0]{\href }%
\providecommand \doibase [0]{https://doi.org/}%
\providecommand \selectlanguage [0]{\@gobble}%
\providecommand \bibinfo  [0]{\@secondoftwo}%
\providecommand \bibfield  [0]{\@secondoftwo}%
\providecommand \translation [1]{[#1]}%
\providecommand \BibitemOpen [0]{}%
\providecommand \bibitemStop [0]{}%
\providecommand \bibitemNoStop [0]{.\EOS\space}%
\providecommand \EOS [0]{\spacefactor3000\relax}%
\providecommand \BibitemShut  [1]{\csname bibitem#1\endcsname}%
\let\auto@bib@innerbib\@empty
\bibitem [{\citenamefont {Kardar}\ \emph {et~al.}(1986)\citenamefont {Kardar},
  \citenamefont {Parisi},\ and\ \citenamefont {Zhang}}]{KPZ}%
  \BibitemOpen
  \bibfield  {author} {\bibinfo {author} {\bibfnamefont {M.}~\bibnamefont
  {Kardar}}, \bibinfo {author} {\bibfnamefont {G.}~\bibnamefont {Parisi}},\
  and\ \bibinfo {author} {\bibfnamefont {Y.~C.}\ \bibnamefont {Zhang}},\
  }\bibfield  {title} {\bibinfo {title} {Dynamic scaling of growing
  interfaces},\ }\href@noop {} {\bibfield  {journal} {\bibinfo  {journal}
  {Phys. Rev. Lett.}\ }\textbf {\bibinfo {volume} {56}},\ \bibinfo {pages}
  {889} (\bibinfo {year} {1986})}\BibitemShut {NoStop}%
\bibitem [{\citenamefont {Huse}\ \emph {et~al.}(1985)\citenamefont {Huse},
  \citenamefont {Henley},\ and\ \citenamefont {Fisher}}]{huse1985huse}%
  \BibitemOpen
  \bibfield  {author} {\bibinfo {author} {\bibfnamefont {D.~A.}\ \bibnamefont
  {Huse}}, \bibinfo {author} {\bibfnamefont {C.~L.}\ \bibnamefont {Henley}},\
  and\ \bibinfo {author} {\bibfnamefont {D.~S.}\ \bibnamefont {Fisher}},\
  }\bibfield  {title} {\bibinfo {title} {Huse, henley, and fisher respond},\
  }\href {https://doi.org/10.1103/PhysRevLett.55.2924} {\bibfield  {journal}
  {\bibinfo  {journal} {Phys. Rev. Lett.}\ }\textbf {\bibinfo {volume} {55}},\
  \bibinfo {pages} {2924} (\bibinfo {year} {1985})}\BibitemShut {NoStop}%
\bibitem [{\citenamefont {Kardar}\ and\ \citenamefont
  {Zhang}(1987)}]{kardar1987scaling}%
  \BibitemOpen
  \bibfield  {author} {\bibinfo {author} {\bibfnamefont {M.}~\bibnamefont
  {Kardar}}\ and\ \bibinfo {author} {\bibfnamefont {Y.~C.}\ \bibnamefont
  {Zhang}},\ }\bibfield  {title} {\bibinfo {title} {Scaling of directed
  polymers in random media},\ }\href
  {https://doi.org/10.1103/PhysRevLett.58.2087} {\bibfield  {journal} {\bibinfo
   {journal} {Phys. Rev. Lett.}\ }\textbf {\bibinfo {volume} {58}},\ \bibinfo
  {pages} {2087} (\bibinfo {year} {1987})}\BibitemShut {NoStop}%
\bibitem [{\citenamefont {Halpin-Healy}\ and\ \citenamefont
  {Zhang}(1995)}]{halpinhealy1995kinetic}%
  \BibitemOpen
  \bibfield  {author} {\bibinfo {author} {\bibfnamefont {T.}~\bibnamefont
  {Halpin-Healy}}\ and\ \bibinfo {author} {\bibfnamefont {Y.~C.}\ \bibnamefont
  {Zhang}},\ }\bibfield  {title} {\bibinfo {title} {Kinetic roughening
  phenomena, stochastic growth, directed polymers and all that. {A}spects of
  multidisciplinary statistical mechanics},\ }\href
  {https://doi.org/https://doi.org/10.1016/0370-1573(94)00087-J} {\bibfield
  {journal} {\bibinfo  {journal} {Phys. Reports}\ }\textbf {\bibinfo {volume}
  {254}},\ \bibinfo {pages} {215} (\bibinfo {year} {1995})}\BibitemShut
  {NoStop}%
\bibitem [{\citenamefont {Edwards}\ and\ \citenamefont
  {Wilkinson}(1982)}]{edwards1982surface}%
  \BibitemOpen
  \bibfield  {author} {\bibinfo {author} {\bibfnamefont {S.~F.}\ \bibnamefont
  {Edwards}}\ and\ \bibinfo {author} {\bibfnamefont {D.~R.}\ \bibnamefont
  {Wilkinson}},\ }\bibfield  {title} {\bibinfo {title} {The surface statistics
  of a granular aggregate},\ }\href@noop {} {\bibfield  {journal} {\bibinfo
  {journal} {Proc. Royal Soc. London. A. Math. Phys. Sc.}\ }\textbf {\bibinfo
  {volume} {381}},\ \bibinfo {pages} {17} (\bibinfo {year} {1982})}\BibitemShut
  {NoStop}%
\bibitem [{\citenamefont {Hammersley}(1967)}]{hammersley1967harnesses}%
  \BibitemOpen
  \bibfield  {author} {\bibinfo {author} {\bibfnamefont {J.~M.}\ \bibnamefont
  {Hammersley}},\ }\bibfield  {title} {\bibinfo {title} {Harnesses},\
  }\href@noop {} {\bibfield  {journal} {\bibinfo  {journal} {Berkeley Symposium
  on Mathematical Statistics and Probability}\ }\textbf {\bibinfo {volume}
  {5.3}} (\bibinfo {year} {1967})}\BibitemShut {NoStop}%
\bibitem [{\citenamefont {Corwin}(2012)}]{corwin2012kardar}%
  \BibitemOpen
  \bibfield  {author} {\bibinfo {author} {\bibfnamefont {I.}~\bibnamefont
  {Corwin}},\ }\bibfield  {title} {\bibinfo {title} {The
  {Kardar--Parisi--Zhang} equation and universality class},\ }\href@noop {}
  {\bibfield  {journal} {\bibinfo  {journal} {Rand. mat.: Theor. appl.}\
  }\textbf {\bibinfo {volume} {1}},\ \bibinfo {pages} {1130001} (\bibinfo
  {year} {2012})},\ \Eprint {https://arxiv.org/abs/1106.1596} {arXiv:1106.1596}
  \BibitemShut {NoStop}%
\bibitem [{\citenamefont {Quastel}\ and\ \citenamefont
  {Spohn}(2015)}]{quastel2015one}%
  \BibitemOpen
  \bibfield  {author} {\bibinfo {author} {\bibfnamefont {J.}~\bibnamefont
  {Quastel}}\ and\ \bibinfo {author} {\bibfnamefont {H.}~\bibnamefont
  {Spohn}},\ }\bibfield  {title} {\bibinfo {title} {The one-dimensional {KPZ}
  equation and its universality class},\ }\href@noop {} {\bibfield  {journal}
  {\bibinfo  {journal} {J. Stat. Phys.}\ }\textbf {\bibinfo {volume} {160}},\
  \bibinfo {pages} {965} (\bibinfo {year} {2015})},\ \Eprint
  {https://arxiv.org/abs/1503.06185} {arXiv:1503.06185} \BibitemShut {NoStop}%
\bibitem [{\citenamefont {Takeuchi}(2018)}]{takeuchi2018appetizer}%
  \BibitemOpen
  \bibfield  {author} {\bibinfo {author} {\bibfnamefont {K.}~\bibnamefont
  {Takeuchi}},\ }\bibfield  {title} {\bibinfo {title} {An appetizer to modern
  developments on the {Kardar--Parisi--Zhang} universality class},\ }\href@noop
  {} {\bibfield  {journal} {\bibinfo  {journal} {Physica A: Stat. Mech. Appl.}\
  }\textbf {\bibinfo {volume} {504}},\ \bibinfo {pages} {77} (\bibinfo {year}
  {2018})},\ \Eprint {https://arxiv.org/abs/1708.06060} {arXiv:1708.06060}
  \BibitemShut {NoStop}%
\bibitem [{\citenamefont {Amir}\ \emph {et~al.}(2011)\citenamefont {Amir},
  \citenamefont {Corwin},\ and\ \citenamefont {Quastel}}]{amir2011probability}%
  \BibitemOpen
  \bibfield  {author} {\bibinfo {author} {\bibfnamefont {G.}~\bibnamefont
  {Amir}}, \bibinfo {author} {\bibfnamefont {I.}~\bibnamefont {Corwin}},\ and\
  \bibinfo {author} {\bibfnamefont {J.}~\bibnamefont {Quastel}},\ }\bibfield
  {title} {\bibinfo {title} {Probability distribution of the free energy of the
  continuum directed random polymer in 1 + 1 dimensions},\ }\href
  {https://doi.org/10.1002/cpa.20347} {\bibfield  {journal} {\bibinfo
  {journal} {Comm. Pure Appl. Math.}\ }\textbf {\bibinfo {volume} {64}},\
  \bibinfo {pages} {466} (\bibinfo {year} {2011})},\ \Eprint
  {https://arxiv.org/abs/1003.0443} {arXiv:1003.0443} \BibitemShut {NoStop}%
\bibitem [{\citenamefont {Dotsenko}(2010)}]{dotsenko2010replica}%
  \BibitemOpen
  \bibfield  {author} {\bibinfo {author} {\bibfnamefont {V.}~\bibnamefont
  {Dotsenko}},\ }\bibfield  {title} {\bibinfo {title} {Replica {B}ethe ansatz
  derivation of the {T}racy–{W}idom distribution of the free energy
  fluctuations in one-dimensional directed polymers},\ }\href@noop {}
  {\bibfield  {journal} {\bibinfo  {journal} {J. Stat. Mech.}\ }\textbf
  {\bibinfo {volume} {2010}},\ \bibinfo {pages} {P07010} (\bibinfo {year}
  {2010})},\ \Eprint {https://arxiv.org/abs/1004.4455} {arXiv:1004.4455}
  \BibitemShut {NoStop}%
\bibitem [{\citenamefont {Calabrese}\ \emph {et~al.}(2010)\citenamefont
  {Calabrese}, \citenamefont {Doussal},\ and\ \citenamefont
  {Rosso}}]{calabrese2010free}%
  \BibitemOpen
  \bibfield  {author} {\bibinfo {author} {\bibfnamefont {P.}~\bibnamefont
  {Calabrese}}, \bibinfo {author} {\bibfnamefont {P.~L.}\ \bibnamefont
  {Doussal}},\ and\ \bibinfo {author} {\bibfnamefont {A.}~\bibnamefont
  {Rosso}},\ }\bibfield  {title} {\bibinfo {title} {Free-energy distribution of
  the directed polymer at high temperature},\ }\href@noop {} {\bibfield
  {journal} {\bibinfo  {journal} {Europhys. Lett.}\ }\textbf {\bibinfo {volume}
  {90}},\ \bibinfo {pages} {20002} (\bibinfo {year} {2010})},\ \Eprint
  {https://arxiv.org/abs/1002.4560} {arXiv:1002.4560} \BibitemShut {NoStop}%
\bibitem [{\citenamefont {Sasamoto}\ and\ \citenamefont
  {Spohn}(2010)}]{sasamoto2010exact}%
  \BibitemOpen
  \bibfield  {author} {\bibinfo {author} {\bibfnamefont {T.}~\bibnamefont
  {Sasamoto}}\ and\ \bibinfo {author} {\bibfnamefont {H.}~\bibnamefont
  {Spohn}},\ }\bibfield  {title} {\bibinfo {title} {Exact height distributions
  for the {KPZ} equation with narrow wedge initial condition},\ }\href@noop {}
  {\bibfield  {journal} {\bibinfo  {journal} {Nuclear Phys. B}\ }\textbf
  {\bibinfo {volume} {834}},\ \bibinfo {pages} {523} (\bibinfo {year}
  {2010})},\ \Eprint {https://arxiv.org/abs/1002.1879} {arXiv:1002.1879}
  \BibitemShut {NoStop}%
\bibitem [{\citenamefont {Calabrese}\ and\ \citenamefont
  {Le~Doussal}(2011)}]{calabrese2011exact}%
  \BibitemOpen
  \bibfield  {author} {\bibinfo {author} {\bibfnamefont {P.}~\bibnamefont
  {Calabrese}}\ and\ \bibinfo {author} {\bibfnamefont {P.}~\bibnamefont
  {Le~Doussal}},\ }\bibfield  {title} {\bibinfo {title} {Exact solution for the
  {Kardar-Parisi-Zhang} equation with flat initial conditions},\ }\href@noop {}
  {\bibfield  {journal} {\bibinfo  {journal} {Phys. Rev. Lett.}\ }\textbf
  {\bibinfo {volume} {106}},\ \bibinfo {pages} {250603} (\bibinfo {year}
  {2011})},\ \Eprint {https://arxiv.org/abs/1104.1993} {arXiv:1104.1993}
  \BibitemShut {NoStop}%
\bibitem [{\citenamefont {Ortmann}\ \emph {et~al.}(2016)\citenamefont
  {Ortmann}, \citenamefont {Quastel},\ and\ \citenamefont
  {Remenik}}]{ortmann2016exact}%
  \BibitemOpen
  \bibfield  {author} {\bibinfo {author} {\bibfnamefont {J.}~\bibnamefont
  {Ortmann}}, \bibinfo {author} {\bibfnamefont {J.}~\bibnamefont {Quastel}},\
  and\ \bibinfo {author} {\bibfnamefont {D.}~\bibnamefont {Remenik}},\
  }\bibfield  {title} {\bibinfo {title} {Exact formulas for random growth with
  half-flat initial data},\ }\href@noop {} {\bibfield  {journal} {\bibinfo
  {journal} {Ann. Appl. Probab.}\ }\textbf {\bibinfo {volume} {26}},\ \bibinfo
  {pages} {507} (\bibinfo {year} {2016})},\ \Eprint
  {https://arxiv.org/abs/1407.8484} {arXiv:1407.8484} \BibitemShut {NoStop}%
\bibitem [{\citenamefont {Corwin}\ and\ \citenamefont
  {Quastel}(2013)}]{corwin2013crossover}%
  \BibitemOpen
  \bibfield  {author} {\bibinfo {author} {\bibfnamefont {I.}~\bibnamefont
  {Corwin}}\ and\ \bibinfo {author} {\bibfnamefont {J.}~\bibnamefont
  {Quastel}},\ }\bibfield  {title} {\bibinfo {title} {Crossover distributions
  at the edge of the rarefaction fan},\ }\href@noop {} {\bibfield  {journal}
  {\bibinfo  {journal} {Ann. Probab.}\ }\textbf {\bibinfo {volume} {41}},\
  \bibinfo {pages} {1243} (\bibinfo {year} {2013})},\ \Eprint
  {https://arxiv.org/abs/1006.1338} {arXiv:1006.1338} \BibitemShut {NoStop}%
\bibitem [{\citenamefont {Imamura}\ and\ \citenamefont
  {Sasamoto}(2011)}]{imamura2011replica}%
  \BibitemOpen
  \bibfield  {author} {\bibinfo {author} {\bibfnamefont {T.}~\bibnamefont
  {Imamura}}\ and\ \bibinfo {author} {\bibfnamefont {T.}~\bibnamefont
  {Sasamoto}},\ }\bibfield  {title} {\bibinfo {title} {Replica approach to the
  {KPZ} equation with the half {Brownian} motion initial condition},\
  }\href@noop {} {\bibfield  {journal} {\bibinfo  {journal} {J. Phys. A: Math.
  Theor.}\ }\textbf {\bibinfo {volume} {44}},\ \bibinfo {pages} {385001}
  (\bibinfo {year} {2011})},\ \Eprint {https://arxiv.org/abs/1105.4659}
  {arXiv:1105.4659} \BibitemShut {NoStop}%
\bibitem [{\citenamefont {Imamura}\ and\ \citenamefont
  {Sasamoto}(2012)}]{imamura2012exact}%
  \BibitemOpen
  \bibfield  {author} {\bibinfo {author} {\bibfnamefont {T.}~\bibnamefont
  {Imamura}}\ and\ \bibinfo {author} {\bibfnamefont {T.}~\bibnamefont
  {Sasamoto}},\ }\bibfield  {title} {\bibinfo {title} {Exact solution for the
  stationary {Kardar-Parisi-Zhang} equation},\ }\href@noop {} {\bibfield
  {journal} {\bibinfo  {journal} {Phys. Rev. Lett.}\ }\textbf {\bibinfo
  {volume} {108}},\ \bibinfo {pages} {190603} (\bibinfo {year} {2012})},\
  \Eprint {https://arxiv.org/abs/1111.4634} {arXiv:1111.4634} \BibitemShut
  {NoStop}%
\bibitem [{\citenamefont {Imamura}\ and\ \citenamefont
  {Sasamoto}(2013)}]{imamura2013stationary}%
  \BibitemOpen
  \bibfield  {author} {\bibinfo {author} {\bibfnamefont {T.}~\bibnamefont
  {Imamura}}\ and\ \bibinfo {author} {\bibfnamefont {T.}~\bibnamefont
  {Sasamoto}},\ }\bibfield  {title} {\bibinfo {title} {Stationary correlations
  for the {1D KPZ} equation},\ }\href@noop {} {\bibfield  {journal} {\bibinfo
  {journal} {J. Stat. Phys.}\ }\textbf {\bibinfo {volume} {150}},\ \bibinfo
  {pages} {908} (\bibinfo {year} {2013})},\ \Eprint
  {https://arxiv.org/abs/1210.4278} {arXiv:1210.4278} \BibitemShut {NoStop}%
\bibitem [{\citenamefont {Borodin}\ \emph {et~al.}(2015)\citenamefont
  {Borodin}, \citenamefont {Corwin}, \citenamefont {Ferrari},\ and\
  \citenamefont {Vet{\H{o}}}}]{borodin2015height}%
  \BibitemOpen
  \bibfield  {author} {\bibinfo {author} {\bibfnamefont {A.}~\bibnamefont
  {Borodin}}, \bibinfo {author} {\bibfnamefont {I.}~\bibnamefont {Corwin}},
  \bibinfo {author} {\bibfnamefont {P.}~\bibnamefont {Ferrari}},\ and\ \bibinfo
  {author} {\bibfnamefont {B.}~\bibnamefont {Vet{\H{o}}}},\ }\bibfield  {title}
  {\bibinfo {title} {Height fluctuations for the stationary {KPZ} equation},\
  }\href@noop {} {\bibfield  {journal} {\bibinfo  {journal} {Math. Phys. Anal.
  Geom.}\ }\textbf {\bibinfo {volume} {18}},\ \bibinfo {pages} {1} (\bibinfo
  {year} {2015})},\ \Eprint {https://arxiv.org/abs/1407.6977} {arXiv:1407.6977}
  \BibitemShut {NoStop}%
\bibitem [{\citenamefont {Kardar}(1985)}]{kardar1985depinning}%
  \BibitemOpen
  \bibfield  {author} {\bibinfo {author} {\bibfnamefont {M.}~\bibnamefont
  {Kardar}},\ }\bibfield  {title} {\bibinfo {title} {Depinning by quenched
  randomness},\ }\href@noop {} {\bibfield  {journal} {\bibinfo  {journal}
  {Phys. Rev. Lett.}\ }\textbf {\bibinfo {volume} {55}},\ \bibinfo {pages}
  {2235} (\bibinfo {year} {1985})}\BibitemShut {NoStop}%
\bibitem [{\citenamefont {De~Gennes}(1985)}]{de1985wetting}%
  \BibitemOpen
  \bibfield  {author} {\bibinfo {author} {\bibfnamefont {P.}~\bibnamefont
  {De~Gennes}},\ }\bibfield  {title} {\bibinfo {title} {Wetting: statics and
  dynamics},\ }\href@noop {} {\bibfield  {journal} {\bibinfo  {journal} {Rev.
  modern phys.}\ }\textbf {\bibinfo {volume} {57}},\ \bibinfo {pages} {827}
  (\bibinfo {year} {1985})}\BibitemShut {NoStop}%
\bibitem [{\citenamefont {Abraham}(1980)}]{abraham1980solvable}%
  \BibitemOpen
  \bibfield  {author} {\bibinfo {author} {\bibfnamefont {D.~B.}\ \bibnamefont
  {Abraham}},\ }\bibfield  {title} {\bibinfo {title} {Solvable model with a
  roughening transition for a planar {I}sing ferromagnet},\ }\href@noop {}
  {\bibfield  {journal} {\bibinfo  {journal} {Phys. Rev. Lett.}\ }\textbf
  {\bibinfo {volume} {44}},\ \bibinfo {pages} {1165} (\bibinfo {year}
  {1980})}\BibitemShut {NoStop}%
\bibitem [{\citenamefont {Tang}\ and\ \citenamefont
  {Lyuksyutov}(1993)}]{tang1993directed}%
  \BibitemOpen
  \bibfield  {author} {\bibinfo {author} {\bibfnamefont {L.-H.}\ \bibnamefont
  {Tang}}\ and\ \bibinfo {author} {\bibfnamefont {I.~F.}\ \bibnamefont
  {Lyuksyutov}},\ }\bibfield  {title} {\bibinfo {title} {Directed polymer
  localization in a disordered medium},\ }\href@noop {} {\bibfield  {journal}
  {\bibinfo  {journal} {Phys. Rev. Lett.}\ }\textbf {\bibinfo {volume} {71}},\
  \bibinfo {pages} {2745} (\bibinfo {year} {1993})}\BibitemShut {NoStop}%
\bibitem [{\citenamefont {Basu}\ \emph {et~al.}(2014)\citenamefont {Basu},
  \citenamefont {Sidoravicius},\ and\ \citenamefont {Sly}}]{basu2014last}%
  \BibitemOpen
  \bibfield  {author} {\bibinfo {author} {\bibfnamefont {R.}~\bibnamefont
  {Basu}}, \bibinfo {author} {\bibfnamefont {V.}~\bibnamefont {Sidoravicius}},\
  and\ \bibinfo {author} {\bibfnamefont {A.}~\bibnamefont {Sly}},\ }\bibfield
  {title} {\bibinfo {title} {Last passage percolation with a defect line and
  the solution of the slow bond problem},\ }\href@noop {} {\bibfield  {journal}
  {\bibinfo  {journal} {arXiv:1408.3464}\ } (\bibinfo {year}
  {2014})}\BibitemShut {NoStop}%
\bibitem [{\citenamefont {Soh}\ \emph {et~al.}(2017)\citenamefont {Soh},
  \citenamefont {Baek}, \citenamefont {Ha},\ and\ \citenamefont
  {Jeong}}]{soh2017effects}%
  \BibitemOpen
  \bibfield  {author} {\bibinfo {author} {\bibfnamefont {H.}~\bibnamefont
  {Soh}}, \bibinfo {author} {\bibfnamefont {Y.}~\bibnamefont {Baek}}, \bibinfo
  {author} {\bibfnamefont {M.}~\bibnamefont {Ha}},\ and\ \bibinfo {author}
  {\bibfnamefont {H.}~\bibnamefont {Jeong}},\ }\bibfield  {title} {\bibinfo
  {title} {Effects of a local defect on one-dimensional nonlinear surface
  growth},\ }\href@noop {} {\bibfield  {journal} {\bibinfo  {journal} {Phys.
  Rev. E}\ }\textbf {\bibinfo {volume} {95}},\ \bibinfo {pages} {042123}
  (\bibinfo {year} {2017})}\BibitemShut {NoStop}%
\bibitem [{\citenamefont {Monthus}(2000)}]{monthus2000localization}%
  \BibitemOpen
  \bibfield  {author} {\bibinfo {author} {\bibfnamefont {C.}~\bibnamefont
  {Monthus}},\ }\bibfield  {title} {\bibinfo {title} {On the localization of
  random heteropolymers at the interface between two selective solvents},\
  }\href@noop {} {\bibfield  {journal} {\bibinfo  {journal} {Eur. Phys. J. B}\
  }\textbf {\bibinfo {volume} {13}},\ \bibinfo {pages} {111} (\bibinfo {year}
  {2000})}\BibitemShut {NoStop}%
\bibitem [{\citenamefont {Giacomin}\ and\ \citenamefont
  {Toninelli}(2006)}]{giacomin2006smoothing}%
  \BibitemOpen
  \bibfield  {author} {\bibinfo {author} {\bibfnamefont {G.}~\bibnamefont
  {Giacomin}}\ and\ \bibinfo {author} {\bibfnamefont {F.~L.}\ \bibnamefont
  {Toninelli}},\ }\bibfield  {title} {\bibinfo {title} {Smoothing of depinning
  transitions for directed polymers with quenched disorder},\ }\href@noop {}
  {\bibfield  {journal} {\bibinfo  {journal} {Phys. Rev. Lett.}\ }\textbf
  {\bibinfo {volume} {96}},\ \bibinfo {pages} {070602} (\bibinfo {year}
  {2006})}\BibitemShut {NoStop}%
\bibitem [{\citenamefont {Toninelli}(2009)}]{toninelli2009localization}%
  \BibitemOpen
  \bibfield  {author} {\bibinfo {author} {\bibfnamefont {F.~L.}\ \bibnamefont
  {Toninelli}},\ }\bibfield  {title} {\bibinfo {title} {Localization transition
  in disordered pinning models},\ }in\ \href@noop {} {\emph {\bibinfo
  {booktitle} {Methods of Contemporary Mathematical Statistical Physics}}}\
  (\bibinfo  {publisher} {Springer},\ \bibinfo {year} {2009})\ pp.\ \bibinfo
  {pages} {129--176}\BibitemShut {NoStop}%
\bibitem [{\citenamefont {Ito}\ and\ \citenamefont
  {Takeuchi}(2018)}]{TakeuchiHalf}%
  \BibitemOpen
  \bibfield  {author} {\bibinfo {author} {\bibfnamefont {Y.}~\bibnamefont
  {Ito}}\ and\ \bibinfo {author} {\bibfnamefont {K.}~\bibnamefont {Takeuchi}},\
  }\bibfield  {title} {\bibinfo {title} {When fast and slow interfaces grow
  together: connection to the half-space problem of the {Kardar-Parisi-Zhang}
  class},\ }\href@noop {} {\bibfield  {journal} {\bibinfo  {journal} {Phys.
  Rev. E}\ }\textbf {\bibinfo {volume} {97}},\ \bibinfo {pages} {040103(R)}
  (\bibinfo {year} {2018})},\ \Eprint {https://arxiv.org/abs/1802.10284}
  {arXiv:1802.10284} \BibitemShut {NoStop}%
\bibitem [{\citenamefont {Iwatsuka}\ \emph {et~al.}(2020)\citenamefont
  {Iwatsuka}, \citenamefont {Fukai},\ and\ \citenamefont
  {Takeuchi}}]{iwatsuka2020direct}%
  \BibitemOpen
  \bibfield  {author} {\bibinfo {author} {\bibfnamefont {T.}~\bibnamefont
  {Iwatsuka}}, \bibinfo {author} {\bibfnamefont {Y.~T.}\ \bibnamefont
  {Fukai}},\ and\ \bibinfo {author} {\bibfnamefont {K.}~\bibnamefont
  {Takeuchi}},\ }\bibfield  {title} {\bibinfo {title} {Direct evidence for
  universal statistics of stationary {Kardar-Parisi-Zhang} interfaces},\
  }\href@noop {} {\bibfield  {journal} {\bibinfo  {journal} {Phys. Rev. Lett.}\
  }\textbf {\bibinfo {volume} {124}},\ \bibinfo {pages} {250602} (\bibinfo
  {year} {2020})},\ \Eprint {https://arxiv.org/abs/2004.11652}
  {arXiv:2004.11652} \BibitemShut {NoStop}%
\bibitem [{\citenamefont {Takeuchi}\ and\ \citenamefont
  {Sano}(2012)}]{takeuchi2012evidence}%
  \BibitemOpen
  \bibfield  {author} {\bibinfo {author} {\bibfnamefont {K.~A.}\ \bibnamefont
  {Takeuchi}}\ and\ \bibinfo {author} {\bibfnamefont {M.}~\bibnamefont
  {Sano}},\ }\bibfield  {title} {\bibinfo {title} {Evidence for
  geometry-dependent universal fluctuations of the {Kardar-Parisi-Zhang}
  interfaces in liquid-crystal turbulence},\ }\href@noop {} {\bibfield
  {journal} {\bibinfo  {journal} {J. Stat. Phys.}\ }\textbf {\bibinfo {volume}
  {147}},\ \bibinfo {pages} {853} (\bibinfo {year} {2012})}\BibitemShut
  {NoStop}%
\bibitem [{\citenamefont {Krug}\ and\ \citenamefont
  {Tang}(1994)}]{krug1994disorder}%
  \BibitemOpen
  \bibfield  {author} {\bibinfo {author} {\bibfnamefont {J.}~\bibnamefont
  {Krug}}\ and\ \bibinfo {author} {\bibfnamefont {L.}~\bibnamefont {Tang}},\
  }\bibfield  {title} {\bibinfo {title} {Disorder-induced unbinding in confined
  geometries},\ }\href@noop {} {\bibfield  {journal} {\bibinfo  {journal}
  {Phys. Rev. E}\ }\textbf {\bibinfo {volume} {50}},\ \bibinfo {pages} {104}
  (\bibinfo {year} {1994})}\BibitemShut {NoStop}%
\bibitem [{\citenamefont {Baik}\ and\ \citenamefont
  {Rains}(2001{\natexlab{a}})}]{baik2001algebraic}%
  \BibitemOpen
  \bibfield  {author} {\bibinfo {author} {\bibfnamefont {J.}~\bibnamefont
  {Baik}}\ and\ \bibinfo {author} {\bibfnamefont {E.~M.}\ \bibnamefont
  {Rains}},\ }\bibfield  {title} {\bibinfo {title} {Algebraic aspects of
  increasing subsequences},\ }\href@noop {} {\bibfield  {journal} {\bibinfo
  {journal} {Duke Math. J.}\ }\textbf {\bibinfo {volume} {109}},\ \bibinfo
  {pages} {1} (\bibinfo {year} {2001}{\natexlab{a}})},\ \Eprint
  {https://arxiv.org/abs/math/9905083} {arXiv:math/9905083} \BibitemShut
  {NoStop}%
\bibitem [{\citenamefont {Baik}\ and\ \citenamefont
  {Rains}(2001{\natexlab{b}})}]{baik2001asymptotics}%
  \BibitemOpen
  \bibfield  {author} {\bibinfo {author} {\bibfnamefont {J.}~\bibnamefont
  {Baik}}\ and\ \bibinfo {author} {\bibfnamefont {E.~M.}\ \bibnamefont
  {Rains}},\ }\bibfield  {title} {\bibinfo {title} {The asymptotics of monotone
  subsequences of involutions},\ }\href@noop {} {\bibfield  {journal} {\bibinfo
   {journal} {Duke Math. J.}\ }\textbf {\bibinfo {volume} {109}},\ \bibinfo
  {pages} {205} (\bibinfo {year} {2001}{\natexlab{b}})},\ \Eprint
  {https://arxiv.org/abs/math/9905084} {arXiv:math/9905084} \BibitemShut
  {NoStop}%
\bibitem [{\citenamefont {Baik}\ and\ \citenamefont
  {Rains}(2001{\natexlab{c}})}]{baik2001symmetrized}%
  \BibitemOpen
  \bibfield  {author} {\bibinfo {author} {\bibfnamefont {J.}~\bibnamefont
  {Baik}}\ and\ \bibinfo {author} {\bibfnamefont {E.~M.}\ \bibnamefont
  {Rains}},\ }\bibfield  {title} {\bibinfo {title} {Symmetrized random
  permutations},\ }in\ \href@noop {} {\emph {\bibinfo {booktitle} {Random
  matrix models and their applications}}},\ \bibinfo {series} {Math. Sci. Res.
  Inst. Publ.}, Vol.~\bibinfo {volume} {40}\ (\bibinfo  {publisher} {Cambridge
  Univ. Press, Cambridge},\ \bibinfo {year} {2001})\ pp.\ \bibinfo {pages}
  {1--19},\ \Eprint {https://arxiv.org/abs/math/9910019} {arXiv:math/9910019}
  \BibitemShut {NoStop}%
\bibitem [{\citenamefont {Baik}\ \emph {et~al.}(2018)\citenamefont {Baik},
  \citenamefont {Barraquand}, \citenamefont {Corwin},\ and\ \citenamefont
  {Suidan}}]{baik2018pfaffian}%
  \BibitemOpen
  \bibfield  {author} {\bibinfo {author} {\bibfnamefont {J.}~\bibnamefont
  {Baik}}, \bibinfo {author} {\bibfnamefont {G.}~\bibnamefont {Barraquand}},
  \bibinfo {author} {\bibfnamefont {I.}~\bibnamefont {Corwin}},\ and\ \bibinfo
  {author} {\bibfnamefont {T.}~\bibnamefont {Suidan}},\ }\bibfield  {title}
  {\bibinfo {title} {Pfaffian {S}chur processes and last passage percolation in
  a half-quadrant},\ }\href@noop {} {\bibfield  {journal} {\bibinfo  {journal}
  {Ann. Probab.}\ }\textbf {\bibinfo {volume} {46}},\ \bibinfo {pages} {3015}
  (\bibinfo {year} {2018})},\ \Eprint {https://arxiv.org/abs/1606.00525}
  {arXiv:1606.00525} \BibitemShut {NoStop}%
\bibitem [{\citenamefont {Barraquand}\ \emph
  {et~al.}(2020{\natexlab{a}})\citenamefont {Barraquand}, \citenamefont
  {Borodin},\ and\ \citenamefont {Corwin}}]{barraquand2018half}%
  \BibitemOpen
  \bibfield  {author} {\bibinfo {author} {\bibfnamefont {G.}~\bibnamefont
  {Barraquand}}, \bibinfo {author} {\bibfnamefont {A.}~\bibnamefont
  {Borodin}},\ and\ \bibinfo {author} {\bibfnamefont {I.}~\bibnamefont
  {Corwin}},\ }\bibfield  {title} {\bibinfo {title} {Half-space {M}acdonald
  processes},\ }\href@noop {} {\bibfield  {journal} {\bibinfo  {journal} {Forum
  Math. Pi}\ }\textbf {\bibinfo {volume} {8}} (\bibinfo {year}
  {2020}{\natexlab{a}})},\ \Eprint {https://arxiv.org/abs/1802.08210}
  {arXiv:1802.08210} \BibitemShut {NoStop}%
\bibitem [{\citenamefont {Gueudr{\'e}}\ and\ \citenamefont
  {Le~Doussal}(2012)}]{gueudre2012directed}%
  \BibitemOpen
  \bibfield  {author} {\bibinfo {author} {\bibfnamefont {T.}~\bibnamefont
  {Gueudr{\'e}}}\ and\ \bibinfo {author} {\bibfnamefont {P.}~\bibnamefont
  {Le~Doussal}},\ }\bibfield  {title} {\bibinfo {title} {Directed polymer near
  a hard wall and {KPZ} equation in the half-space},\ }\href@noop {} {\bibfield
   {journal} {\bibinfo  {journal} {EuroPhys. Lett.}\ }\textbf {\bibinfo
  {volume} {100}},\ \bibinfo {pages} {26006} (\bibinfo {year} {2012})},\
  \Eprint {https://arxiv.org/abs/1208.5669} {arXiv:1208.5669} \BibitemShut
  {NoStop}%
\bibitem [{\citenamefont {Borodin}\ \emph
  {et~al.}(2016{\natexlab{a}})\citenamefont {Borodin}, \citenamefont
  {Bufetov},\ and\ \citenamefont {Corwin}}]{borodin2016directed}%
  \BibitemOpen
  \bibfield  {author} {\bibinfo {author} {\bibfnamefont {A.}~\bibnamefont
  {Borodin}}, \bibinfo {author} {\bibfnamefont {A.}~\bibnamefont {Bufetov}},\
  and\ \bibinfo {author} {\bibfnamefont {I.}~\bibnamefont {Corwin}},\
  }\bibfield  {title} {\bibinfo {title} {Directed random polymers via nested
  contour integrals},\ }\href@noop {} {\bibfield  {journal} {\bibinfo
  {journal} {Ann. Phys.}\ }\textbf {\bibinfo {volume} {368}},\ \bibinfo {pages}
  {191} (\bibinfo {year} {2016}{\natexlab{a}})},\ \Eprint
  {https://arxiv.org/abs/1511.07324} {arXiv:1511.07324} \BibitemShut {NoStop}%
\bibitem [{\citenamefont {Barraquand}\ \emph {et~al.}(2018)\citenamefont
  {Barraquand}, \citenamefont {Borodin}, \citenamefont {Corwin},\ and\
  \citenamefont {Wheeler}}]{barraquand2018stochastic}%
  \BibitemOpen
  \bibfield  {author} {\bibinfo {author} {\bibfnamefont {G.}~\bibnamefont
  {Barraquand}}, \bibinfo {author} {\bibfnamefont {A.}~\bibnamefont {Borodin}},
  \bibinfo {author} {\bibfnamefont {I.}~\bibnamefont {Corwin}},\ and\ \bibinfo
  {author} {\bibfnamefont {M.}~\bibnamefont {Wheeler}},\ }\bibfield  {title}
  {\bibinfo {title} {Stochastic six-vertex model in a half-quadrant and
  half-line open asymmetric simple exclusion process},\ }\href@noop {}
  {\bibfield  {journal} {\bibinfo  {journal} {Duke Math. J.}\ }\textbf
  {\bibinfo {volume} {167}},\ \bibinfo {pages} {2457} (\bibinfo {year}
  {2018})},\ \Eprint {https://arxiv.org/abs/1704.04309} {arXiv:1704.04309}
  \BibitemShut {NoStop}%
\bibitem [{\citenamefont {Krajenbrink}\ and\ \citenamefont
  {Le~Doussal}(2020)}]{AlexLD}%
  \BibitemOpen
  \bibfield  {author} {\bibinfo {author} {\bibfnamefont {A.}~\bibnamefont
  {Krajenbrink}}\ and\ \bibinfo {author} {\bibfnamefont {P.}~\bibnamefont
  {Le~Doussal}},\ }\bibfield  {title} {\bibinfo {title} {Replica bethe ansatz
  solution to the {Kardar-Parisi-Zhang} equation on the half-line},\
  }\href@noop {} {\bibfield  {journal} {\bibinfo  {journal} {SciPost Phys}\
  }\textbf {\bibinfo {volume} {8}},\ \bibinfo {pages} {035} (\bibinfo {year}
  {2020})},\ \Eprint {https://arxiv.org/abs/1905.05718} {arXiv:1905.05718}
  \BibitemShut {NoStop}%
\bibitem [{\citenamefont {Barraquand}\ \emph
  {et~al.}(2020{\natexlab{b}})\citenamefont {Barraquand}, \citenamefont
  {Krajenbrink},\ and\ \citenamefont {Le~Doussal}}]{barraquand2020half}%
  \BibitemOpen
  \bibfield  {author} {\bibinfo {author} {\bibfnamefont {G.}~\bibnamefont
  {Barraquand}}, \bibinfo {author} {\bibfnamefont {A.}~\bibnamefont
  {Krajenbrink}},\ and\ \bibinfo {author} {\bibfnamefont {P.}~\bibnamefont
  {Le~Doussal}},\ }\bibfield  {title} {\bibinfo {title} {Half-space stationary
  {Kardar--Parisi--Zhang} equation},\ }\href@noop {} {\bibfield  {journal}
  {\bibinfo  {journal} {J. Stat. Phys.}\ }\textbf {\bibinfo {volume} {181}},\
  \bibinfo {pages} {1149} (\bibinfo {year} {2020}{\natexlab{b}})},\ \Eprint
  {https://arxiv.org/abs/2003.03809} {arXiv:2003.03809} \BibitemShut {NoStop}%
\bibitem [{\citenamefont {De~Nardis}\ \emph {et~al.}(2020)\citenamefont
  {De~Nardis}, \citenamefont {Krajenbrink}, \citenamefont {Le~Doussal},\ and\
  \citenamefont {Thiery}}]{deNardisPLDTT}%
  \BibitemOpen
  \bibfield  {author} {\bibinfo {author} {\bibfnamefont {J.}~\bibnamefont
  {De~Nardis}}, \bibinfo {author} {\bibfnamefont {A.}~\bibnamefont
  {Krajenbrink}}, \bibinfo {author} {\bibfnamefont {P.}~\bibnamefont
  {Le~Doussal}},\ and\ \bibinfo {author} {\bibfnamefont {T.}~\bibnamefont
  {Thiery}},\ }\bibfield  {title} {\bibinfo {title} {Delta-bose gas on a
  half-line and the kpz equation: boundary bound states and unbinding
  transitions (2019)},\ }\href {https://doi.org/10.1088/1742-5468/ab7751}
  {\bibfield  {journal} {\bibinfo  {journal} {J. Stat. Mech.: Theor. Exp.}\
  }\textbf {\bibinfo {volume} {2020}},\ \bibinfo {pages} {043207} (\bibinfo
  {year} {2020})},\ \Eprint {https://arxiv.org/abs/1911.06133}
  {arXiv:1911.06133} \BibitemShut {NoStop}%
\bibitem [{\citenamefont {Baik}\ \emph {et~al.}(2005)\citenamefont {Baik},
  \citenamefont {Ben~Arous},\ and\ \citenamefont
  {P{\'e}ch{\'e}}}]{baik2005phase}%
  \BibitemOpen
  \bibfield  {author} {\bibinfo {author} {\bibfnamefont {J.}~\bibnamefont
  {Baik}}, \bibinfo {author} {\bibfnamefont {G.}~\bibnamefont {Ben~Arous}},\
  and\ \bibinfo {author} {\bibfnamefont {S.}~\bibnamefont {P{\'e}ch{\'e}}},\
  }\bibfield  {title} {\bibinfo {title} {Phase transition of the largest
  eigenvalue for nonnull complex sample covariance matrices},\ }\href@noop {}
  {\bibfield  {journal} {\bibinfo  {journal} {Ann. Probab.}\ }\textbf {\bibinfo
  {volume} {33}},\ \bibinfo {pages} {1643} (\bibinfo {year} {2005})},\ \Eprint
  {https://arxiv.org/abs/math/0403022} {arXiv:math/0403022} \BibitemShut
  {NoStop}%
\bibitem [{\citenamefont {Borodin}\ and\ \citenamefont
  {Corwin}(2014)}]{borodin2014macdonald}%
  \BibitemOpen
  \bibfield  {author} {\bibinfo {author} {\bibfnamefont {A.}~\bibnamefont
  {Borodin}}\ and\ \bibinfo {author} {\bibfnamefont {I.}~\bibnamefont
  {Corwin}},\ }\bibfield  {title} {\bibinfo {title} {Macdonald processes},\
  }\href@noop {} {\bibfield  {journal} {\bibinfo  {journal} {Probab. Theory
  Rel. Fields}\ }\textbf {\bibinfo {volume} {158}},\ \bibinfo {pages} {225}
  (\bibinfo {year} {2014})},\ \Eprint {https://arxiv.org/abs/1111.4408}
  {arXiv:1111.4408} \BibitemShut {NoStop}%
\bibitem [{\citenamefont {Le~Doussal}\ and\ \citenamefont
  {Calabrese}(2012)}]{le2012kpz}%
  \BibitemOpen
  \bibfield  {author} {\bibinfo {author} {\bibfnamefont {P.}~\bibnamefont
  {Le~Doussal}}\ and\ \bibinfo {author} {\bibfnamefont {P.}~\bibnamefont
  {Calabrese}},\ }\bibfield  {title} {\bibinfo {title} {The kpz equation with
  flat initial condition and the directed polymer with one free end},\
  }\href@noop {} {\bibfield  {journal} {\bibinfo  {journal} {J. Stat. Mech.:
  Theor. Exp.}\ }\textbf {\bibinfo {volume} {2012}},\ \bibinfo {pages} {P06001}
  (\bibinfo {year} {2012})},\ \Eprint {https://arxiv.org/abs/1204.2607}
  {arXiv:1204.2607} \BibitemShut {NoStop}%
\bibitem [{\citenamefont {Lee}(2010)}]{lee2010distribution}%
  \BibitemOpen
  \bibfield  {author} {\bibinfo {author} {\bibfnamefont {E.}~\bibnamefont
  {Lee}},\ }\bibfield  {title} {\bibinfo {title} {Distribution of a
  particle’s position in the {ASEP} with the alternating initial condition},\
  }\href@noop {} {\bibfield  {journal} {\bibinfo  {journal} {J. Stat. Phys.}\
  }\textbf {\bibinfo {volume} {140}},\ \bibinfo {pages} {635} (\bibinfo {year}
  {2010})},\ \Eprint {https://arxiv.org/abs/1004.1470} {arXiv:1004.1470}
  \BibitemShut {NoStop}%
\bibitem [{\citenamefont {Borodin}\ \emph {et~al.}(2014)\citenamefont
  {Borodin}, \citenamefont {Corwin},\ and\ \citenamefont
  {Ferrari}}]{borodin2012free}%
  \BibitemOpen
  \bibfield  {author} {\bibinfo {author} {\bibfnamefont {A.}~\bibnamefont
  {Borodin}}, \bibinfo {author} {\bibfnamefont {I.}~\bibnamefont {Corwin}},\
  and\ \bibinfo {author} {\bibfnamefont {P.}~\bibnamefont {Ferrari}},\
  }\bibfield  {title} {\bibinfo {title} {Free energy fluctuations for directed
  polymers in random media in 1+ 1 dimension},\ }\href@noop {} {\bibfield
  {journal} {\bibinfo  {journal} {Comm. Pure Appl. Math.}\ }\textbf {\bibinfo
  {volume} {67}},\ \bibinfo {pages} {1129} (\bibinfo {year} {2014})},\ \Eprint
  {https://arxiv.org/abs/1204.1024} {arXiv:1204.1024} \BibitemShut {NoStop}%
\bibitem [{\citenamefont {Baik}(2006)}]{baik2006painleve}%
  \BibitemOpen
  \bibfield  {author} {\bibinfo {author} {\bibfnamefont {J.}~\bibnamefont
  {Baik}},\ }\bibfield  {title} {\bibinfo {title} {Painlev{\'e} formulas of the
  limiting distributions for nonnull complex sample covariance matrices},\
  }\href@noop {} {\bibfield  {journal} {\bibinfo  {journal} {Duke Math. J.}\
  }\textbf {\bibinfo {volume} {133}},\ \bibinfo {pages} {205} (\bibinfo {year}
  {2006})},\ \Eprint {https://arxiv.org/abs/math/0504606} {arXiv:math/0504606}
  \BibitemShut {NoStop}%
\bibitem [{\citenamefont {Barraquand}(2015)}]{barraquand2014phase}%
  \BibitemOpen
  \bibfield  {author} {\bibinfo {author} {\bibfnamefont {G.}~\bibnamefont
  {Barraquand}},\ }\bibfield  {title} {\bibinfo {title} {A phase transition for
  q-{TASEP} with a few slower particles},\ }\href@noop {} {\bibfield  {journal}
  {\bibinfo  {journal} {Stoch. Proc. Appl.}\ }\textbf {\bibinfo {volume}
  {125}},\ \bibinfo {pages} {2674 } (\bibinfo {year} {2015})},\ \Eprint
  {https://arxiv.org/abs/1404.7409} {arXiv:1404.7409} \BibitemShut {NoStop}%
\bibitem [{\citenamefont {Krajenbrink}\ \emph {et~al.}(2020)\citenamefont
  {Krajenbrink}, \citenamefont {Le~Doussal},\ and\ \citenamefont
  {O'Connell}}]{krajenbrink2020tilted}%
  \BibitemOpen
  \bibfield  {author} {\bibinfo {author} {\bibfnamefont {A.}~\bibnamefont
  {Krajenbrink}}, \bibinfo {author} {\bibfnamefont {P.~L.}\ \bibnamefont
  {Le~Doussal}},\ and\ \bibinfo {author} {\bibfnamefont {N.}~\bibnamefont
  {O'Connell}},\ }\bibfield  {title} {\bibinfo {title} {Tilted elastic lines
  with columnar and point disorder, non-{H}ermitian quantum mechanics and
  spiked random matrices: pinning and localization},\ }\href@noop {} {\bibfield
   {journal} {\bibinfo  {journal} {preprint}\ } (\bibinfo {year} {2020})},\
  \Eprint {https://arxiv.org/abs/2009.11284} {arXiv:2009.11284} \BibitemShut
  {NoStop}%
\bibitem [{\citenamefont {Monthus}\ and\ \citenamefont
  {Le~Doussal}(2004)}]{monthus2004low}%
  \BibitemOpen
  \bibfield  {author} {\bibinfo {author} {\bibfnamefont {C.}~\bibnamefont
  {Monthus}}\ and\ \bibinfo {author} {\bibfnamefont {P.}~\bibnamefont
  {Le~Doussal}},\ }\bibfield  {title} {\bibinfo {title} {Low-temperature
  properties of some disordered systems from the statistical properties of
  nearly degenerate two-level excitations},\ }\href@noop {} {\bibfield
  {journal} {\bibinfo  {journal} {Europ. Phys. J. B - Cond. Matt. Complex
  Sys.}\ }\textbf {\bibinfo {volume} {41}},\ \bibinfo {pages} {535} (\bibinfo
  {year} {2004})}\BibitemShut {NoStop}%
\bibitem [{\citenamefont {Bouchaud}\ \emph {et~al.}(1990)\citenamefont
  {Bouchaud}, \citenamefont {Comtet}, \citenamefont {Georges},\ and\
  \citenamefont {Doussal}}]{BouchaudPLD1990}%
  \BibitemOpen
  \bibfield  {author} {\bibinfo {author} {\bibfnamefont {J.~P.}\ \bibnamefont
  {Bouchaud}}, \bibinfo {author} {\bibfnamefont {A.}~\bibnamefont {Comtet}},
  \bibinfo {author} {\bibfnamefont {A.}~\bibnamefont {Georges}},\ and\ \bibinfo
  {author} {\bibfnamefont {P.~L.}\ \bibnamefont {Doussal}},\ }\bibfield
  {title} {\bibinfo {title} {Classical diffusion of a particle in a
  one-dimensional random force field},\ }\href@noop {} {\bibfield  {journal}
  {\bibinfo  {journal} {Ann. Phys.}\ }\textbf {\bibinfo {volume} {201}},\
  \bibinfo {pages} {285} (\bibinfo {year} {1990})}\BibitemShut {NoStop}%
\bibitem [{\citenamefont {Dufresne}(1990)}]{Dufresne1990}%
  \BibitemOpen
  \bibfield  {author} {\bibinfo {author} {\bibfnamefont {D.}~\bibnamefont
  {Dufresne}},\ }\bibfield  {title} {\bibinfo {title} {The distribution of a
  perpetuity, with applications to risk theory and pension funding},\
  }\href@noop {} {\bibfield  {journal} {\bibinfo  {journal} {Scand. Actuar.
  J.}\ }\textbf {\bibinfo {volume} {1990}},\ \bibinfo {pages} {39} (\bibinfo
  {year} {1990})}\BibitemShut {NoStop}%
\bibitem [{Pro()}]{ProbaMax}%
  \BibitemOpen
  \href@noop {} {}\bibinfo {note} {See e.g. Chapter IV, item 32 in A.N. Borodin
  and E Salminen. Handbook of Brownian Motion: Facts and Formulae. Birkhauser,
  Basel, 1996, or taking the limit $T \to +\infty$ in Shepp, Lawrence A. The
  joint density of the maximum and its location for a Wiener process with
  drift. Journal of Applied Probability 16.2 (1979): 423-427, or in Eq. (30) in
  S. N. Majumdar and J.-P. Bouchaud. Optimal time to sell a stock in the
  black-scholes model: comment on ‘thou shalt buy and hold’, by A.
  Shiryaev, Z. XU and X.Y. Zhou. Quantitative Finance, 8(8):753-760,
  2008.}\BibitemShut {Stop}%
\bibitem [{\citenamefont {Comtet}\ and\ \citenamefont
  {Texier}(1998)}]{TexierComtetSUSY}%
  \BibitemOpen
  \bibfield  {author} {\bibinfo {author} {\bibfnamefont {A.}~\bibnamefont
  {Comtet}}\ and\ \bibinfo {author} {\bibfnamefont {C.}~\bibnamefont
  {Texier}},\ }\bibfield  {title} {\bibinfo {title} {One-dimensional disordered
  supersymmetric quantum mechanics: a brief survey},\ }\href@noop {} {\bibfield
   {journal} {\bibinfo  {journal} {Supersymmetry and Integrable Models}\ ,\
  \bibinfo {pages} {313}} (\bibinfo {year} {1998})},\ \Eprint
  {https://arxiv.org/abs/cond-mat/9707313} {arXiv:cond-mat/9707313}
  \BibitemShut {NoStop}%
\bibitem [{\citenamefont {Monthus}\ and\ \citenamefont
  {Comtet}(1994)}]{monthus1994flux}%
  \BibitemOpen
  \bibfield  {author} {\bibinfo {author} {\bibfnamefont {C.}~\bibnamefont
  {Monthus}}\ and\ \bibinfo {author} {\bibfnamefont {A.}~\bibnamefont
  {Comtet}},\ }\bibfield  {title} {\bibinfo {title} {On the flux distribution
  in a one dimensional disordered system},\ }\href@noop {} {\bibfield
  {journal} {\bibinfo  {journal} {Journal de Physique I}\ }\textbf {\bibinfo
  {volume} {4}},\ \bibinfo {pages} {635} (\bibinfo {year} {1994})}\BibitemShut
  {NoStop}%
\bibitem [{\citenamefont {Broderix}\ and\ \citenamefont
  {Kree}(1995)}]{broderix1995thermal}%
  \BibitemOpen
  \bibfield  {author} {\bibinfo {author} {\bibfnamefont {K.}~\bibnamefont
  {Broderix}}\ and\ \bibinfo {author} {\bibfnamefont {R.}~\bibnamefont
  {Kree}},\ }\bibfield  {title} {\bibinfo {title} {Thermal equilibrium with the
  {W}iener potential: testing the replica variational approximation},\
  }\href@noop {} {\bibfield  {journal} {\bibinfo  {journal} {EuroPhys. Lett.}\
  }\textbf {\bibinfo {volume} {32}},\ \bibinfo {pages} {343} (\bibinfo {year}
  {1995})}\BibitemShut {NoStop}%
\bibitem [{\citenamefont {Comtet}\ \emph {et~al.}(1998)\citenamefont {Comtet},
  \citenamefont {Monthus},\ and\ \citenamefont {Yor}}]{comtet1998exponential}%
  \BibitemOpen
  \bibfield  {author} {\bibinfo {author} {\bibfnamefont {A.}~\bibnamefont
  {Comtet}}, \bibinfo {author} {\bibfnamefont {C.}~\bibnamefont {Monthus}},\
  and\ \bibinfo {author} {\bibfnamefont {M.}~\bibnamefont {Yor}},\ }\bibfield
  {title} {\bibinfo {title} {Exponential functionals of {B}rownian motion and
  disordered systems},\ }\href@noop {} {\bibfield  {journal} {\bibinfo
  {journal} {J. Appl. Probab.}\ }\textbf {\bibinfo {volume} {35}},\ \bibinfo
  {pages} {255} (\bibinfo {year} {1998})},\ \Eprint
  {https://arxiv.org/abs/cond-mat/9601014} {arXiv:cond-mat/9601014}
  \BibitemShut {NoStop}%
\bibitem [{\citenamefont {Monthus}\ and\ \citenamefont
  {Le~Doussal}(2002)}]{monthus2002localization}%
  \BibitemOpen
  \bibfield  {author} {\bibinfo {author} {\bibfnamefont {C.}~\bibnamefont
  {Monthus}}\ and\ \bibinfo {author} {\bibfnamefont {P.}~\bibnamefont
  {Le~Doussal}},\ }\bibfield  {title} {\bibinfo {title} {Localization of
  thermal packets and metastable states in the sinai model},\ }\href@noop {}
  {\bibfield  {journal} {\bibinfo  {journal} {Phys. Rev. E}\ }\textbf {\bibinfo
  {volume} {65}},\ \bibinfo {pages} {066129} (\bibinfo {year} {2002})},\
  \Eprint {https://arxiv.org/abs/cond-mat/0202295} {arXiv:cond-mat/0202295}
  \BibitemShut {NoStop}%
\bibitem [{Note1()}]{Note1}%
  \BibitemOpen
  \bibinfo {note} {$x(2 b t)$ for any fixed $0<b<1$ has the same distribution
  for $t \to +\infty $.}\BibitemShut {Stop}%
\bibitem [{\citenamefont {Liggett}(1975)}]{liggett1975ergodic}%
  \BibitemOpen
  \bibfield  {author} {\bibinfo {author} {\bibfnamefont {T.~M.}\ \bibnamefont
  {Liggett}},\ }\bibfield  {title} {\bibinfo {title} {Ergodic theorems for the
  asymmetric simple exclusion process},\ }\href@noop {} {\bibfield  {journal}
  {\bibinfo  {journal} {Trans. Amer. Math. Soc.}\ }\textbf {\bibinfo {volume}
  {213}},\ \bibinfo {pages} {237} (\bibinfo {year} {1975})}\BibitemShut
  {NoStop}%
\bibitem [{\citenamefont {Derrida}\ \emph {et~al.}(1993)\citenamefont
  {Derrida}, \citenamefont {Evans}, \citenamefont {Hakim},\ and\ \citenamefont
  {Pasquier}}]{derrida1993exact}%
  \BibitemOpen
  \bibfield  {author} {\bibinfo {author} {\bibfnamefont {B.}~\bibnamefont
  {Derrida}}, \bibinfo {author} {\bibfnamefont {M.~R.}\ \bibnamefont {Evans}},
  \bibinfo {author} {\bibfnamefont {V.}~\bibnamefont {Hakim}},\ and\ \bibinfo
  {author} {\bibfnamefont {V.}~\bibnamefont {Pasquier}},\ }\bibfield  {title}
  {\bibinfo {title} {Exact solution of a {1D} asymmetric exclusion model using
  a matrix formulation},\ }\href@noop {} {\bibfield  {journal} {\bibinfo
  {journal} {J. Phys. A: Math. Gen.}\ }\textbf {\bibinfo {volume} {26}},\
  \bibinfo {pages} {1493} (\bibinfo {year} {1993})}\BibitemShut {NoStop}%
\bibitem [{Note2()}]{Note2}%
  \BibitemOpen
  \bibinfo {note} {Assuming universality, both identities would imply in the
  large time limit the same relation between full-space and half-space KPZ
  fixed point.}\BibitemShut {Stop}%
\bibitem [{\citenamefont {Parekh}(2019)}]{parekh2019positive}%
  \BibitemOpen
  \bibfield  {author} {\bibinfo {author} {\bibfnamefont {S.}~\bibnamefont
  {Parekh}},\ }\bibfield  {title} {\bibinfo {title} {Positive random walks and
  an identity for half-space {SPDEs}},\ }\href@noop {} {\bibfield  {journal}
  {\bibinfo  {journal} {arXiv preprint arXiv:1901.09449}\ } (\bibinfo {year}
  {2019})},\ \Eprint {https://arxiv.org/abs/1901.09449} {arXiv:1901.09449}
  \BibitemShut {NoStop}%
\bibitem [{\citenamefont {Imamura}\ and\ \citenamefont
  {Sasamoto}(2004)}]{imamura2004fluctuations}%
  \BibitemOpen
  \bibfield  {author} {\bibinfo {author} {\bibfnamefont {T.}~\bibnamefont
  {Imamura}}\ and\ \bibinfo {author} {\bibfnamefont {T.}~\bibnamefont
  {Sasamoto}},\ }\bibfield  {title} {\bibinfo {title} {Fluctuations of the
  one-dimensional polynuclear growth model with external sources},\ }\href@noop
  {} {\bibfield  {journal} {\bibinfo  {journal} {Nucl. Phys. B}\ }\textbf
  {\bibinfo {volume} {699}},\ \bibinfo {pages} {503} (\bibinfo {year}
  {2004})},\ \Eprint {https://arxiv.org/abs/math-ph/0406001}
  {arXiv:math-ph/0406001} \BibitemShut {NoStop}%
\bibitem [{\citenamefont {Corwin}\ \emph {et~al.}(2010)\citenamefont {Corwin},
  \citenamefont {Ferrari},\ and\ \citenamefont
  {P{\'e}ch{\'e}}}]{corwin2010limit}%
  \BibitemOpen
  \bibfield  {author} {\bibinfo {author} {\bibfnamefont {I.}~\bibnamefont
  {Corwin}}, \bibinfo {author} {\bibfnamefont {P.~L.}\ \bibnamefont
  {Ferrari}},\ and\ \bibinfo {author} {\bibfnamefont {S.}~\bibnamefont
  {P{\'e}ch{\'e}}},\ }\bibfield  {title} {\bibinfo {title} {Limit processes for
  {TASEP} with shocks and rarefaction fans},\ }\href@noop {} {\bibfield
  {journal} {\bibinfo  {journal} {J. Stat. Phys.}\ }\textbf {\bibinfo {volume}
  {140}},\ \bibinfo {pages} {232} (\bibinfo {year} {2010})},\ \Eprint
  {https://arxiv.org/abs/1002.3476} {arXiv:1002.3476} \BibitemShut {NoStop}%
\bibitem [{\citenamefont {Corwin}\ and\ \citenamefont
  {Knizel}(2021)}]{corwin2021stationary}%
  \BibitemOpen
  \bibfield  {author} {\bibinfo {author} {\bibfnamefont {I.}~\bibnamefont
  {Corwin}}\ and\ \bibinfo {author} {\bibfnamefont {A.}~\bibnamefont
  {Knizel}},\ }\bibfield  {title} {\bibinfo {title} {Stationary measure for the
  open {KPZ} equation},\ }\href@noop {} {\bibfield  {journal} {\bibinfo
  {journal} {preprint}\ } (\bibinfo {year} {2021})},\ \Eprint
  {https://arxiv.org/abs/2103.12253} {arXiv:2103.12253} \BibitemShut {NoStop}%
\bibitem [{\citenamefont {Sepp{\"a}l{\"a}inen}(2012)}]{seppalainen2012scaling}%
  \BibitemOpen
  \bibfield  {author} {\bibinfo {author} {\bibfnamefont {T.}~\bibnamefont
  {Sepp{\"a}l{\"a}inen}},\ }\bibfield  {title} {\bibinfo {title} {Scaling for a
  one-dimensional directed polymer with boundary conditions},\ }\href@noop {}
  {\bibfield  {journal} {\bibinfo  {journal} {Ann. Probab.}\ }\textbf {\bibinfo
  {volume} {40}},\ \bibinfo {pages} {19} (\bibinfo {year} {2012})},\ \Eprint
  {https://arxiv.org/abs/0911.2446} {arXiv:0911.2446} \BibitemShut {NoStop}%
\bibitem [{\citenamefont {Borodin}\ \emph {et~al.}(2013)\citenamefont
  {Borodin}, \citenamefont {Corwin},\ and\ \citenamefont
  {Remenik}}]{borodin2013log}%
  \BibitemOpen
  \bibfield  {author} {\bibinfo {author} {\bibfnamefont {A.}~\bibnamefont
  {Borodin}}, \bibinfo {author} {\bibfnamefont {I.}~\bibnamefont {Corwin}},\
  and\ \bibinfo {author} {\bibfnamefont {D.}~\bibnamefont {Remenik}},\
  }\bibfield  {title} {\bibinfo {title} {Log-gamma polymer free energy
  fluctuations via a fredholm determinant identity},\ }\href@noop {} {\bibfield
   {journal} {\bibinfo  {journal} {Comm. Math. Phys.}\ }\textbf {\bibinfo
  {volume} {324}},\ \bibinfo {pages} {215} (\bibinfo {year} {2013})},\ \Eprint
  {https://arxiv.org/abs/1206.4573} {arXiv:1206.4573} \BibitemShut {NoStop}%
\bibitem [{\citenamefont {Barraquand}\ \emph
  {et~al.}(2020{\natexlab{c}})\citenamefont {Barraquand}, \citenamefont
  {Corwin},\ and\ \citenamefont {Dimitrov}}]{barraquand2020fluctuations}%
  \BibitemOpen
  \bibfield  {author} {\bibinfo {author} {\bibfnamefont {G.}~\bibnamefont
  {Barraquand}}, \bibinfo {author} {\bibfnamefont {I.}~\bibnamefont {Corwin}},\
  and\ \bibinfo {author} {\bibfnamefont {E.}~\bibnamefont {Dimitrov}},\
  }\bibfield  {title} {\bibinfo {title} {Fluctuations of the log-gamma polymer
  free energy with general parameters and slopes},\ }\href@noop {} {\bibfield
  {journal} {\bibinfo  {journal} {arXiv preprint arXiv:2012.12316}\ } (\bibinfo
  {year} {2020}{\natexlab{c}})},\ \Eprint {https://arxiv.org/abs/2012.12316}
  {arXiv:2012.12316} \BibitemShut {NoStop}%
\bibitem [{\citenamefont {Corwin}\ \emph {et~al.}(2014)\citenamefont {Corwin},
  \citenamefont {O'Connell}, \citenamefont {Sepp{\"a}l{\"a}inen},\ and\
  \citenamefont {Zygouras}}]{corwin2014tropical}%
  \BibitemOpen
  \bibfield  {author} {\bibinfo {author} {\bibfnamefont {I.}~\bibnamefont
  {Corwin}}, \bibinfo {author} {\bibfnamefont {N.}~\bibnamefont {O'Connell}},
  \bibinfo {author} {\bibfnamefont {T.}~\bibnamefont {Sepp{\"a}l{\"a}inen}},\
  and\ \bibinfo {author} {\bibfnamefont {N.}~\bibnamefont {Zygouras}},\
  }\bibfield  {title} {\bibinfo {title} {Tropical combinatorics and {W}hittaker
  functions},\ }\href@noop {} {\bibfield  {journal} {\bibinfo  {journal} {Duke
  Math. J.}\ }\textbf {\bibinfo {volume} {163}},\ \bibinfo {pages} {513}
  (\bibinfo {year} {2014})},\ \Eprint {https://arxiv.org/abs/1110.3489}
  {arXiv:1110.3489} \BibitemShut {NoStop}%
\bibitem [{\citenamefont {Borodin}\ \emph
  {et~al.}(2016{\natexlab{b}})\citenamefont {Borodin}, \citenamefont {Corwin},
  \citenamefont {Gorin},\ and\ \citenamefont
  {Shakirov}}]{borodin2016observables}%
  \BibitemOpen
  \bibfield  {author} {\bibinfo {author} {\bibfnamefont {A.}~\bibnamefont
  {Borodin}}, \bibinfo {author} {\bibfnamefont {I.}~\bibnamefont {Corwin}},
  \bibinfo {author} {\bibfnamefont {V.}~\bibnamefont {Gorin}},\ and\ \bibinfo
  {author} {\bibfnamefont {S.}~\bibnamefont {Shakirov}},\ }\bibfield  {title}
  {\bibinfo {title} {Observables of {M}acdonald processes},\ }\href@noop {}
  {\bibfield  {journal} {\bibinfo  {journal} {Trans. Amer. Math. Soc.}\
  }\textbf {\bibinfo {volume} {368}},\ \bibinfo {pages} {1517} (\bibinfo {year}
  {2016}{\natexlab{b}})},\ \Eprint {https://arxiv.org/abs/1306.0659}
  {arXiv:1306.0659} \BibitemShut {NoStop}%
\bibitem [{\citenamefont {Alberts}\ \emph {et~al.}(2010)\citenamefont
  {Alberts}, \citenamefont {Khanin},\ and\ \citenamefont
  {Quastel}}]{alberts2012intermediate}%
  \BibitemOpen
  \bibfield  {author} {\bibinfo {author} {\bibfnamefont {T.}~\bibnamefont
  {Alberts}}, \bibinfo {author} {\bibfnamefont {K.}~\bibnamefont {Khanin}},\
  and\ \bibinfo {author} {\bibfnamefont {J.}~\bibnamefont {Quastel}},\
  }\bibfield  {title} {\bibinfo {title} {Intermediate disorder regime for
  directed polymers in dimension $1+1$},\ }\href
  {https://doi.org/10.1103/PhysRevLett.105.090603} {\bibfield  {journal}
  {\bibinfo  {journal} {Phys. Rev. Lett.}\ }\textbf {\bibinfo {volume} {105}},\
  \bibinfo {pages} {090603} (\bibinfo {year} {2010})}\BibitemShut {NoStop}%
\bibitem [{\citenamefont {Alberts}\ \emph {et~al.}(2014)\citenamefont
  {Alberts}, \citenamefont {Khanin},\ and\ \citenamefont
  {Quastel}}]{alberts2014intermediate}%
  \BibitemOpen
  \bibfield  {author} {\bibinfo {author} {\bibfnamefont {T.}~\bibnamefont
  {Alberts}}, \bibinfo {author} {\bibfnamefont {K.}~\bibnamefont {Khanin}},\
  and\ \bibinfo {author} {\bibfnamefont {J.}~\bibnamefont {Quastel}},\
  }\bibfield  {title} {\bibinfo {title} {The intermediate disorder regime for
  directed polymers in dimension $1+ 1$},\ }\href@noop {} {\bibfield  {journal}
  {\bibinfo  {journal} {Ann. Probab.}\ }\textbf {\bibinfo {volume} {42}},\
  \bibinfo {pages} {1212} (\bibinfo {year} {2014})},\ \Eprint
  {https://arxiv.org/abs/1202.4398} {arXiv:1202.4398} \BibitemShut {NoStop}%
\bibitem [{\citenamefont {Corwin}\ and\ \citenamefont
  {Nica}(2017)}]{corwin2017intermediate}%
  \BibitemOpen
  \bibfield  {author} {\bibinfo {author} {\bibfnamefont {I.}~\bibnamefont
  {Corwin}}\ and\ \bibinfo {author} {\bibfnamefont {M.}~\bibnamefont {Nica}},\
  }\bibfield  {title} {\bibinfo {title} {Intermediate disorder limits for
  multi-layer semi-discrete directed polymers},\ }\href@noop {} {\bibfield
  {journal} {\bibinfo  {journal} {Electron. J. Probab}\ } (\bibinfo {year}
  {2017})},\ \Eprint {https://arxiv.org/abs/1609.00298} {arXiv:1609.00298}
  \BibitemShut {NoStop}%
\bibitem [{\citenamefont {Schulz}\ \emph {et~al.}(1988)\citenamefont {Schulz},
  \citenamefont {Villain}, \citenamefont {Br{\'e}zin},\ and\ \citenamefont
  {Orland}}]{schulz1988thermal}%
  \BibitemOpen
  \bibfield  {author} {\bibinfo {author} {\bibfnamefont {U.}~\bibnamefont
  {Schulz}}, \bibinfo {author} {\bibfnamefont {J.}~\bibnamefont {Villain}},
  \bibinfo {author} {\bibfnamefont {E.}~\bibnamefont {Br{\'e}zin}},\ and\
  \bibinfo {author} {\bibfnamefont {H.}~\bibnamefont {Orland}},\ }\bibfield
  {title} {\bibinfo {title} {Thermal fluctuations in some random field
  models},\ }\href@noop {} {\bibfield  {journal} {\bibinfo  {journal} {J. Stat.
  Phys.}\ }\textbf {\bibinfo {volume} {51}},\ \bibinfo {pages} {1} (\bibinfo
  {year} {1988})}\BibitemShut {NoStop}%
\bibitem [{\citenamefont {Fisher}\ and\ \citenamefont
  {Huse}(1991)}]{fisher1991directed}%
  \BibitemOpen
  \bibfield  {author} {\bibinfo {author} {\bibfnamefont {D.~S.}\ \bibnamefont
  {Fisher}}\ and\ \bibinfo {author} {\bibfnamefont {D.~A.}\ \bibnamefont
  {Huse}},\ }\bibfield  {title} {\bibinfo {title} {Directed paths in a random
  potential},\ }\href@noop {} {\bibfield  {journal} {\bibinfo  {journal} {Phys.
  Rev. B}\ }\textbf {\bibinfo {volume} {43}},\ \bibinfo {pages} {10728}
  (\bibinfo {year} {1991})}\BibitemShut {NoStop}%
\bibitem [{\citenamefont {Le~Doussal}\ \emph {et~al.}(2016)\citenamefont
  {Le~Doussal}, \citenamefont {Majumdar},\ and\ \citenamefont
  {Schehr}}]{le2016large}%
  \BibitemOpen
  \bibfield  {author} {\bibinfo {author} {\bibfnamefont {P.}~\bibnamefont
  {Le~Doussal}}, \bibinfo {author} {\bibfnamefont {S.~N.}\ \bibnamefont
  {Majumdar}},\ and\ \bibinfo {author} {\bibfnamefont {G.}~\bibnamefont
  {Schehr}},\ }\bibfield  {title} {\bibinfo {title} {Large deviations for the
  height in {1D Kardar-Parisi-Zhang} growth at late times},\ }\href@noop {}
  {\bibfield  {journal} {\bibinfo  {journal} {Europhys. Lett.}\ }\textbf
  {\bibinfo {volume} {113}},\ \bibinfo {pages} {60004} (\bibinfo {year}
  {2016})},\ \Eprint {https://arxiv.org/abs/1601.05957} {arXiv:1601.05957}
  \BibitemShut {NoStop}%
\bibitem [{\citenamefont {De~Luca}\ and\ \citenamefont
  {Le~Doussal}(2017)}]{de2017mutually}%
  \BibitemOpen
  \bibfield  {author} {\bibinfo {author} {\bibfnamefont {A.}~\bibnamefont
  {De~Luca}}\ and\ \bibinfo {author} {\bibfnamefont {P.}~\bibnamefont
  {Le~Doussal}},\ }\bibfield  {title} {\bibinfo {title} {Mutually avoiding
  paths in random media and largest eigenvalues of random matrices},\
  }\href@noop {} {\bibfield  {journal} {\bibinfo  {journal} {Phys. Rev. E}\
  }\textbf {\bibinfo {volume} {95}},\ \bibinfo {pages} {030103(R)} (\bibinfo
  {year} {2017})},\ \Eprint {https://arxiv.org/abs/1606.08509}
  {arXiv:1606.08509} \BibitemShut {NoStop}%
\bibitem [{\citenamefont {Shelton}\ and\ \citenamefont
  {Tsvelik}(1998)}]{shelton1998effective}%
  \BibitemOpen
  \bibfield  {author} {\bibinfo {author} {\bibfnamefont {D.~G.}\ \bibnamefont
  {Shelton}}\ and\ \bibinfo {author} {\bibfnamefont {A.~M.}\ \bibnamefont
  {Tsvelik}},\ }\bibfield  {title} {\bibinfo {title} {Effective theory for
  midgap states in doped spin-ladder and spin-peierls systems: {L}iouville
  quantum mechanics},\ }\href@noop {} {\bibfield  {journal} {\bibinfo
  {journal} {Phys. Rev. B}\ }\textbf {\bibinfo {volume} {57}},\ \bibinfo
  {pages} {14242} (\bibinfo {year} {1998})}\BibitemShut {NoStop}%
\bibitem [{\citenamefont {Nagar}\ \emph {et~al.}(2006)\citenamefont {Nagar},
  \citenamefont {Majumdar},\ and\ \citenamefont {Barma}}]{nagar2006strong}%
  \BibitemOpen
  \bibfield  {author} {\bibinfo {author} {\bibfnamefont {A.}~\bibnamefont
  {Nagar}}, \bibinfo {author} {\bibfnamefont {S.~N.}\ \bibnamefont
  {Majumdar}},\ and\ \bibinfo {author} {\bibfnamefont {M.}~\bibnamefont
  {Barma}},\ }\bibfield  {title} {\bibinfo {title} {Strong clustering of
  noninteracting, sliding passive scalars driven by fluctuating surfaces},\
  }\href@noop {} {\bibfield  {journal} {\bibinfo  {journal} {Phys. Rev. E}\
  }\textbf {\bibinfo {volume} {74}},\ \bibinfo {pages} {021124} (\bibinfo
  {year} {2006})}\BibitemShut {NoStop}%
\bibitem [{\citenamefont {Quinn}\ \emph {et~al.}(2015)\citenamefont {Quinn},
  \citenamefont {Cope}, \citenamefont {Bardarson},\ and\ \citenamefont
  {Ossipov}}]{quinn2015scaling}%
  \BibitemOpen
  \bibfield  {author} {\bibinfo {author} {\bibfnamefont {E.}~\bibnamefont
  {Quinn}}, \bibinfo {author} {\bibfnamefont {T.}~\bibnamefont {Cope}},
  \bibinfo {author} {\bibfnamefont {J.~H.}\ \bibnamefont {Bardarson}},\ and\
  \bibinfo {author} {\bibfnamefont {A.}~\bibnamefont {Ossipov}},\ }\bibfield
  {title} {\bibinfo {title} {Scaling of critical wave functions at topological
  anderson transitions in one dimension},\ }\href@noop {} {\bibfield  {journal}
  {\bibinfo  {journal} {Phys. Rev. B}\ }\textbf {\bibinfo {volume} {92}},\
  \bibinfo {pages} {104204} (\bibinfo {year} {2015})}\BibitemShut {NoStop}%
\bibitem [{\citenamefont {Andrews}\ \emph {et~al.}(1999)\citenamefont
  {Andrews}, \citenamefont {Askey},\ and\ \citenamefont
  {Roy}}]{andrews1999special}%
  \BibitemOpen
  \bibfield  {author} {\bibinfo {author} {\bibfnamefont {G.~E.}\ \bibnamefont
  {Andrews}}, \bibinfo {author} {\bibfnamefont {R.}~\bibnamefont {Askey}},\
  and\ \bibinfo {author} {\bibfnamefont {R.}~\bibnamefont {Roy}},\ }\href@noop
  {} {\emph {\bibinfo {title} {Special Functions}}}\ (\bibinfo  {publisher}
  {Cambridge University Press, Cambridge},\ \bibinfo {year} {1999})\BibitemShut
  {NoStop}%
\bibitem [{\citenamefont {Laloux}\ and\ \citenamefont
  {Le~Doussal}(1998)}]{laloux1998aging}%
  \BibitemOpen
  \bibfield  {author} {\bibinfo {author} {\bibfnamefont {L.}~\bibnamefont
  {Laloux}}\ and\ \bibinfo {author} {\bibfnamefont {P.}~\bibnamefont
  {Le~Doussal}},\ }\bibfield  {title} {\bibinfo {title} {Aging and diffusion in
  low dimensional environments},\ }\href@noop {} {\bibfield  {journal}
  {\bibinfo  {journal} {Phys. Rev. E}\ }\textbf {\bibinfo {volume} {57}},\
  \bibinfo {pages} {6296} (\bibinfo {year} {1998})}\BibitemShut {NoStop}%
\end{thebibliography}%
\end{document}